\documentclass[11pt, aps, prd, preprintnumbers, eqsecnum, nofootinbib]{revtex4-1}

\usepackage{epsf}
\usepackage{epsfig}
\usepackage{graphics}
\usepackage{graphicx}
\usepackage{amssymb}
\usepackage{mathtools}
\usepackage{color}
\usepackage{hyperref}
\usepackage{multirow}
\usepackage{feynmp-auto}
\usepackage[normalem]{ulem}

\usepackage{relsize}

\usepackage{amsmath, latexsym, amssymb, hyperref, graphicx, slashed, color, multirow}
\definecolor{nicered}{rgb}{0.0,.7,.3}
\definecolor{nicegreen}{rgb}{.1,.5,.1}
\definecolor{darkblue}{rgb}{0,.1,.9}
\hypersetup{colorlinks, citecolor=nicered ,linkcolor=darkblue, urlcolor=nicered}
\usepackage{cancel}

\def\lapp{\mathrel{\rlap{\raise.5ex\hbox{$<$}}
                    {\lower.5ex\hbox{$\sim$}}}}
\def\gapp{\mathrel{\rlap{\raise.5ex\hbox{$>$}}
                    {\lower.5ex\hbox{$\sim$}}}}

\newcommand{\beq}{\begin {equation}}  
\newcommand{\eeq}{\end   {equation}} 
\newcommand{\bea}{\begin {eqnarray}} 
\newcommand{\eea}{\end   {eqnarray}}  
\newcommand{\baa}{\begin {array}   } 
\newcommand{\eaa}{\end   {array}   }     
\newcommand{\bit}{\begin {itemize} }
\newcommand{\eit}{\end   {itemize} }
\newcommand{\be }{\begin {equation}} 
\newcommand{\ee }{\end   {equation}}
\newcommand{\nn }{\nonumber        }

\def\be{\begin{equation}}
\def\ee{\end{equation}}
\def\bea{\begin{eqnarray}}
\def\eea{\end{eqnarray}}

\def\beq{\begin{equation}}
\def\eeq{\end{equation}}
\newcommand{\beqa}{\begin{eqnarray}} 
\newcommand{\eeqa}{\end{eqnarray}}
\newcommand{\barr}{\begin{array}}
\newcommand{\earr}{\end{array}}

\def\gs{\mathrel{
   \rlap{\raise 0.511ex \hbox{$>$}}{\lower 0.511ex \hbox{$\sim$}}}}
\def\ls{\mathrel{
   \rlap{\raise 0.511ex \hbox{$<$}}{\lower 0.511ex \hbox{$\sim$}}}}

\newcommand{\braket}[1]{\left<#1\right>}

\newcommand{\met} {{E\!\!\!\!/_{\rm T}}}

\begin{document}

\preprint{UH-511-1288-2017}
\preprint{OSU-HEP-18-01}

\title{\Large Phenomenology of the Higgs sector of a Dimension-7 Neutrino Mass Generation Mechanism}

\author{\textbf{Tathagata Ghosh}$^{1,2}$ \footnote{Email: tghosh@hawaii.edu} , \textbf{Sudip Jana}$^{2,3}$ \footnote{Email: sudip.jana@okstate.edu} and  \textbf{S. Nandi}$^{2}$ \footnote{Email: s.nandi@okstate.edu}
}

\affiliation{$^1$Department of Physics \& Astronomy, University of Hawaii, Honolulu, HI 96822, USA
\\
$^2$Department of Physics and Oklahoma Center for High Energy Physics,
Oklahoma State University, Stillwater, OK 74078-3072, USA
\\
$^{3}$Theoretical Physics Department, Fermilab, Batavia, IL 60510, USA
}


\begin{abstract}

\section*{Abstract}

In this paper, we revisit the dimension-7 neutrino mass generation mechanism based on the addition of an isospin $3/2$ scalar quadruplet and two vector-like iso-triplet leptons to the standard model. We discuss the LHC phenomenology of the charged scalars of this model, complemented by the electroweak precision and lepton flavor violation constraints. We pay particular attention to the triply charged and doubly charged components. We focus on the same-sign-tri-lepton signatures originating from the triply-charged scalars and find a discovery reach of 600 - 950 GeV at 3 ab$^{-1}$ of integrated luminosity at the LHC. On the other hand, doubly charged Higgs has been an object of collider searches for a long time, and we show how the present bounds on its mass depend on the particle spectrum of the theory. Strong constraint on the model parameter space can arise from the measured decay rate of the Standard Model Higgs to a pair of photons as well.

\end{abstract}


\maketitle

\newpage
\section{Introduction}
\label{sec:intro}

The remarkable discovery of the 125 GeV scalar particle by CMS and ATLAS collaborations~\cite{:2012gk, :2012gu} is the crowning achievement of the Run-I of the Large Hadron Collider (LHC). The data collected by the LHC experiments so far indicate that the discovered particle is the final piece of the standard model (SM) -- the Higgs boson, which provides mass to the fermions and gauge bosons of the SM via spontaneous symmetry breaking. At the same time, any signature beyond the SM remains elusive at the LHC. Notwithstanding many successes of the SM, it fails to answer many critical questions. Hence, the pursuit of unearthing signals of new physics is at the forefront of particle physics experiments for many decades.  

One of the most robust evidence that points out to an important inadequacy of the SM is the existence of non-zero tiny masses of neutrinos. The neutrinos are the only class of fermion within the SM, whose mass cannot be generated by the Higgs mechanism, due to the absence of right-handed neutrinos. However, various neutrino oscillation experiments have long established the fact that not only neutrinos possess small masses [$\mathcal{O}(0.01 - 0.1$ eV)], but also they mix between flavors. In addition, the Planck collaboration constrains the sum of neutrino masses to be $\sum m_i \lesssim 0.23$ eV~\cite{Ade:2013zuv}, which again emphasizes the fact that neutrino masses are many order of magnitude smaller than their charged lepton counterparts. This drastic departure of neutrino masses and mixings from charged leptons poses a fundamental question, how such tiny neutrino masses are generated?  

The simplest way to achieve that goal is via an effective dimension-5 operator, $LLHH/M$~\cite{Weinberg:1979sa}, where $H$ is the SM Higgs doublet, $L$ is the left-handed lepton doublet, and $M$ is the scale of new physics. Under this mechanism, neutrinos acquire a mass $m_{\nu} \sim v^2/M$, with $v$ being the vacuum expectation value (VEV) of $H$.  There have been many realizations of such dimension-5 operator in the literature, namely, Type-I see-saw~\cite{Type-I}, Type-II see-saw~\cite{Type-II}, Type-III see-saw~\cite{Type-III}, loop induced~\cite{Zee} \emph{etc.}, with all new particles are at the mass order $M$. From the above formula of neutrino masses, one can notice that neutrino oscillation data, combined with cosmological constraint, will force $M \sim \mathcal{O}(10^{14} - 10^{15})$ GeV with $\mathcal{O}(1)$ Yukawa couplings. Alternatively one needs an unusually small Yukawa coupling, $Y_{\nu} \sim 10^{-6}$ for TeV scale $M$. In either case, the LHC is unlikely to probe any signature of such particles. Instead, we focus on a model proposed by Babu, Nandi, and Tavartkiladze (BNT)~\cite{Babu:2009aq}, where neutrino masses are generated at tree level by an effective dimension-7 operator, $LLHH(H^{\dagger}H)/M^3$, resulting in a neutrino mass formula, $m_{\nu} \sim v^4/M^3$. Owing to the increased suppression factor $M^3$ in the denominator, one can easily lower the scale of new physics in this model to TeV without introducing minuscule Yukawa couplings. The above model contains two vector-like lepton triplets ($\Sigma_{1,2}$) and an isospin $\dfrac{3}{•2}$ scalar quadruplet ($\Delta$) on top of the SM fields. Hence, this model predicts striking same-sign multi-lepton signatures at the LHC due to the presence of multi-charged scalars and vector-like leptons.

The goal of our paper is twofold. First, we present a detailed analysis of electroweak precision test (EWPT) constraints on the Higgs spectrum of the model for the first time. Next, we investigate the latest LHC and lepton flavor violation (LFV) bounds on the Higgs sector, not ruled out by the  EWPT, and further project future LHC reach of the triply-charged Higgs boson for definitive validation/falsification of the model. 

Refs.~\cite{earlylhc, Jana, Bhatta} have studied the BNT model in the context of the LHC and dark matter previously. Nonetheless, the LHC experiments have accumulated a significant volume of data since then, and a revision of those constraints from the new data is warranted at this point. In addition, a loop-induced dimension-5 operator is also present in the model, which contributes to the neutrino mass generation at a comparable rate w.r.t the dimension-7 operator for $M_{\Sigma} \gtrsim \mathcal{O}$(TeV). Although, the existence of this dimension-5 operator is well-known ~\cite{Babu:2009aq,earlylhc}, the impact of their interplay with the dimension-7 operator on the LHC searches were not taken into account in previous studies at a quantitative level. 

In addition, we would like to point out that the LHC experiments traditionally show their bound on doubly-charged Higgs particle mass in same-sign dilepton final states assuming a 100$\%$ branching ratio (BR) for particular flavor combinations. Instead, we reinterpret their results using realistic benchmark points (BP), consistent with neutrino oscillation data and show that the constraints on doubly charged Higgs mass can be relaxed. Also, we demonstrate that for our realistic BPs, the proper decay length of doubly and triply charged Higgs bosons are quite large in regions of the parameter space and discuss when they will be beyond the scope of prompt lepton searches performed at the LHC. 

LFV constraints on the model were previously discussed in Ref.~\cite{Liao:2010rx} for very light $\Sigma_{1,2}$ ($\sim 200$ GeV) and they did not take into account the contribution of multi-charged scalars on LFV processes. In contrast, we derive relevant LFV constraints due to light scalars ($M_{\Delta} \lesssim 1$ TeV). In our chosen benchmark scenarios $\Sigma_{1,2}$ are much heavier ($\sim$ 5 TeV) than $H, \Delta$, which in turn force their contribution to LFV processes negligible.  Using the current most stringent bound by the MEG Collaboration~\cite{Adam:2013mnn}, a lower bound on induced VEV $v_{\Delta}$ as a function of mass $M_{\Delta^{}}$ has been derived.

Lastly, we search for triply-charged Higgs boson at the LHC in same-sign three leptons final state. A potential discovery of $\Delta^{\pm \pm \pm}$ at the LHC will shed some light on the possible mechanism of neutrino mass generation.

The paper is organized as follows. In Section~\ref{sec:model} we present a brief overview of the BNT model and the neutrino mass generation mechanisms within the model, along with our choice of neutrino oscillation parameters for subsequent calculations. In Sections~\ref{sec:EWPT} and \ref{sec:LFV} we discuss EWPT and LFV constraints, respectively, on the Higgs sector of the model. Updated constraints form various LHC searches relevant to the Higgs sector of this model are discussed in Section~\ref{sec:collider}. We also outline the projected reach at the LHC for triply-charged Higgs in the same section, in association with detailed discussion on their relevant production and decay mechanisms. Finally, we conclude in Section~\ref{sec:conclusions}.

\section{Model and Formalism}
\label{sec:model}

In this section, we present a brief overview of the BNT model~\cite{Babu:2009aq}. The chief goal of the model is to develop light neutrino masses with new physics at TeV scale without introducing unnaturally small Yukawa couplings or fine-tuned cancellations. The BNT model is based on the SM symmetry group $SU(3)_{C}\times SU(2)_{L}\times U(1)_{Y}$. The enlarged particle content of the model includes an isospin $\dfrac{3}{2}$ scalar quadruplet, $\Delta$, and a pair of vector-like fermion triplets, $\Sigma_{1,2}$. We use $H$ to denote the SM-like Higgs doublet. The particle contents along with their quantum numbers are shown in the Table~\ref{Table1} below.

\begin{table}[htp]
\begin{center}
\begin{tabular}{|c|c|c|}
\hline 
&$SU(3)_{C} \times SU(2)_{L} \times U(1)_{Y}$\\\hline
\small Matter &${\begin{pmatrix} u \\ d \end{pmatrix}}_L\sim(3,2,\frac{1}{3}), u_R\sim (3,1,\frac{4}{3}), d_R \sim (3,1,-\frac{2}{3})$ \\
&$ {\begin{pmatrix} \nu_e \\ e \end{pmatrix}}_L\sim (1,2,-1), e_R\sim (1,1,-2)$ \\&$ \Sigma_2 \equiv {\begin{pmatrix} \Sigma^{++}_2 \\ \Sigma^{+}_2 \\ \Sigma^{0}_2 \end{pmatrix}}\sim (1,3,2)$, $\Sigma_1 \equiv {\begin{pmatrix} {\Sigma}^{++}_1 \\ {\Sigma}^{+}_1 \\ {\Sigma}^{0}_1 \end{pmatrix}}\sim (1,3,2)$ \\
\hline
\small Gauge & $G^\mu_{a,a=1-8}, A^\mu_{i, i=1-3}, B^\mu$ \\
\hline
\small Higgs & ${H \equiv \begin{pmatrix} \phi^{+} \\ \phi^{0} \end{pmatrix}}\sim(1,2,1)$, $\Delta \equiv {\begin{pmatrix} \Delta^{+++} \\ \Delta^{++} \\ \Delta^{+} \\\Delta^{0}\end{pmatrix}}\sim(1,4,3)$\\
\hline
\end{tabular}
\\
\end{center}
\caption{Matter, gauge and Higgs contents of the BNT model.}
\label{Table1}
\end{table}

\subsection{Higgs sector of the model}

The scalar kinetic and potential terms of the model is given by :
\begin{eqnarray}
{\cal{L}}^{Kin}_{Scalar}  = 
(D^{\mu} \Delta )^{\dag }(D_{\mu} \Delta) + 
(D^{\mu} H )^{\dag }(D_{\mu} H) + V ( H, \Delta),
\end{eqnarray}
with the covariant derivatives 
\begin{equation}
\begin{split}
D_{\mu} H = \left(\partial_{\mu} -i g \vec{\tau} {\bf .}  \vec{W}_{\mu} - i g' \frac{Y}{2} B_{\mu}\right) H ,
\\
D_{\mu} \Delta = \left(\partial_{\mu} -i g \vec{T} {\bf .}  \vec{W}_{\mu} - i g' \frac{Y}{2} B_{\mu}\right) \Delta ,
\end{split}
\end{equation} 
where $\vec{\tau}$ are standard Pauli matrices and $\vec{T}$ are $SU (2)$ generators in the isospin $\dfrac{3}{2}$ representation~\cite{earlylhc}. The interactions of the new scalar field $\Delta$ with the gauge bosons originate from the above term.   
The most general renormalizable  scalar potential involving the Higgs fields of the model is given by,
\begin{equation}
\begin{split}
V ( H, \Delta)= - \mu_H^{2}H^{\dagger}H + \mu_\Delta^{2}\Delta^{\dagger}\Delta +   {\lambda_{1}}(H^{\dagger}H)^{2} + {\lambda_{2}}(\Delta^{\dagger}\Delta)^{2} \\+  \lambda_{3}(H^{\dagger}H)(\Delta^{\dagger}\Delta) + \lambda_{4}(H^{\dagger}\tau_{a}H)(\Delta^{\dagger}T_{a}\Delta)  + \lbrace\lambda_{5}H^{3}\Delta^{\star} + h.c. \rbrace .
\end{split}
\label{V_H_Delta}
\end{equation}


We assume $\mu_{\Delta}^2 > 0$ and thus $\Delta$ can not initiate any spontaneous symmetry breaking. Hence similar to the SM, the electroweak (EW) symmetry is broken spontaneously once the Higgs doublet, $H$, acquires a VEV, $\braket{H} = \dfrac{v_{H}}{\sqrt{2}}$. Interestingly, even with a positive ${\mu_{\Delta}}^2$, due to the presence of the $\lambda_5$ term in the potential the neutral component of $\Delta$ acquires an induced VEV at the tree level, 
\beq
 \braket{\Delta} = \dfrac{v_{\Delta}}{\sqrt{2}} = -\dfrac{ \lambda_5 v_H^3} {2 \sqrt{2} M_{\Delta}^2}.
\label{vDelta}
\eeq 
However, $v_{\Delta}$ suffers from strong bounds coming from the EW $\rho$ parameter. In the BNT model the analytical form of the $\rho$ parameter is  $\rho \approx (1-6 v_{\Delta}^{2}/v_{H}^{2})$. In order to  satisfy the experimentally observed value, $\rho= 1.00037^{+0.00023}_{-0.00023}$ \cite{PDG} at $2 \sigma$, $v_\Delta $ is constrained to be $v_{\Delta} \lesssim 1 $ GeV. In the above equation $M_{\Delta}$ denotes the mass of the neutral scalar $\Delta^0$, which can be expressed as
\beq
M^2_{\Delta} = \mu^2_{\Delta} + \dfrac{v^2_H}{8} (4 \lambda_3 + 3 \lambda_4).
\eeq 
On the other hand, masses of other members of $\Delta$ quadruplet are given by
\begin{equation}
M_{i}^{2} = M_{\Delta}^{2} - Q_{i}\frac{\lambda_{4}}{4}v_{H}^{2},
\label{spectrum}
\end{equation}
where $Q_{i}$ is the (non-negative) electric charge of the respective field. We neglect small corrections proportional to $v_{\Delta}$ in the above expressions since $v_{\Delta} \ll v_H$. The mass gaps are equally spaced. Also, two mass orderings are possible here. For $\lambda_{4}$ positive,
we have the ordering $M_{\Delta^{+++}} < M_{\Delta^{++}} < M_{\Delta^{+}} < M_{\Delta^{0}}$ and for $\lambda_{4}$ negative, we have the ordering $M_{\Delta^{+++}} > M_{\Delta^{++}} > M_{\Delta^{+}} > M_{\Delta^{0}}$. Clearly, large mass-gaps between the constituents of the quadruplet can be developed by choosing a large value of $\lambda_4$ that is allowed by perturbativity. These mass-splittings are integral part of our present analysis. We shall see in subsequent sections that not only they play a pivotal role in EW precision constraints but also LHC mass-reaches are highly dependent on them.

\subsection{Generation of neutrino mass}
\label{sec:Nu_Mass}

\begin{figure}[!htp]
\includegraphics[scale=0.3]{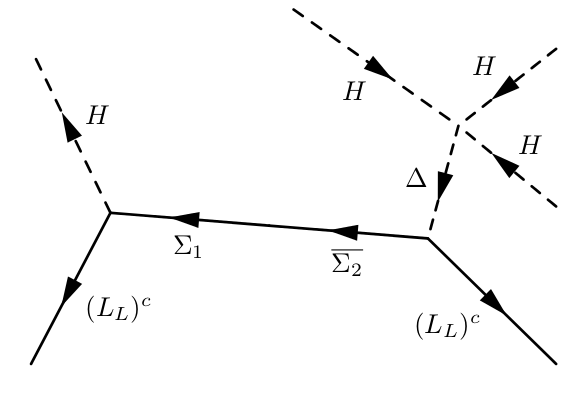}
\caption{Tree level diagram that generates dimension-7 operator for neutrino mass.}
\label{fig:d7}
\end{figure}

Neutrino masses arise in the model from the renormalizable Lagrangian~\cite{Babu:2009aq}
\beq 
{\cal L_{\nu- {\rm mass}}} = Y_i \overline{{L_{iL}}^c} H^* \Sigma_1 + Y'_i \overline{\Sigma_2} \Delta L_{iL} + M_{\Sigma} \overline{\Sigma_2} \Sigma_1+ h.c., 
\label{Lsig} 
\eeq
where $Y_i,~{Y'}_i$ are Yukawa couplings and $i$ is the generation index. Integrating out the $\Sigma_{1,2}$ fermions, one obtains an effective dimension-5 neutrino mass
operator 
\bea 
 {\cal L}_{\rm eff} = -{(Y_i Y'_j +  Y_j Y'_i) \overline{{L_{iL}}^c} L_{jL} H^* \Delta \over M_\Sigma} + h.c. \, .
\label{Leff}
\eea 
The tree level diagram generating this operator is shown in Fig.~\ref{fig:d7} \cite{Babu:2009aq}. The detailed structure of the Yukawa interactions are given in \cite{earlylhc}.

\begin{figure}[!htp]
\includegraphics[scale=0.3]{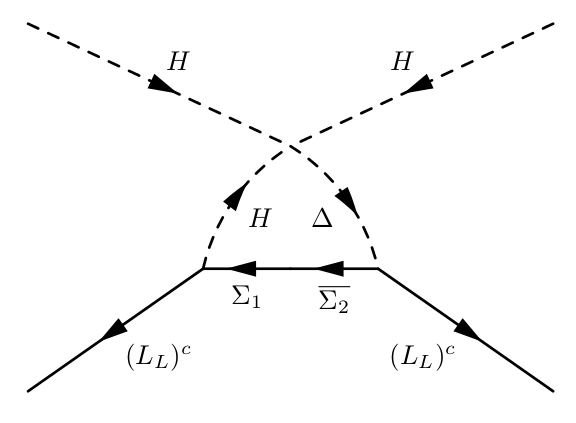}
\caption{Loop level diagram that generates dimension-5 operator for neutrino mass.}
\label{fig:d5}
\end{figure}

We have already seen from the analysis of the Higgs potential that $\Delta^0$ acquires an induced VEV $v_{\Delta} = - \lambda_5 v^3 /2 M_{\Delta}^2$.  When this value is substituted in Eq.~\ref{Leff}, to the leading order, we obtain the neutrino masses at tree level, $(m_{\nu})^{\text{tree}}$, which can be written as \cite{Babu:2009aq},
\beqa 
(m_{\nu})_{ij}^{\text{tree}}  =  -
\frac{(Y_i Y_j^{\prime} +  {Y_i^\prime} Y_j) v_\Delta v_{H}}{M_\Sigma} =
\frac{\lambda_5(Y_i Y_j^\prime +  {Y_i^\prime} Y_j) v_{H}^4}{2 M_{\Sigma}
M_{\Delta^0}^2 }.
\label{eq:mnu2} 
\eeqa
This provides us with a tree level dimension-7 neutrino mass generation mechanism. Clearly the particle content of the model prevents it from developing a dimension-5 operator at the tree level. Nevertheless, there is no mechanism present in the model that prevents generating a dimension-5 operator at the loop level. For the diagram that generates the loop-level dimension-5 operator we refer the reader to Fig.~\ref{fig:d5}~\cite{earlylhc,Jana}. The loop contribution to the neutrino mass, $(m_{\nu})^{\text{loop}}$, can be computed at the leading order [$\mathcal{O}(v_H^2)$] as \cite{earlylhc} : 
\begin{equation} 
\label{m_loop}
(m_{\nu})_{ij}^{\text{loop}}=\frac{\left(3+\sqrt{3}\right) \lambda _5 v_H^2 M_{\Sigma } 
\left(Y_i Y_j^{'}+ Y_i^{'} Y_j \right)}{32 \pi ^2 \left(M_{\Delta }^2-M_H^2\right)}
\left(
\frac{M_{\Delta}^2 \log \left(\frac{M_{\Sigma }^2}{M_{\Delta }^2}\right)} {M_{\Sigma }^2-M_{\Delta }^2}-
\frac{M_H^2 \log \left( \frac{M_{\Sigma }^2}{M_H^2}\right)} {M_{\Sigma}^2-M_H^2}
\right).
\end{equation}

\begin{figure}[!htp]
\includegraphics[width=3.2 in, height=3 in]{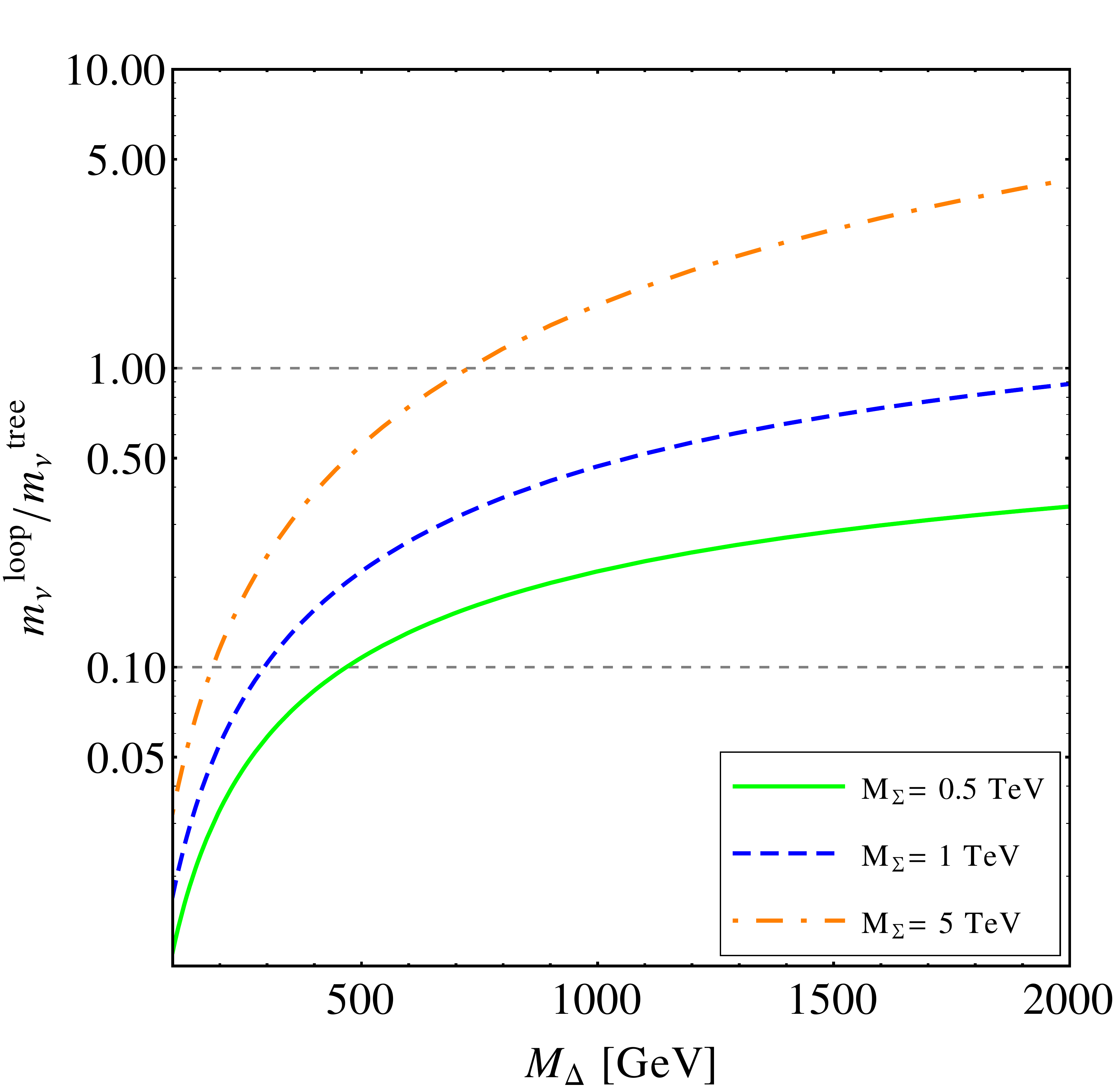}
\caption{$(m_{\nu})_{ij}^{\text{loop}}/(m_{\nu})_{ij}^{\text{tree}}$ as a function of $M_{\Delta}$ for different values of $M_{\Sigma}$.}
\label{fig:loopVtree}
\end{figure}

It is important to examine what are the relevant masses $M_\Delta$ and $M_{\Sigma}$ that determine the relative contribution of the loop level dimension-5 operator in comparison with the tree level dimension-7 operator. In Fig.~\ref{fig:loopVtree} we plot $(m_{\nu})_{ij}^{\text{loop}}/(m_{\nu})_{ij}^{\text{tree}}$ as a function of $M_{\Delta}$ for three different values of $M_{\Sigma}$. We should mention here that both $\Delta^0$ and $\Delta^{\pm}$ enters the loop level dimension-5 operator of Eq.~\ref{m_loop}~\cite{earlylhc} but they are assumed to be the same in the computation of Fig.~\ref{fig:loopVtree} for simplicity. For $M_{\Sigma} = 0.5, \, 1 $ TeV $(m_{\nu})_{ij}^{\text{tree}}$ dominates over $(m_{\nu})_{ij}^{\text{loop}}$ in the range of $M_{\Delta} \lesssim 2$ TeV. In contrast, for $M_{\Sigma} = 5$ TeV, $(m_{\nu})_{ij}^{\text{loop}}$ catches up with $(m_{\nu})_{ij}^{\text{tree}}$ at $M_{\Delta} \sim 0.75 $ TeV. Thus, it is desirable to set $M_{\Sigma} \lesssim 1$ TeV to test purely dimension-7 generation of neutrino mass at the LHC. However, such a choice of the parameter will significantly increase the difficulty of signal simulation for LHC searches. This is due to the fact that in the aforesaid scenario we shall not be able to integrate out $M_{\Sigma}$ and a very careful and tedious treatment is needed regarding the charged lepton mass matrix without any significant phenomenological gain at the LHC. On the other hand, for $M_{\Sigma} \sim 5$ TeV we can avoid this complexity and perform relevant collider simulations with ease. In addition, the range of $M_{\Delta}$ that is accessible for the ongoing run of the LHC, as will be shown in Section~\ref{sec:H+++_LHC}, dimension-7 operator is still dominant with $M_{\Sigma} \sim 5$ TeV. Also, we should emphasize here that our main goal in this paper is to study multiple aspects of the Higgs sector of the BNT model. Various Higgs analyses performed in this paper are, to a large extent, not sensitive to dimension-7 or dimension-5 neutrino mass generation operators. They can only alter the leptonic decay BRs of Higgs bosons marginally and will not qualitatively impact the important conclusions of this study. Henceforth, we set $M_{\Sigma} = 5$ TeV for the rest of the paper.

We conclude this subsection by a brief remark on possible extensions of the BNT model, available in the literature, that can potentially prevent the appearance of a dimension-5 operator via loops. One way to achieve that is to impose a symmetry that forbids the generation of neutrino masses at dimensions $d < 7$. In effective field theory language the dimension-5 and dimension-7 operators can be written as, $\mathcal{O}^5 = LLHH$ and $\mathcal{O}^7 = LLHH (H^{\dagger} H)$, respectively. Similarly, one can expand it further by adding higher powers of the combination $(H^{\dagger} H)$ to generate $d > 7$ dimension operators. The shortcoming of this approach is that $(H^{\dagger} H)$ is a singlet under any symmetry and does not carry any charge. Thus, one can not avert the problem and all powers of $(H^{\dagger} H)$ is allowed. Therefore, we need to add new Higgs field(s) to the theory and charge it under some $U(1)$ or discrete symmetry that allows dimension-7 operator but not any operator of lower dimensions.

In the context of the BNT model, one can add another Higgs doublet to the field, similar to the Two Higgs Doublet Model~\cite{Giudice, Carena}, leading to the following effective Lagrangian in the n-th dimension
\beq
\mathcal{L}^{d=2n+5}_{\rm \text{eff}} = \dfrac{1}{\Lambda^{d-4}_{\text{NP}}} \, (LL H_u H_u) \, (H_u H_d)^n, \, \, \, \, \, \, \, n=1,2,3,...\, \, \, . 
\label{Leff_2HDM}
\eeq
The simplest pure dimension-7 model can be constructed from this effective Lagrangian by introducing a $Z_5$ symmetry and assigning the following charges~\cite{Ota},
\beq
q_{H_u}=0, \, q_{H_d}=3, \, q_L =1, \, q_E=1, \, q_Q=0, \, q_U=0, \, q_D=2. 
\eeq
One can also attain the same goal by using one Higgs doublet only and a singlet scalar~\cite{Ota}. A more complex solution is realized within the next-to-minimal SUSY standard model, which contain two Higgs doublets and a singlet~\cite{Ilia}. Finally, if one is interested in pure dimension-7 loop induced neutrino mass generation, he/she can take a look at at Ref.~\cite{Cepedello:2017eqf}.

\subsection{Neutrino mass hierarchies and Yukawa couplings}
\label{sec:Nu_Mass_fit}

Next, we discuss the benchmark Yukawa couplings we used in our paper, consistent with all neutrino mass and mixing data. In a basis, where the charged lepton mass matrix is diagonal, the light neutrino matrix ($m_{\nu}$) can be diagonalized as,
\beq
(m_{\nu})_{}^{\text{diag}} = diag (m_1,m_2,m_3) = U^T_{PMNS} \, m_{\nu} \, U_{PMNS} ,
\label{Mnu_diag}
\eeq
where $ U_{PMNS}$ is the neutrino mixing matrix. $ U_{PMNS}$ is  parametrized by three mixing angles $\theta_{ij} \, (ij=12,13,23)$, one Dirac phase ($\delta$) and two Majorana phases ($\alpha_{1,2}$) as 
\beq
 U_{PMNS} = \begin{pmatrix}
 c_{12} c_{13} & s_{12} c_{13} & s_{13} e^{-i \delta}\\
 - c_{23} s_{12} - s_{23} s_{13} c_{12} e^{i \delta} &   c_{23} c_{12} - s_{23} s_{13} s_{12} e^{i \delta} & s_{23} c_{13}\\
 s_{23} s_{12} - c_{23} s_{13} c_{12} e^{i \delta} &   - s_{23} c_{12} - c_{23} s_{13} s_{12} e^{i \delta} & c_{23} c_{13}
 \end{pmatrix} \, P,
\label{U_PMNS}
\eeq 
with $c_{ij} \, (s_{ij}) = \cos \theta_{ij} \, (\sin \theta_{ij})$ and $P = diag(1, e^{i \alpha_1}, e^{i \alpha_2})$.

In the BNT model, due to the presence of two vector-like lepton triplets, the neutral lepton mass matrix is $5 \times 5$ with rank 4~\cite{earlylhc}. Therefore, the neutrino mass spectrum consists of one massless neutrino, two massive light neutrinos, and two heavy neutrinos, which are nearly degenerate. Since the lightest neutrino in the model is massless, we can express the mass eigenvalues of two light massive neutrinos in terms of the solar and atmospheric mass-squared differences as
\begin{itemize}
\item Normal Hierarchy (NH) : $m_1 \ll m_2 \approx m_3$\\
      \beq
      m_1 =0, \, \, \, \, \, m_2 = \sqrt{\Delta m^2_{21}}, \, \, \, \, \, m_3=\sqrt{\Delta m^2_{32}+\Delta m^2_{21}}\, ,
      \eeq
\item Inverted Hierarchy (IH) : $m_3 \ll m_1 \approx m_2$\\
      \beq
      m_3 =0, \, \, \, \, \, m_1 = \sqrt{\Delta m^2_{13}}, \, \, \, \, \, m_2=\sqrt{\Delta m^2_{13}+\Delta m^2_{21}}\, ,
      \eeq
\end{itemize}
where $\Delta m^2_{ij} \equiv m^2_j - m^2_i$. The best-fit values and $3\sigma$ ranges of oscillation parameters, extracted from~\cite{NuOsc_Global}, are tabulated in Table~\ref{tab:nuOsc}. We also show, in the same table, the benchmark values of these parameters that we shall use for the rest of the paper. We set all $CP$-violating phases to be 0, for simplicity, in our analysis.    

\begin{table}[!htp]
\begin{tabular}{|c| c| c|c| }
\hline
Oscillation parameter & Best-fit & $3 \sigma$ range & Our benchmark
\\
\hline
$\Delta m^2_{21} \, [10^{-5}$ eV$^2$] & 7.50 & $7.02 \rightarrow 8.09$ & 7.50 \\
\hline
\multirow{2}{*}{$\Delta m^2_{3l}\, [10^{-3}$ eV$^2$]} & 2.457 [NH] & $2.317 \rightarrow 2.607$ [NH] & {2.50}\\ 
     & -2.449 [IH] & $-2.590 \rightarrow -2.307$ [IH] & -2.50 \\
\hline 
$\sin^2 \theta_{12}$ & 0.304 & $0.270 \rightarrow 0.344$ & 0.320 \\
\hline
\multirow{2}{*}{$\sin^2 \theta_{23}$} & 0.452 [NH] & $0.382 \rightarrow 0.643$ [NH] & \multirow{2}{*}{0.500} \\
     & 0.579 [IH]  & $0.389 \rightarrow 0.644$ [IH] & \\
\hline
\multirow{2}{*}{$\sin^2 \theta_{13}$} & 0.0218 [NH] &  $0.0186 \rightarrow 0.0250$ [NH] & \multirow{2}{*}{0.0250} \\
     & 0.0219 [IH] & $0.0188 \rightarrow 0.0251$ [IH] & \\
\hline
\multirow{2}{*}{$\delta$} & 0.85$\pi$ [NH] & \multirow{2}{*}{$0 \rightarrow 2\pi$} & \multirow{2}{*}{0} \\
     & 0.71$\pi$ [IH]  &  & \\
\hline

\end{tabular}
\caption{The best-fit values and $3 \sigma$ ranges of neutrino oscillation parameters, extracted from the global analysis of~\cite{NuOsc_Global}. We show our choice of these parameters, used for the rest of the paper, in the last column. Please note that $\Delta m^2_{3l} \equiv \Delta m^2_{32} > 0$ for NH and $\Delta m^2_{3l} \equiv \Delta m^2_{31} < 0$ for IH.   }
\label{tab:nuOsc}
\end{table}

\section{Electroweak precision tests}
\label{sec:EWPT}

In this section we put our effective theory, after integrating out $M_{\Sigma}$, under the microscope of high precision EW observables measured at the LEP and SLC. For heavy $\Sigma_{1,2}$ the Higgs quadruplet, $\Delta$, only contributes to processes that can distort successful EW predictions of the SM. The principal effect of the $SU(2)_L$ quadruplet on the EW observables enter by means of oblique parameters, which are nothing but the gauge boson vaccuum polarization correlations~\cite{oblique}. The oblique parameters are parametrized by three independent parameters $S, \, T$ and $U$ defined as~\cite{oblique}
\bea
\alpha S & \equiv & 4 e^2 [\Pi'_{33}(0) - \Pi'_{3Q}(0)]
\\ \nn
\alpha T & \equiv & \dfrac{e^2}{s_W^2 c_W^2 M_Z^2} [\Pi_{11}(0) - \Pi_{33}(0)]
\\ \nn
\alpha U & \equiv & 4 e^2 [\Pi'_{11}(0) - \Pi'_{33}(0)],
\eea
where $\alpha$ is the fine structure constant and $s_W \, (c_W)$ are sine (cosine) of the EW mixing angle. $\Pi_{XY} \, (X, \, Y = 1, \, 3, \, Q )$ represents the vacuum polarization amplitudes and $\Pi'_{XY} = \dfrac{d}{dq^2} \Pi_{XY} (q^2)$.

Here, we make use of the general formulae of Ref.~\cite{Lavoura:1993nq} to the quadruplet. Two important assumptions made in the calculation of Ref.~~\cite{Lavoura:1993nq} are --(i) the complex scalar multiplet of interest does not acquire any VEV, and (ii) it's members do not mix with themselves or any other scalar. In the BNT model we have already seen that $v_{\Delta} \ll v_H$ is a necessary condition from EW $\rho$ parameter. So, we can safely work in a $v_{\Delta} \rightarrow 0$ paradigm. In addition, the mixing terms between the SM-like Higgs $h \approx \text{Re}(\phi^0)$ and $\text{Re}(\Delta^{0})$ are proportional to either $v_{\Delta}$ or $\lambda_5$. For $v_{\Delta} \ll v_H$ Eq.~\ref{vDelta} tells us that we require $\bigg|\dfrac{\lambda_5}{v_{\Delta}}\bigg| \ll 1 $ GeV$^{-1}$ to achieve $\mathcal{O}$(100-1000 GeV) mass for $\Delta$. Hence, applying the generic treatment of Ref.~\cite{Lavoura:1993nq} is apt for our study.

\begin{figure}[!htp]
\includegraphics[scale=0.25]{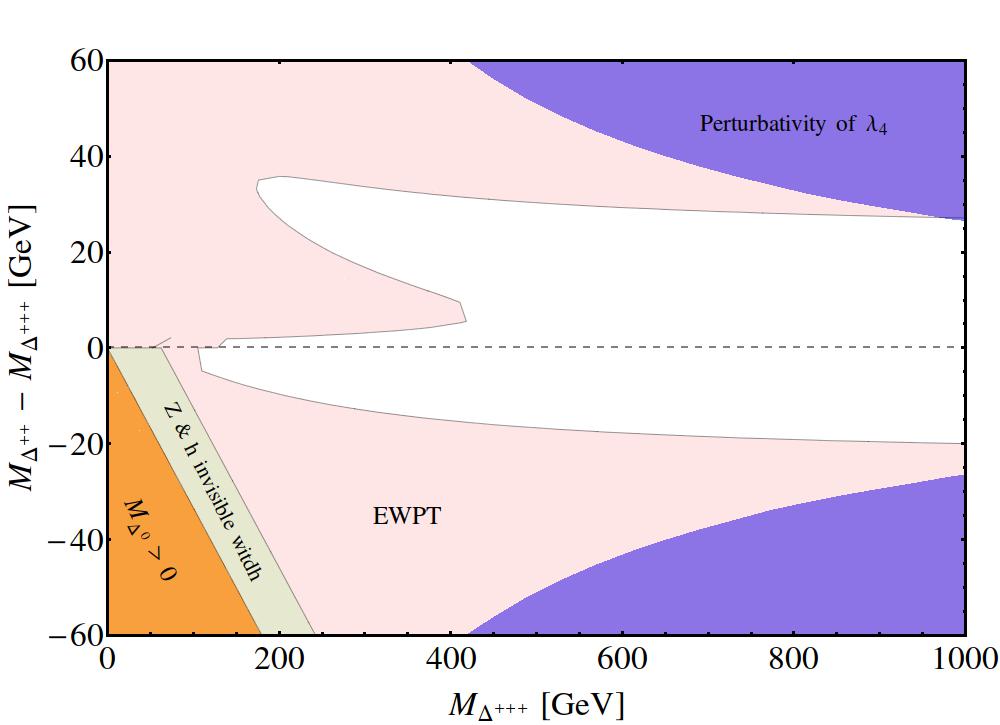}
\caption{Summary of few experimental and theoretical constraints in the $M_{\Delta^{++}} - M_{\Delta^{+++}}$ parameter space. The pink contour excluded by EWPT at $95\%$ C.L., the green region bounded by the measured $Z$ and $h$ invisible widths. On the other hand, the blue and orange regions are excluded by perturbativity of $\lambda_4 \, (\leq \sqrt{4 \pi})$ and positivity of $M_{\Delta^0}$ respectively. }
\label{fig:EWPT}
\end{figure}

The constraints on $S, \, T$ and $U$ are extracted from the global fit of the EW precision data. We use the fit results from the {\tt GFitter} collaboration~\cite{Baak:2012kk} for the reference SM parameters $m_h = 126$ GeV and $m_t=173$ GeV. The latest constraints are 
\beq
S_{exp} = 0.03 \pm 0.10, \, \,\,\, T_{exp}=0.05\pm0.12, \,\,\,\, U_{exp}=0.03\pm 0.10, 
\label{STU_value}
\eeq
with relative correlations
\beq
\rho_{ST} = 0.89, \, \,\,\, \rho_{TU}=-0.83, \,\,\,\, \rho_{SU}=-0.54 \, .
\label{STU_corr}
\eeq
Using the above experimental values we constrain $M_{\Delta^{\pm \pm \pm}}$ and $\lambda_4$ by means of a two parameter $\chi^2$ analysis. In Fig.~\ref{fig:EWPT} we show $95 \%$ C.L. limits EW precision test (EWPT) bounds on $\Delta M - M_{\Delta^{\pm \pm \pm}}$ plane by the pink shaded region, with $\Delta M \equiv M_{\Delta^{\pm \pm }} - M_{\Delta^{\pm \pm \pm}} \approx  \dfrac{\lambda_4}{8}\dfrac{v_H^2}{ M_{\Delta^{\pm \pm \pm}}}$. Additionally, we also present limits from perturbativity of $\lambda_4 \, (\leq \sqrt{4 \pi})$ by the blue shaded region in Fig.~\ref{fig:EWPT}. For large negative value of $\Delta M$, lighter members of the quadruplet will have negative masses. We constrain such scenarios by the orange shaded region. Also for $\Delta M < 0$ scenarios $Z$ or $h$ bosons can decay to neutral quadruplet members (which are the lightest) in pair and which will in turn decay to a pair of neutrinos resulting in large invisible decay width of $Z$ and $h$ boson measured at the LEP and LHC respectively. The constrain on the above cases from the measured $Z$ and $h$ invisible decay widths are shown by the green shaded region in Fig.~\ref{fig:EWPT}.

From Fig.~\ref{fig:EWPT} we can infer that at low $M_{\Delta^{\pm \pm \pm}}$ the bounds are dominated by the $S$ parameter. For larger $M_{\Delta^{\pm \pm \pm}} \gtrsim 200$ GeV the limits form $T$ parameter takes over but for very large value of $M_{\Delta^{\pm \pm \pm}} > 1$ TeV the perturbativity limit of $\lambda_4$ impose the most stringent constraint on $\Delta M$. One important observation from the  above figure is that EWPT limit the mass-splitting of the quadruplets to be $\lesssim 30$ GeV. This poses serious problems  for collider searches of $\Delta^{\pm \pm \pm}$ (when it is the heaviest member of the quadruplet) or $\Delta^{\pm \pm}$ (all cases). For $\Delta M \gtrsim 10$ GeV cascade decay always dominates and with $\Delta M \lesssim 30$ GeV the decay products will be too soft to pass LHC thresholds, as we shall demonstrate in Section~\ref{sec:H++_bound}.

\section{Constraints from LFV experiments}
\label{sec:LFV}
\begin{figure}[!htb]
\includegraphics[width=0.5\textwidth]{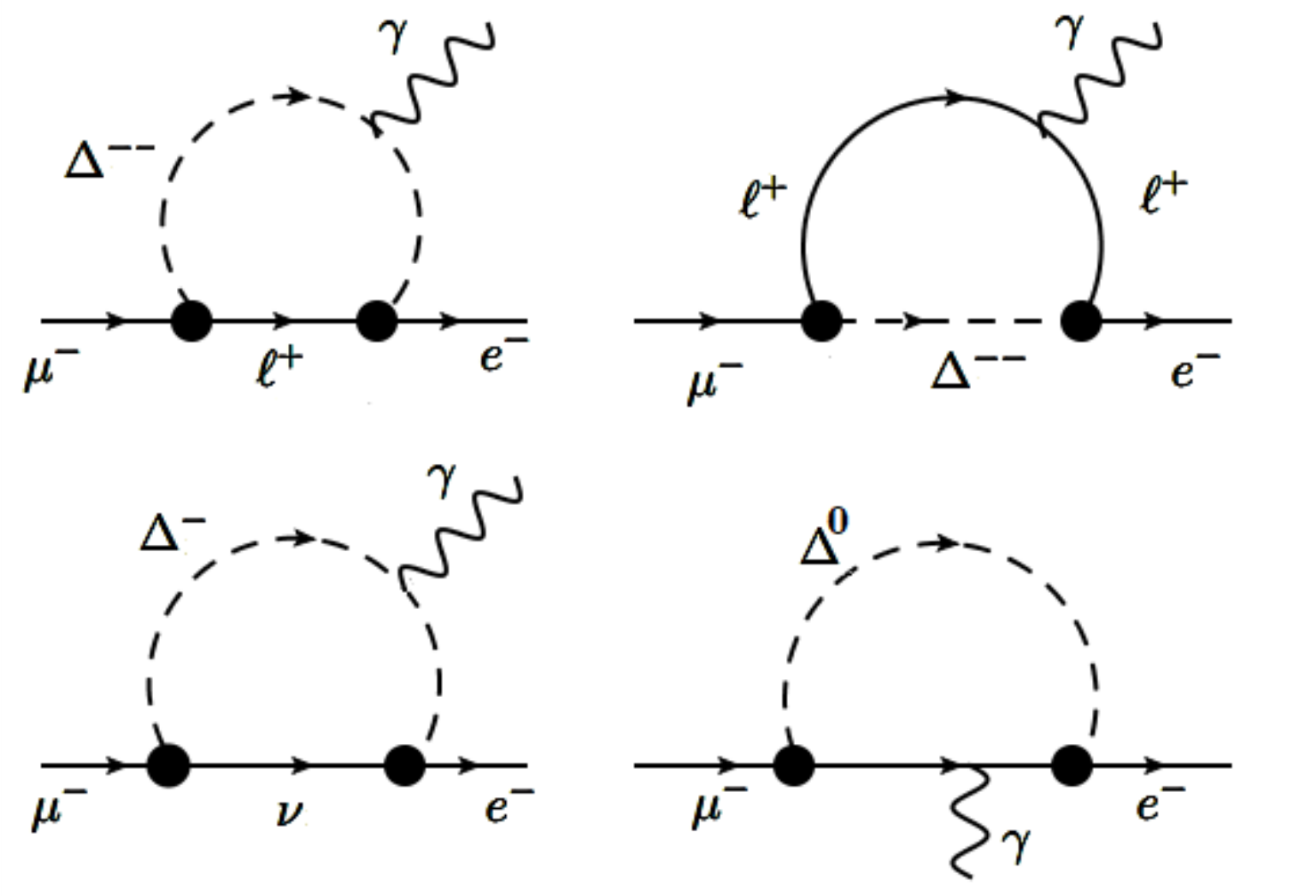}
\caption{Leading representative Feynman diagrams for $\mu \rightarrow e \gamma$ process. }
\label{fig:mu1}
\end{figure}

As it is well-known that experimental upper limits on lepton flavor violating decays provide important constraints on TeV-scale extensions of the standard model and thus it puts constraints on the free parameters of our model also.  In the canonical SM seesaw, the LFV decay rates induced by the neutrino mixings are highly suppressed by the requirement that the scale of new physics is at $10^{15}$ GeV, and hence, are well below the current experimental bounds. On the other hand, in the TeV scale BNT model, several new contributions appear due to the additional contributions from scalar quadruplet and triplet vector-like lepton members, which could lead to sizeable LFV rates. Since we are concentrating on the scenario where vector-like leptons $\Sigma'$s are heavy enough ($\sim 5$ TeV), whereas scalar quadruplet members are as light as less than a TeV, the contribution of vector-like leptons ($\Sigma'$s) to the lepton flavor violating process $\mu \rightarrow e \gamma$ is negligible compared to the contribution from the $\Delta$ members. We refer the reader to ref.~\cite{Liao:2010rx} for the complementary scenario. Leading representative Feynman diagrams for $\mu \rightarrow e \gamma$ process is shown in Fig.~\ref{fig:mu1}. Here Charged scalars ($\Delta^{\pm \pm},\Delta^{\pm}$) contribute more dominantly than the neutral one. 

Then, LFV $\mu \to e \gamma$ decay branching ratio can be easily calculated by
\begin{align}\label{eq:mue}
B(\mu \to e \gamma) = \frac{\alpha_{QED} \mid \left( M_{\nu}^2 \right)_{e \mu} \mid ^2}{108 \, \pi \, G_F^2 \, D^4} \left[\frac{1}{M_{\Delta^{++}}^2} + \frac{1}{4 M_{\Delta^{+}}^2} \right]^2,
\end{align}
where $M_{\nu} = (m_{\nu})^{\text{tot}}$, and $D$ is defined in Eq.~\ref{deno}.
\begin{figure}[!htp]
\includegraphics[scale=0.25]{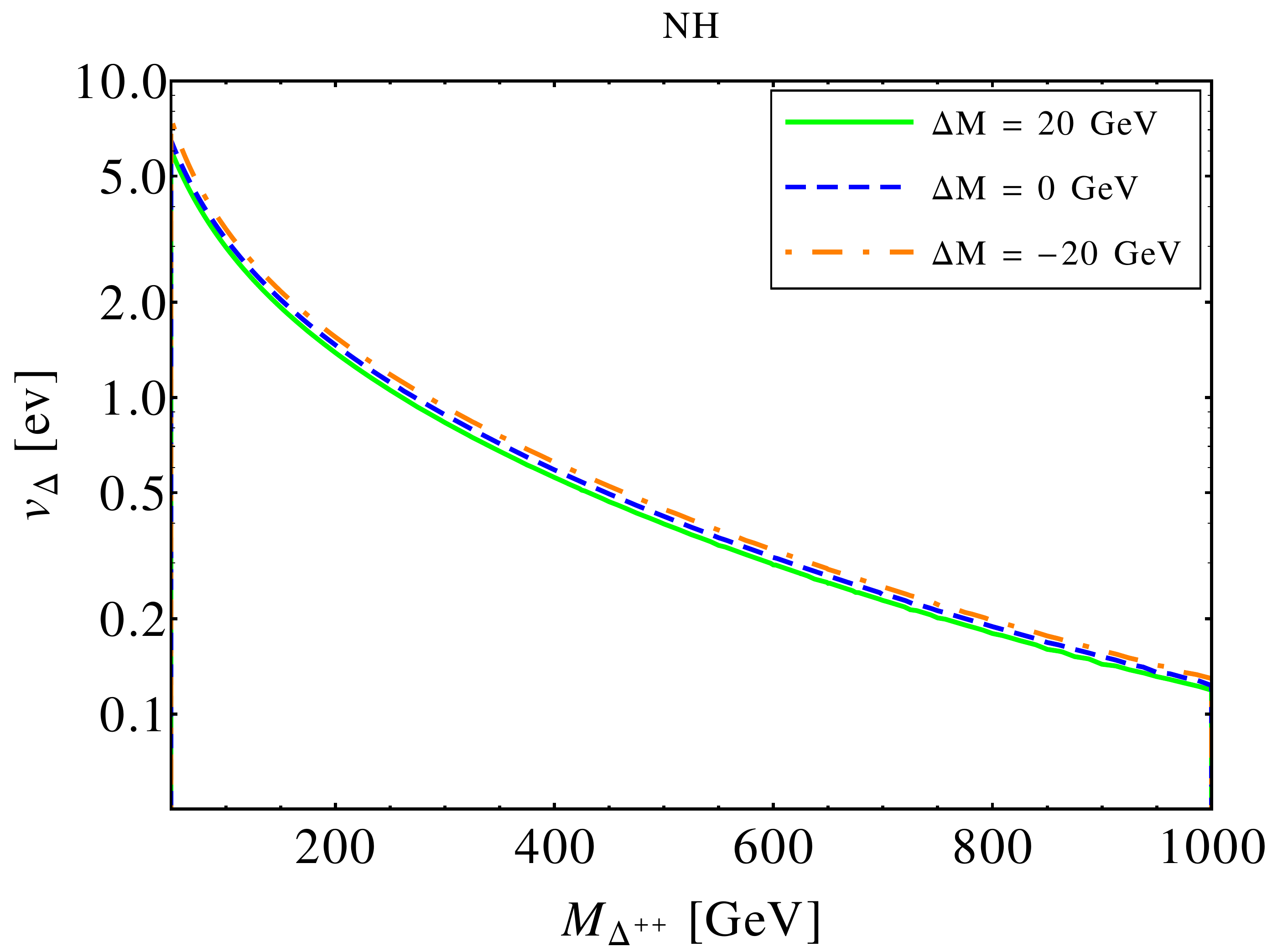}
\includegraphics[scale=0.25]{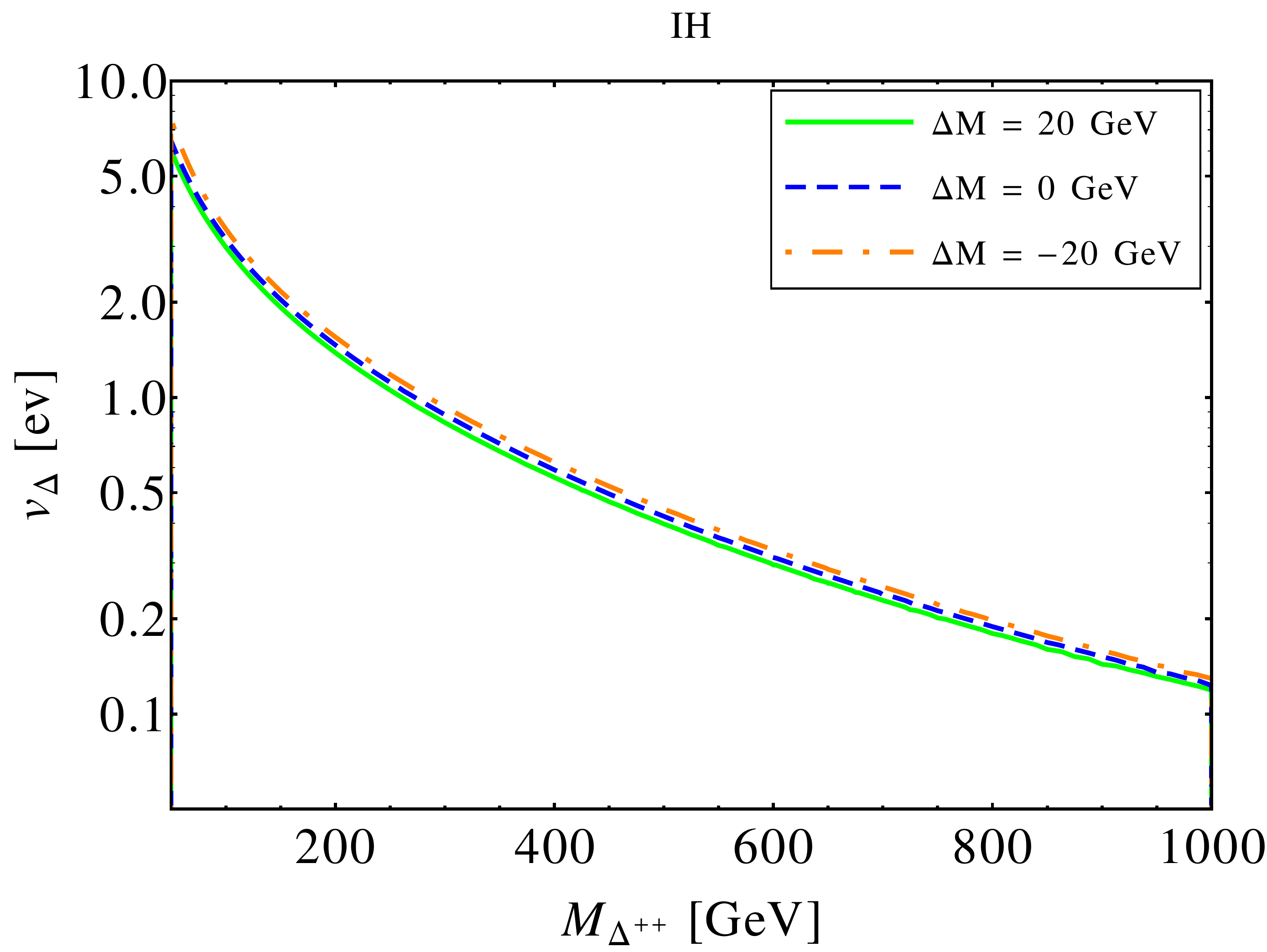}
\caption{Bounds on $v_{\Delta}-M_{\Delta^{++}}$ plane from lepton flavor vioilating $\mu \rightarrow e \gamma$ processes at 90$\%$ C.L. for both NH [$Left$] and IH [$Right$] of neutrino masses. The area below the curves are ruled out.}
\label{fig:mu2}
\end{figure}

We have used the currently most stringent bound by the MEG
Collaboration, BR ($\mu \rightarrow e \gamma$) $<$ ($5.7 \times 10^{-13}$) at 90$\%$ C.L. \cite{Adam:2013mnn}, and the bound on VEV $v_{\Delta}$ as a function of $M_{\Delta^{++}}$ for a given mass splitting of the charged scalars is shown in Fig.~\ref{fig:mu2} for both NH [Left] and IH [Right]. The region below respective lines are ruled out and $\mu \rightarrow e \gamma$ essentially provides a lower bound on $v_{\Delta}$. As we can see from Eq.~\ref{eq:mue}, the contribution from the doubly charged Higgs is the most dominant one. Mass splitting between $\Delta$ members has no significant impact in $\mu \rightarrow e \gamma$ limits. Also, the above limits are not sensitive to the mass ordering of neutrinos. 
However, in this model there exists a tree level diagram for $\mu \rightarrow 3e$ mediated by the doubly charged scalar. It is worth to mention that the constraints from $\mu \rightarrow 3e$ is less stringent~\cite{bhu} than the corresponding of $\mu \rightarrow e \gamma$ process. We do not explicitly discuss here other LFV processes, such as $\mu \rightarrow e$ conversion in nuclei, or electric dipole moments \cite{he}, which are left for future studies in detail since they also impose weaker bounds on our parameter space compared to $\mu \rightarrow e \gamma$.

\section{Collider Implications}
\label{sec:collider}

This model provides an interesting avenue to test the neutrino mass generation mechanism at the LHC. The presence of the isospin 3/2 scalar multiplet can give rise to rich phenomenology at the LHC. The collider signatures of the BNT model has been studied in the literature~\cite{earlylhc, Jana}.
However, there is not only new data made public by the LHC experiments since then, resulting in updated constraints, but also few subtle points regarding the phenomenology of multi-charged Higgs particles needs to be clarified that were absent in previous analyses. In this section, we try to investigate the limits on the $\Delta$ masses from the recent experimental data. 

\subsection{Constraints from $h \rightarrow \gamma \gamma$ at the LHC}
\begin{figure}[!htp]
\includegraphics[scale=0.3]{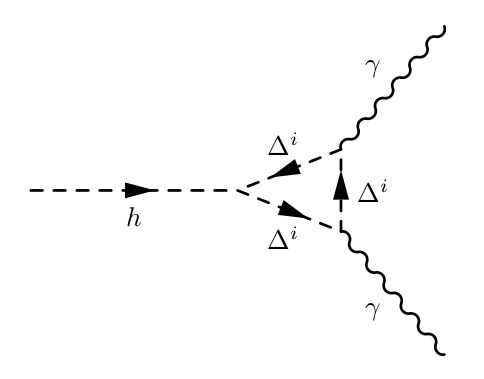}
\caption{Triangle diagrams that mediate $h \rightarrow \gamma \gamma$ decay in the BNT model. Here $\Delta^{i}$ stands for singly, doubly and triply charged Higgs.}
\label{fig:hdiphoton}
\end{figure}

The BNT model is rich in multi-charged scalars. These multi-charged scalars can mediate SM-like Higgs decay to a pair of photons in addition to $t$ and $W$ loops. A representative triangle loop diagram for these processes is shown in Fig.~\ref{fig:hdiphoton}. In fact, the $\Delta$ mediated processes can both augment or suppress the SM predicted $h \rightarrow \gamma \gamma$ rate at the LHC depending on the signs and relative strengths of $\lambda_3$ and $\lambda_4$.  This is because the coupling between the SM-like Higgs $h$ and a pair of singly, doubly and triply charged Higgs are 
\bea
\tilde{\lambda}_1 & =& v_H \bigg(\lambda_3 + \dfrac{\lambda_4}{4}\bigg)
\\ \nn
\tilde{\lambda}_{2} & = & v_H \bigg(\lambda_3 - \dfrac{\lambda_4}{4}\bigg) 
\\ \nn
\tilde{\lambda}_{3} & = & v_H  \bigg(\lambda_3 - \dfrac{3\lambda_4}{4}\bigg),
\label{hDeltacoupling}
\eea
 respectively.

For a given production process of a Higgs, denoted by $X$, and the subsequent decay into  final state $Y$ the signal strength parameter, normalized to the SM values, is defined as
\beq
\mu_{Y} = \dfrac{\sigma_{X}}{\sigma^{\text{SM}}_{X}} \dfrac{\Gamma_{h \rightarrow Y}}{\Gamma^{\text{SM}}_{h \rightarrow Y}} \dfrac{\Gamma^{\text{SM}}_{h,\text{tot}}}{\Gamma^{}_{h,\text{tot}}} \, .
\eeq
In our study the new physics can influence only the total decay width, $\Gamma_{h, \text{tot}}$, and the partial decay rate, $\Gamma_{h \rightarrow Y}$. We formulate this change in the $h \gamma \gamma$  coupling as
\beq
g_{h\gamma\gamma} = \kappa_{\gamma} \, g^{\text{SM}}_{h\gamma\gamma},
\eeq
where~\cite{Djouadi:2005gi,Knapen:2015dap, Jana:2017hqg} 
\beq
\kappa_{\gamma} =  \dfrac{\Bigg| \dfrac{N^c_t Q^2_t}{v_H}A_{\frac{1}{2}}(\tau_t) + \dfrac{1}{v_H} A_{1}(\tau_W) + \mathlarger{\mathlarger{ \mathlarger{\sum}_{i=1}^{3}}} \dfrac{\tilde{\lambda}_i Q^2_{i}}{2 M_{i}} A_0(\tau_{i})\Bigg|^2}{\Bigg| \dfrac{N^c_t Q^2_t}{v_H}A_{\frac{1}{2}}(\tau_t) + \dfrac{1}{v_H} A_{1}(\tau_W) \Bigg|^2}\, .
\eeq
Here, the loop functions are given by~\cite{Djouadi:2005gi},
\bea
A_0 & = & - \tau + \tau^2 f(\tau),
\\ \nn
A_{\frac{1}{2}}(\tau) &=& 2 \tau [1+ (1-\tau) f(\tau)],
\\ \nn
A_1 & = & -2 - 3 \tau (1 + (2 - \tau) f(\tau) ), 
\label{loop_function_1}
\eea
with
\bea
f(x)= 
	\begin{cases}
		arcsin^2[1/\sqrt{x}],  & \mbox{if } x \geq 1 \\
		-\dfrac{1}{4}[ln \dfrac{1+\sqrt{1-x}}{1-\sqrt{1-x}} - i \pi]^2, & \mbox{if } x < 1 \,.
	\end{cases}
\label{loop_function_2}
\eea
The parameters $\tau_i=4 M^2_i/M^2_h$ are defined by the corresponding masses of the heavy loop particles. Thus, the partial decay width of the SM-like Higgs to $\gamma\gamma$ can be written as 
\beq
\dfrac{\Gamma_{h \rightarrow \gamma\gamma}}{\Gamma^{\text{SM}}_{h \rightarrow \gamma\gamma}} = \kappa^2_{\gamma} \,.
\eeq
Consequently the total decay width of $h$ in terms of the rescaling factor $\kappa_{\gamma}$ is~\cite{Denner:2011mq, Heinemeyer:2013tqa} 
\beq
\dfrac{\Gamma_{h, \text{tot}}}{\Gamma^{\text{SM}}_{h, \text{tot}}} \approx 0.9977 + 0.0023 \, \kappa^2_{\gamma},
\eeq 
with $\Gamma^{\text{SM}}_{h \text{tot}} = 4.07$ MeV~\cite{Denner:2011mq}.
 
\begin{figure}[!htp]
\includegraphics[scale=0.2]{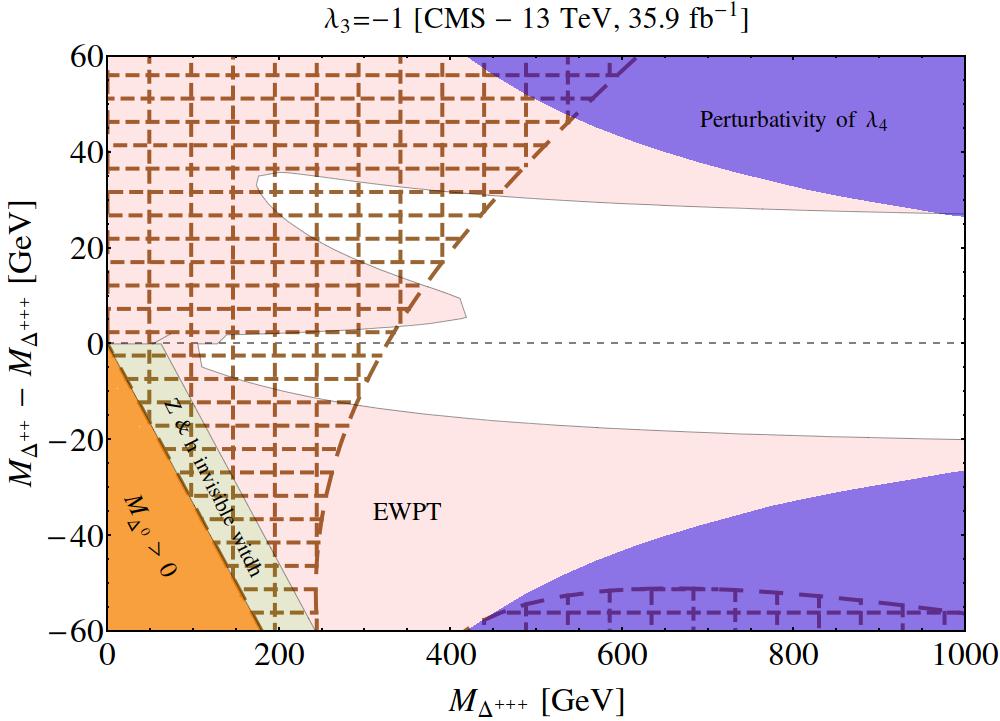}
\includegraphics[scale=0.2]{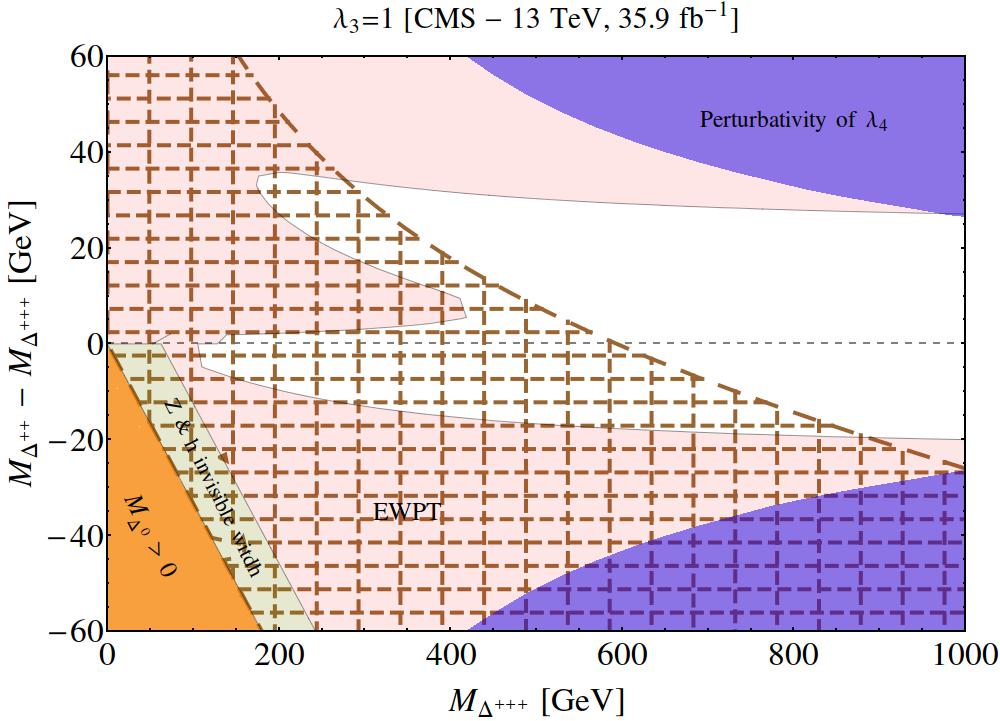}
\includegraphics[scale=0.2]{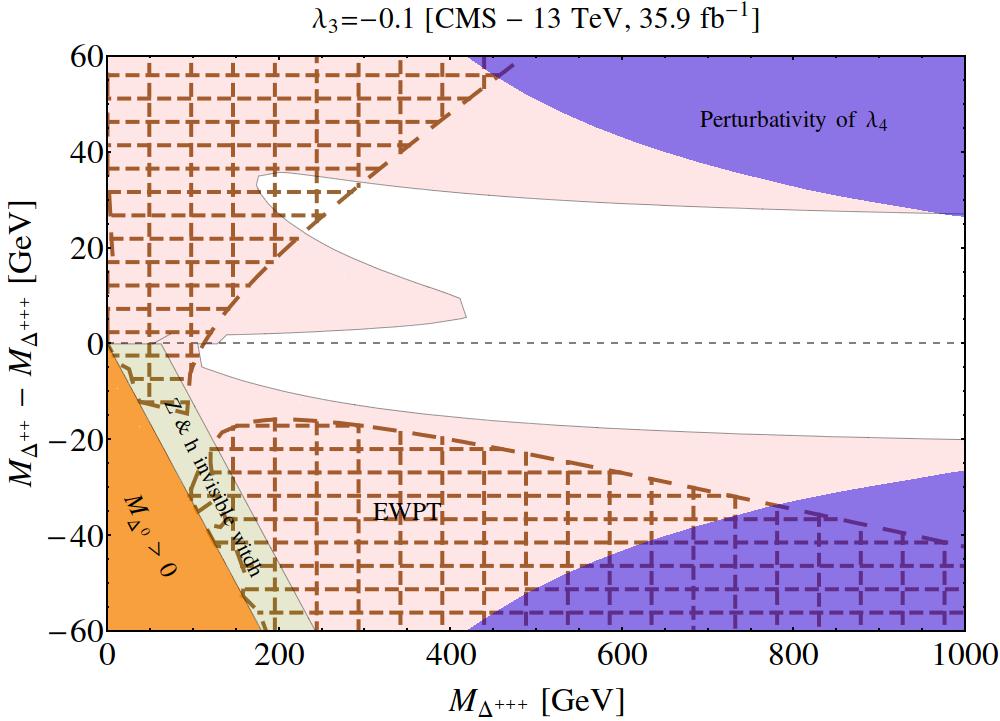}
\includegraphics[scale=0.2]{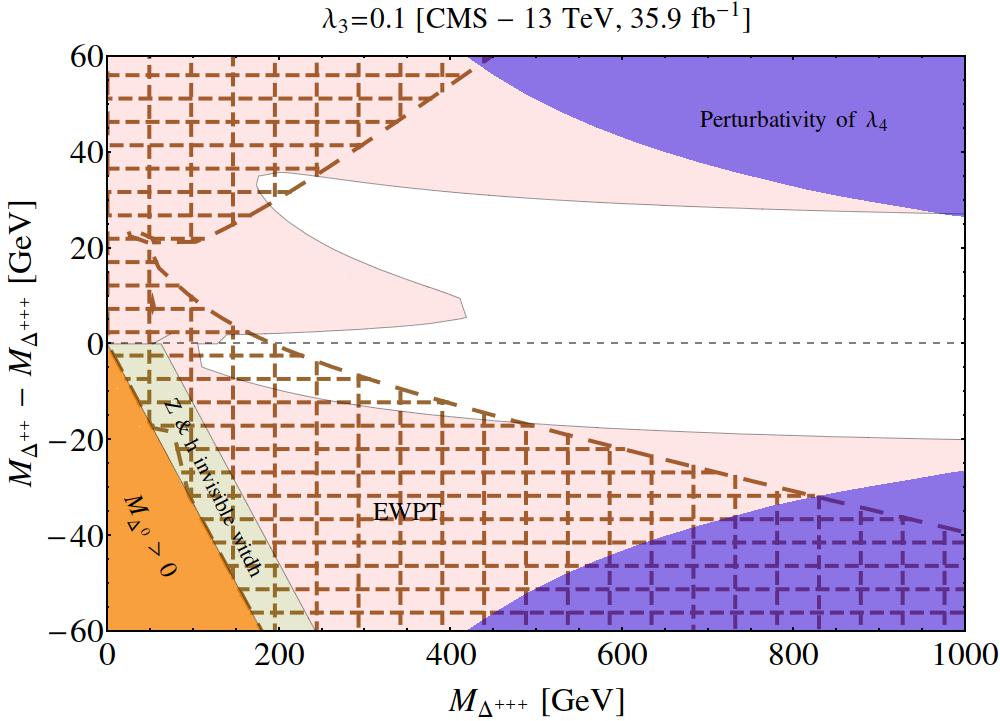}
\caption{Constraints form $h \rightarrow \gamma \gamma$ decay rate measured by CMS in the $M_{\Delta}-M_{\Delta^{\pm \pm \pm}}$ plane is shown by the brown shaded region. We plot the limits for $\lambda_3 = -1$ [top left], 1 [top right], -0.1 [bottom left] and 1 [bottom right]. The other colored regions has the same meaning as Fig.~\ref{fig:EWPT}.  }
\label{fig:mugammaCMS}
\end{figure}

\begin{figure}[!htp]
\includegraphics[scale=0.2]{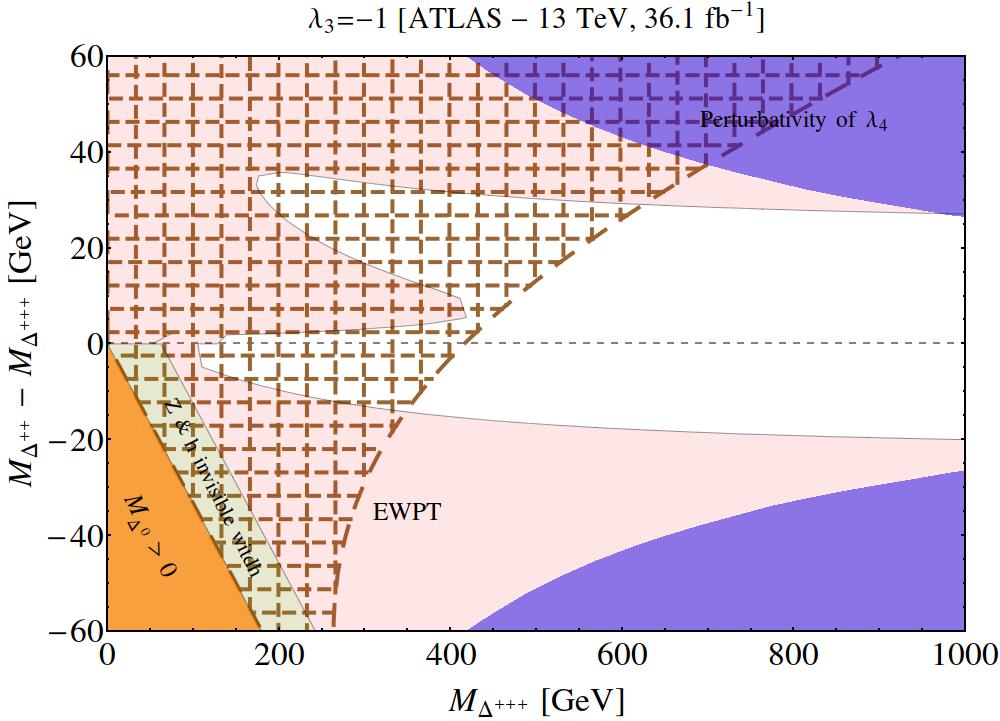}
\includegraphics[scale=0.2]{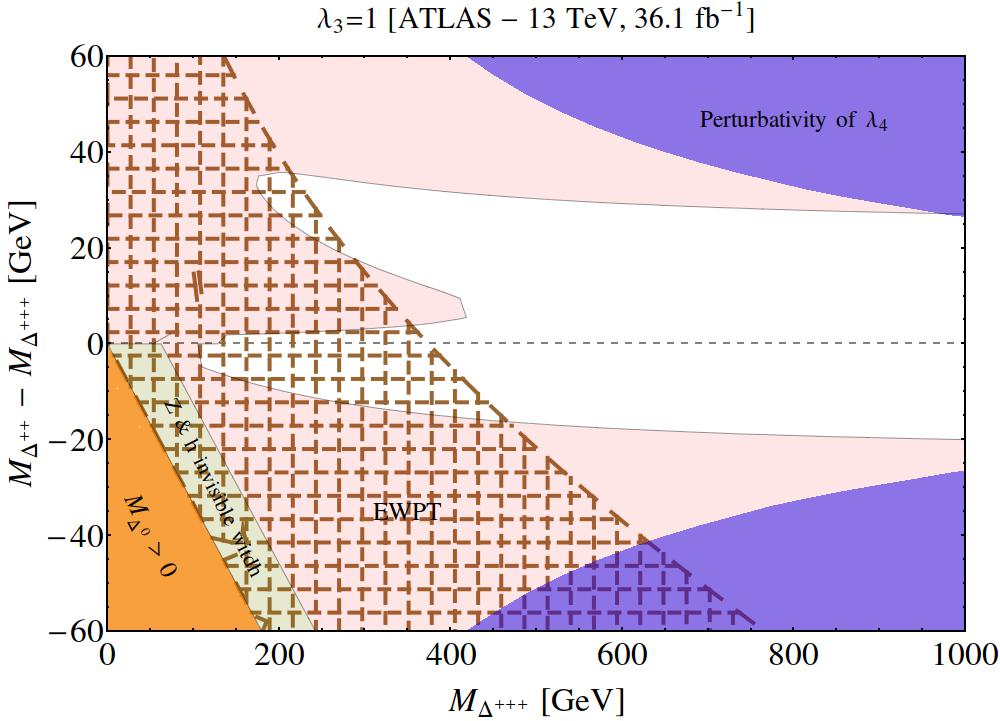}
\includegraphics[scale=0.2]{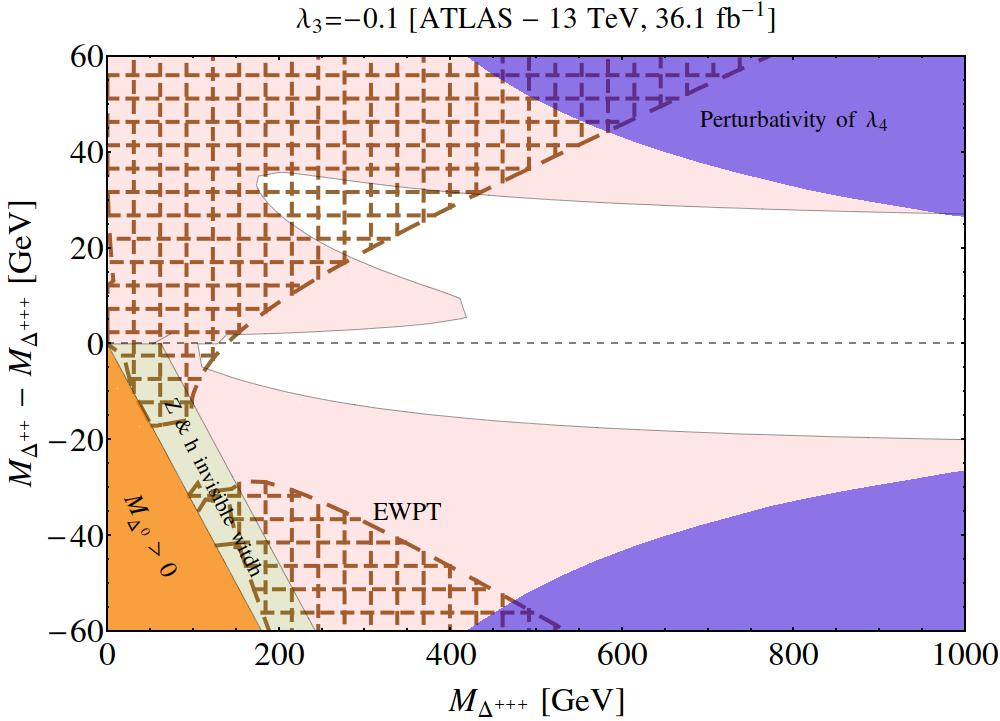}
\includegraphics[scale=0.2]{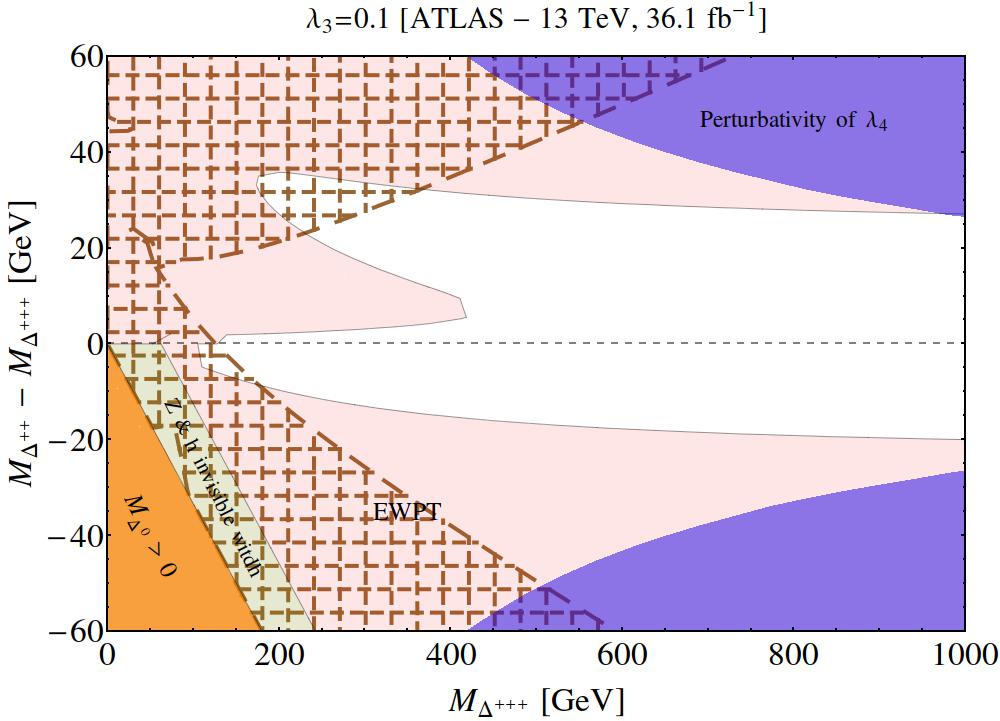}
\caption{Constraints form $h \rightarrow \gamma \gamma$ decay rate measured by ATLAS in the $M_{\Delta}-M_{\Delta^{\pm \pm \pm}}$ plane is shown by the brown shaded region. We plot the limits for $\lambda_3 = -1$ [top left], 1 [top right], -0.1 [bottom left] and 1 [bottom right]. The other colored regions has the same meaning as Fig.~\ref{fig:EWPT}}
\label{fig:mugammaATLAS}
\end{figure}

 CMS and ATLAS both recently made public their $h \rightarrow \gamma \gamma$ analysis, combining all production channels, based on $\sim 36$ fb$^{-1}$ of data at 13 TeV center of mass energy. The measured strength ($\mu_{\gamma}$) of the above decay rate by CMS~\cite{CMS:2017rli} and ATLAS~\cite{ATLAS:2017myr} are  $\mu^{\text{CMS}}_{\gamma}=1.16^{+0.15}_{-0.14}$ and  $\mu^{\text{ATLAS}}_{\gamma} = 0.99 \pm 0.14$ respectively. In Fig.~\ref{fig:mugammaCMS} we overlay the limits obtained from $\mu^{\text{CMS}}_{\gamma}$, shown by brown shaded regions, on top of EWPT excluded regions in $\Delta M - M_{\Delta^{\pm \pm \pm}}$ plane. From Eq.~\ref{hDeltacoupling} we can notice that the strength of $\mu_{\gamma}$ in the BNT model is controlled by a combination of $\lambda_3$ and $\lambda_4$. In the results of Fig.~\ref{fig:mugammaCMS} $\lambda_4$ is fixed by $\Delta M$. So, we show our results in the above figure for four values of $\lambda_3 = \pm 1, \, \pm0.1$. In Fig.~\ref{fig:mugammaATLAS} we plot the same bounds from $\mu^{\text{ATLAS}}_{\gamma}$. The shape of exclusion contours from CMS and ATLAS differ marginally for the same value of $\lambda_3$ since the measured $\mu_{\gamma}$ by them are not the same.   
 
We notice from Figs.~\ref{fig:mugammaCMS} and \ref{fig:mugammaATLAS} that $h \rightarrow \gamma \gamma$ limits depend strongly on the magnitude of $\lambda_3$. For $|\lambda_3| \gtrsim 1$, $h \rightarrow \gamma \gamma$ excludes a relatively large fraction of the parameter space that is not ruled out by EWPT. In contrast, if $|\lambda_3|$ assumes a small value ($\lesssim 0.1$) it will hardly add anything on top of EWPT bounds.

\subsection{Production of $\Delta^{\pm \pm}$ and $\Delta^{\pm \pm \pm}$ at the LHC}
\label{sec:prod}

\begin{figure}[!htp]
\includegraphics[scale=0.30]{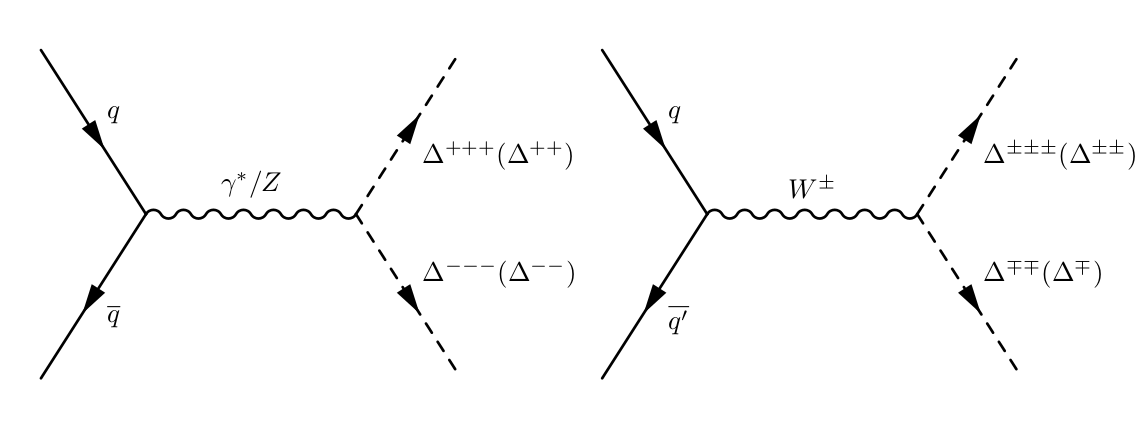}
\caption{Pair production [$Left$] and associated production [$Right$] of $\Delta^{\pm \pm \pm} \, (\Delta^{\pm \pm })$ via DY processes.}
\label{fig:DY}
\end{figure}

A pair of $\Delta^{\pm \pm \pm} \, (\Delta^{\pm \pm })$ can be produced at the LHC by Drell-Yan (DY) process via $s$-channel $\gamma^*/Z$ boson exchange. Also, associated production of $\Delta^{\pm \pm \pm}\Delta^{\mp \mp } \, (\Delta^{\pm \pm}\Delta^{\mp })$ is possible via $s$-channel $W$ exchange. The relevant diagrams for such processes are shown in Fig.~\ref{fig:DY}.  Being $s$-channel, DY pair production cross-sections are significantly suppressed for large $\Delta^{\pm \pm \pm} \, (\Delta^{\pm \pm })$ masses. Additionally, due to large electromagnetic charges carried by $\Delta^{\pm \pm \pm} \, (\Delta^{\pm \pm })$ they can be pair produced by photon fusion (PF) as well. We refer the reader to Ref.~\cite{Jana} for Feynman diagrams relevant for the above process. In comparison with DY, photo-production of these multi-charged scalars takes place via $t$ and $u$-channel processes mediated by charged scalars and hence falls less sharply for higher $\Delta$ masses. Although the photo-production cross-section of triply and doubly charged scalars benefit from enhancements by a factor of $3^4$ and $2^4$, respectively, due to their large electric charges but it is suppressed, at the same time, by the tiny parton density of photon inside a proton. For a detailed discussion on parton density function of photons from different collaborations we refer the reader to Refs.~\cite{Jana, Babu-Jana}. In this study, we use the {\tt NNPDF23$\_$lo$\_$as$\_$130} PDF set~\cite{NNPDF}  which contains photon PDF. It is important to point out that although including PF boosts the production cross-section for heavier masses, they also suffer from large uncertainties. In this analysis, we build on the work of the above references and include the errors associated with using all the available eigenvector sets of a given PDF.

\begin{figure}[!htp]
\includegraphics[scale=0.3]{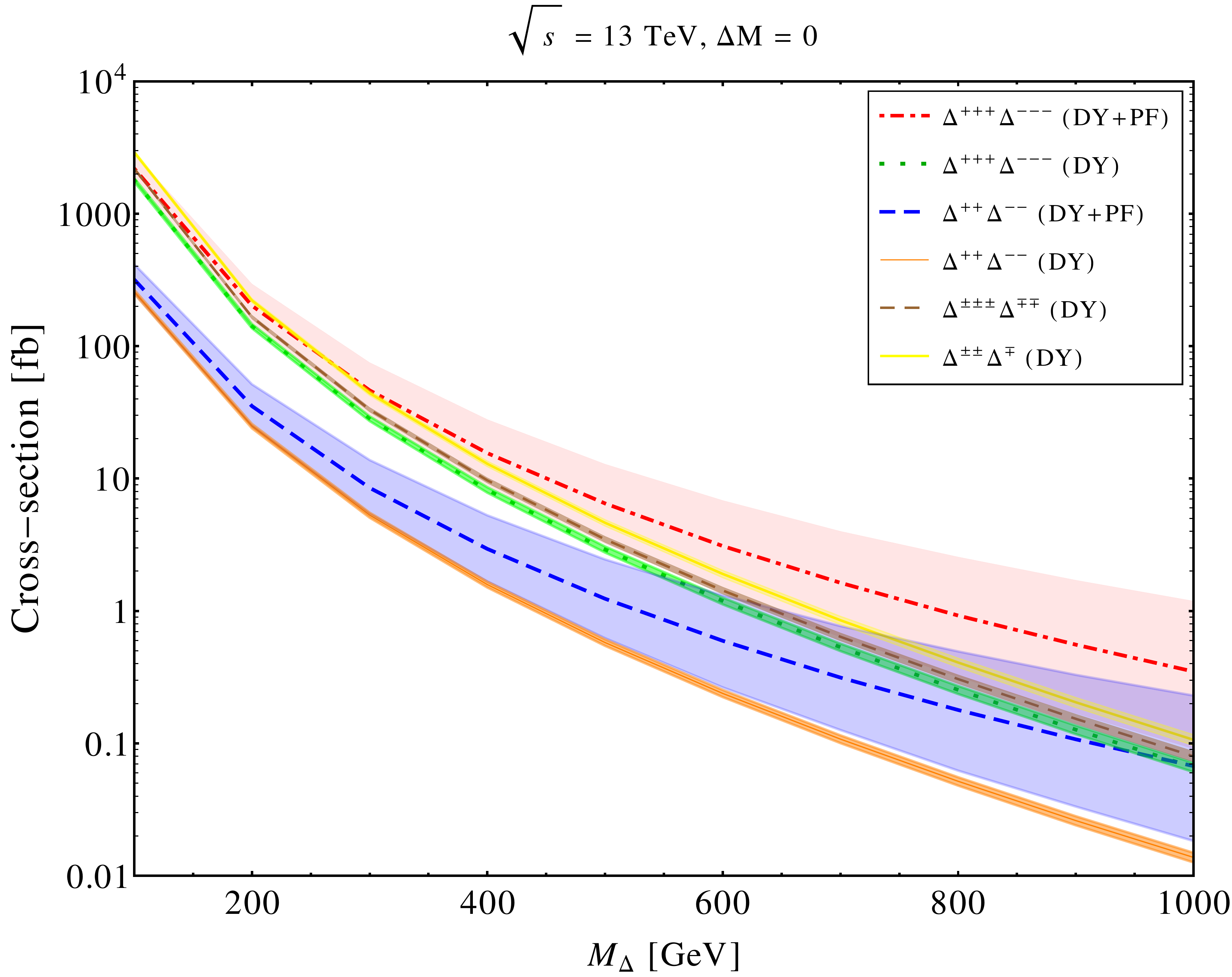}
\caption{The cross-sections of various PP and AP channels for $\sqrt{s} = 13$ TeV. No mass-splitting between the quadruplet components are considered here. Large mass-splittings will change AP cross-sections. The uncertainties associated with the variation of PDF eigenvector sets are shown by bands of the same color as the cross-section curves.}
\label{fig:xsec13}
\end{figure}

In Fig.~\ref{fig:xsec13} we present cross-sections of various pair-production and associated production processes. We employ {\tt MadGraph5$\_$aMC@NLO$\_$v2.5.4} code~\cite{MG5} for our calculation, where the BNT model is implemented using {\tt FeynRules$\_$v2.0}~\cite{FR}. We have not used any K-factor in above computations. Pair production of $\Delta^{\pm \pm \pm}$ and $\Delta^{\pm \pm}$ via DY mechanism are shown by green and thin orange lines respectively. The same for the above two particles in a combination of DY and PF are depicted by dot-dashed red and dashed blue lines. In contrast, dashed brown and thick yellow lines represent associated production cross-sections for the same two particles. The uncertainties related to each process due to PDF variation are encoded within a band of the same color as the respective cross-section curve. As expected, the presence of $t$-channel diagrams of PF enhances pair production cross-sections of both doubly and triply charged bosons significantly for masses above 500 GeV. However, while errors of DY processes are tiny ($\sim 5 \%$), the large error bands of the two channels that include PF will not escape the reader's attention. In fact, the error of DY+PF channels are $> 100\%$ for $M_{\Delta} \gtrsim 500 $ GeV. So, one can infer from the results of  Fig.~\ref{fig:xsec13} that although adding PF to DY production provides an apparent enhancement in pair production cross-section, but one can not be certain about such increase in cross-section due to enormous PDF uncertainty associated with PF. Hence, we ignore the inclusion of PF in this paper.  

 
\subsection{Decay of $\Delta^{\pm \pm}$ and $\Delta^{\pm \pm \pm}$}
\label{sec:decay}

In this section we discuss the decay of doubly and triply charged Higgs bosons of the BNT model in details. Especially, we shall pay particular attention to proper decay length of these particles and the corresponding implications for their LHC detection. Another point we want to emphasize is that for our choice of $M_{\Sigma} = 5$ TeV, $(m_{\nu})_{ij}^{\text{tree}} \sim (m_{\nu})_{ij}^{\text{loop}}$ for a  range of $M_{\Delta}$ that is accessible to the future high luminosity LHC run. The interplay between these two contributions should reflect in the leptonic branching ratios (BR) of the quadruplet components. This point was not considered by previous LHC studies~\cite{earlylhc, Jana} of the BNT model. The inclusion of dimension-5 loop contribution to the Yukawa couplings changes the value of $v_{\Delta}$ where the cross-over from leptonic to bosonic decay channels takes place.    

First, let us quantify the impact of the inclusion of dimension-5 contribution to the Yukawa couplings. In the absence of dmension-5 operator, from Eq.~\ref{Leff} and \ref{eq:mnu2} one can deduce the Feynman rule corresponding to the coupling of lepton doublets with the Higgs qudruplet $-\dfrac{2}{\sqrt{3}}\dfrac{(Y_iY'_j+ Y_j Y'_i) v_H}{\sqrt{2} M_{\Sigma}} = \dfrac{2}{\sqrt{6}} \dfrac{ (m_{\nu})_{ij}^{\text{tree}}}{v_{\Delta}}$, where the pre-factor 2 in the numerator arises since the coupling can come from two vertices and the other factor $1/ \sqrt{3}$ comes from Clebsch-Gordon coefficient related to the interaction of Eq.~\ref{Leff}, as described in Appendix~\ref{sec:AA}. Now, if we include the loop contribution the above Feynman rule modifies to 
\beq
h_{ij} =-\dfrac{2}{\sqrt{3}}\dfrac{(Y_iY'_j+ Y_j Y'_i) v_H}{\sqrt{2} M_{\Sigma}} = \dfrac{2}{\sqrt{6}} \dfrac{ (m_{\nu})_{ij}^{\text{tot}}}{D},
\label{FR_Yuk_tot}
\eeq
where $(m_{\nu})_{ij}^{\text{tot}} = (m_{\nu})_{ij}^{\text{tree}}+(m_{\nu})_{ij}^{\text{loop}}$ and $D$ is given by
\beq
D =  v_{\Delta} \, - \, \frac{\left(3+\sqrt{3}\right) \lambda _5 v_H M_{\Sigma }^2 
\left(Y_i Y_j^{'}+ Y_i^{'} Y_j \right)}{32 \pi ^2 \left(M_{\Delta }^2-M_H^2\right)}
\left(
\frac{M_{\Delta}^2 \log \left(\frac{M_{\Sigma }^2}{M_{\Delta }^2}\right)} {M_{\Sigma }^2-M_{\Delta }^2}-
\frac{M_H^2 \log \left( \frac{M_{\Sigma }^2}{M_H^2}\right)} {M_{\Sigma}^2-M_H^2}
\right).
\label{deno}
\eeq

\begin{figure}[!htp]
\includegraphics[scale=0.25]{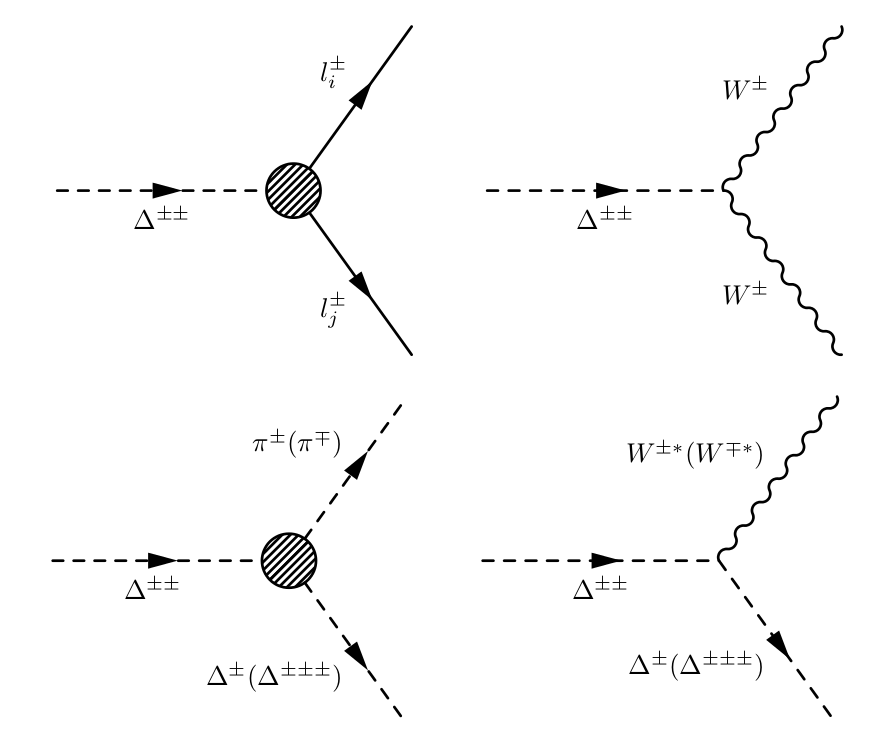}
\caption{Feynman diagrams for decay of $\Delta^{\pm \pm}$.}
\label{fig:H++_decay}
\end{figure}

Next, we list the decay widths of doubly-charged Higgs in various channels. The corresponding Feynman diagrams are shown in Fig.~\ref{fig:H++_decay}. The decay of $\Delta^{\pm \pm}$ can happen in four possible channels. While $l^{\pm}_i l^{\pm}_j$ and $W^{\pm} W^{\pm}$ final states are always accessible, cascade decays $\Delta^{} \pi^{}$ and $\Delta^{} W^{*}$ open up only when the mass-splitting between quadruplet members are non-zero. We should note that $\Delta^{\pm \pm}$ can not be either lightest or heaviest member of the $\Delta$ multiplet under any circumstances. Hence, for non-zero mass-gap it can decay in cascades via $\Delta^{\pm} X^{\pm}$ or $\Delta^{\pm \pm \pm} X^{\mp}$ (where $X=\pi, W^*$) depending on whether $\Delta M < 0$ or $\Delta M> 0$. So, the relevant decay width formulas of $\Delta^{\pm \pm}$ are~\cite{Aoki,Perez}
\bea
\Gamma(\Delta^{\pm \pm} \rightarrow l^{\pm}_i l^{\pm}_j ) & = & \dfrac{|h_{ij}|^2 M_{\Delta^{\pm \pm}}}{4 \pi (1+\delta_{ij})} \, \bigg(1 - \dfrac{m^2_i}{M_{\Delta^{\pm \pm}}^2}-\dfrac{m^2_j}{M_{\Delta^{\pm \pm}}^2}\bigg) \, \bigg[\lambda(\dfrac{m^2_i}{M_{\Delta^{\pm \pm}}^2},\dfrac{m^2_j}{M_{\Delta^{\pm \pm}}^2})\bigg]^{1/2},
 \nn \\
\Gamma(\Delta^{\pm \pm} \rightarrow W^{\pm} W^{\pm} ) &=& S_{W^{\pm}W^{\pm}}^2 \, \dfrac{g^4 v_{\Delta}^2  M_{\Delta^{\pm \pm}}^3}{16 \pi M_W^4} \, \bigg(  \dfrac{3 M^4_W}{ M_{\Delta^{\pm \pm}}^4} \dfrac{M^2_W}{ M_{\Delta^{\pm \pm}}^2} + \dfrac{1}{4}\bigg) \, \beta\bigg( \dfrac{M_W^2}{ M_{\Delta^{\pm \pm}}^2} \bigg),
\nn \\
\Gamma(\Delta^{\pm \pm} \rightarrow \Delta^{\pm} \pi^{\pm}) &=& S_{\Delta^{\pm}W^{\pm}}^2 \, \dfrac{g^4 |V_{ud}|^2 \Delta M^3 f^2_{\pi}}{16 \pi M^4_W},
\nn \\
\Gamma(\Delta^{\pm \pm} \rightarrow \Delta^{\pm} l^{\pm} \nu_l) & = & S_{\Delta^{\pm}W^{\pm}}^2 \, \dfrac{ g^4 \Delta M^5}{240 \pi^3 M_W^4} ,
\nn \\
\Gamma(\Delta^{\pm \pm} \rightarrow \Delta^{\pm} q \overline{q'}) & = & 3 \, \Gamma(\Delta^{\pm \pm} \rightarrow \Delta^{\pm} l^{\pm} \nu_l) ,
\nn \\
\Gamma(\Delta^{\pm \pm} \rightarrow W^{\pm} W^{\pm*} ) &=& S_{W^{\pm}W^{\pm}}^2 \,  \dfrac{3 g^6 M_{\Delta^{\pm \pm}}}{512 \pi^3 } \, \dfrac{v_{\Delta}^2}{M_W^2} \, F\bigg( \dfrac{M_W^2}{ M_{\Delta^{\pm \pm}}^2} \bigg),
\label{H++_decay}
\eea
where $S_{W^{\pm}W^{\pm}} = \sqrt{3}$ and $S_{\Delta^{\pm}W^{\pm}} =\sqrt{2}$ are scale factors that we use to convert the expressions of decay widths given in Refs.~\cite{Aoki,Perez} for $SU(2)$ triplet to quadruplet. Here, $V_{ud}$ is the $ud$ element of the CKM matrix and $f_{\pi }=131$ MeV is the pion decay constant. One can easily use the results of $\Gamma(\Delta^{\pm \pm} \rightarrow \Delta^{\pm} X^{\pm})$ to derive $\Gamma(\Delta^{\pm \pm} \rightarrow \Delta^{\pm \pm \pm} X^{\mp})$ decay widths by changing the scale factor from $S_{\Delta^{\pm}W^{\pm}}$ to $S_{\Delta^{\pm \pm \pm }W^{\mp}} = \sqrt{3/2}$. The kinematic functions are given by
\bea
\lambda(x,y) & = & 1+x^2+y^2-2xy-2x-2z,
\nn \\
\beta(x) & = & \sqrt{1-4x},
\nn \\
F(x) &=&  -|1-x| \bigg(\dfrac{47}{2}x - \dfrac{13}{2} + \dfrac{1}{x}\bigg ) + 3(1 - 6x + 4x^2) |\log\sqrt{x}| 
\nn \\
& & + \dfrac{3(1-8x+20 x^2)}{\sqrt{4x-1}} \arccos \Big(\dfrac{3x-1}{2 x^{3/2}} \Big)\, .
\eea    

\begin{figure}[!htp]
\includegraphics[scale=0.35]{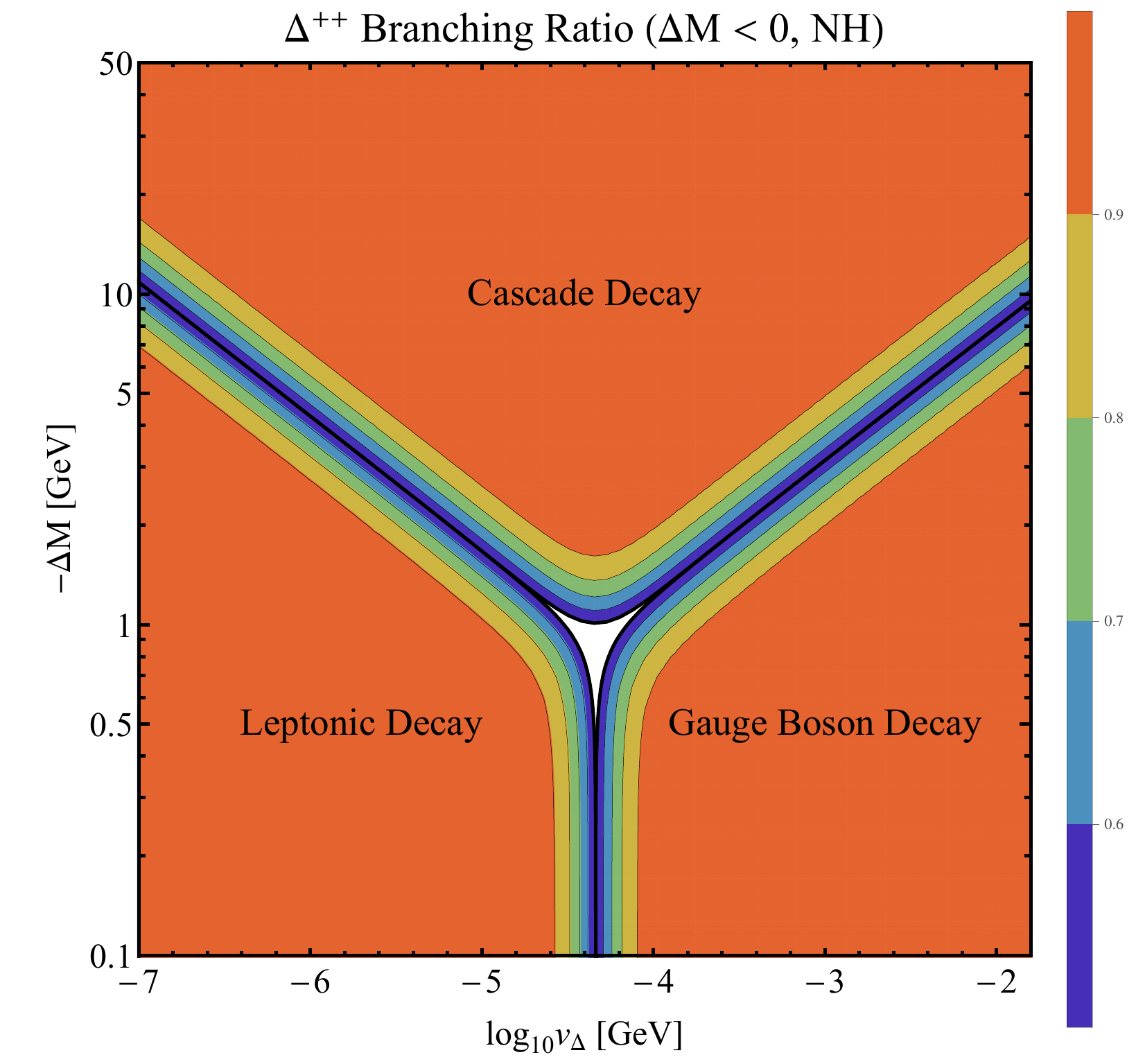}
\includegraphics[scale=0.35]{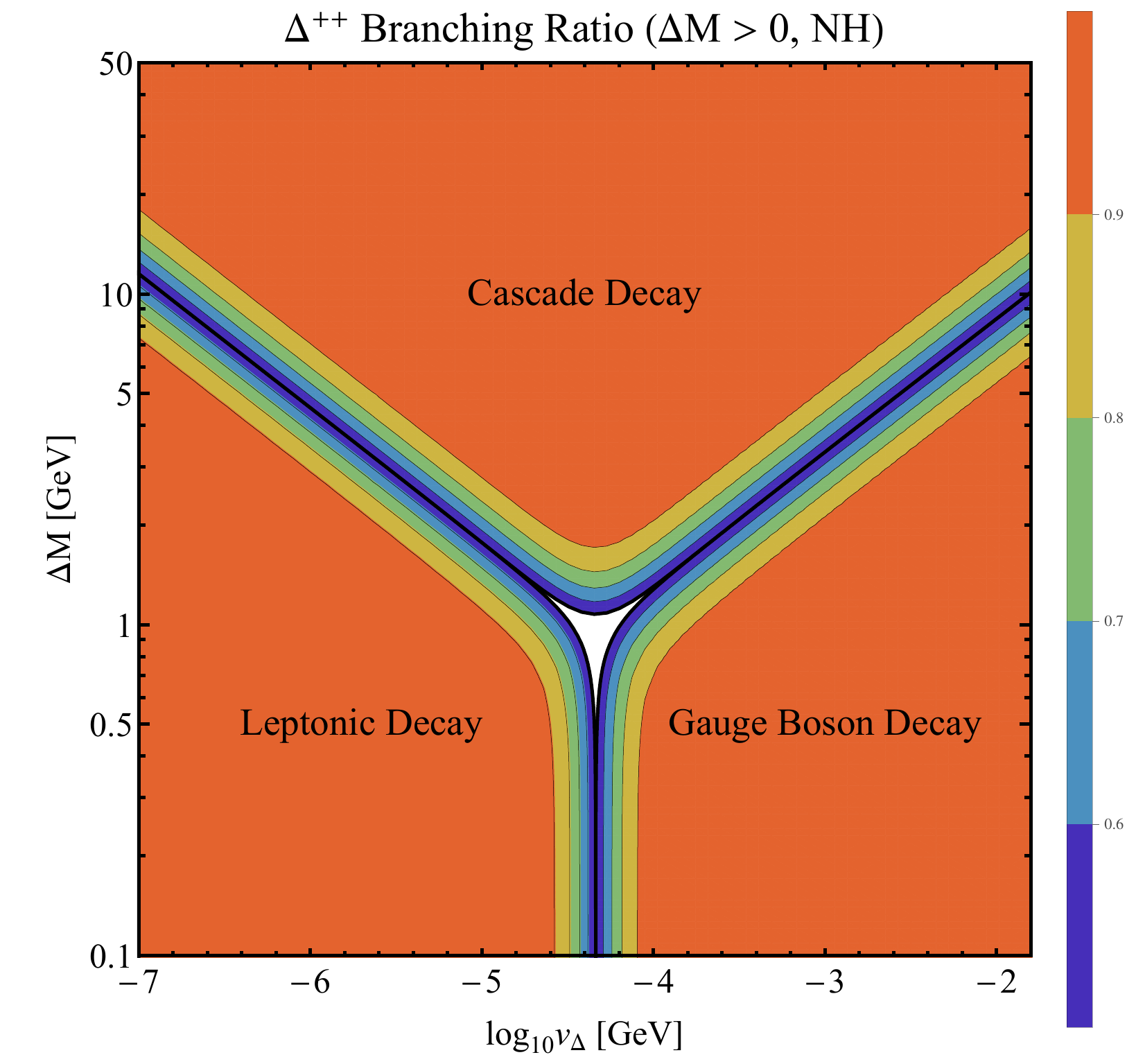}
\includegraphics[scale=0.35]{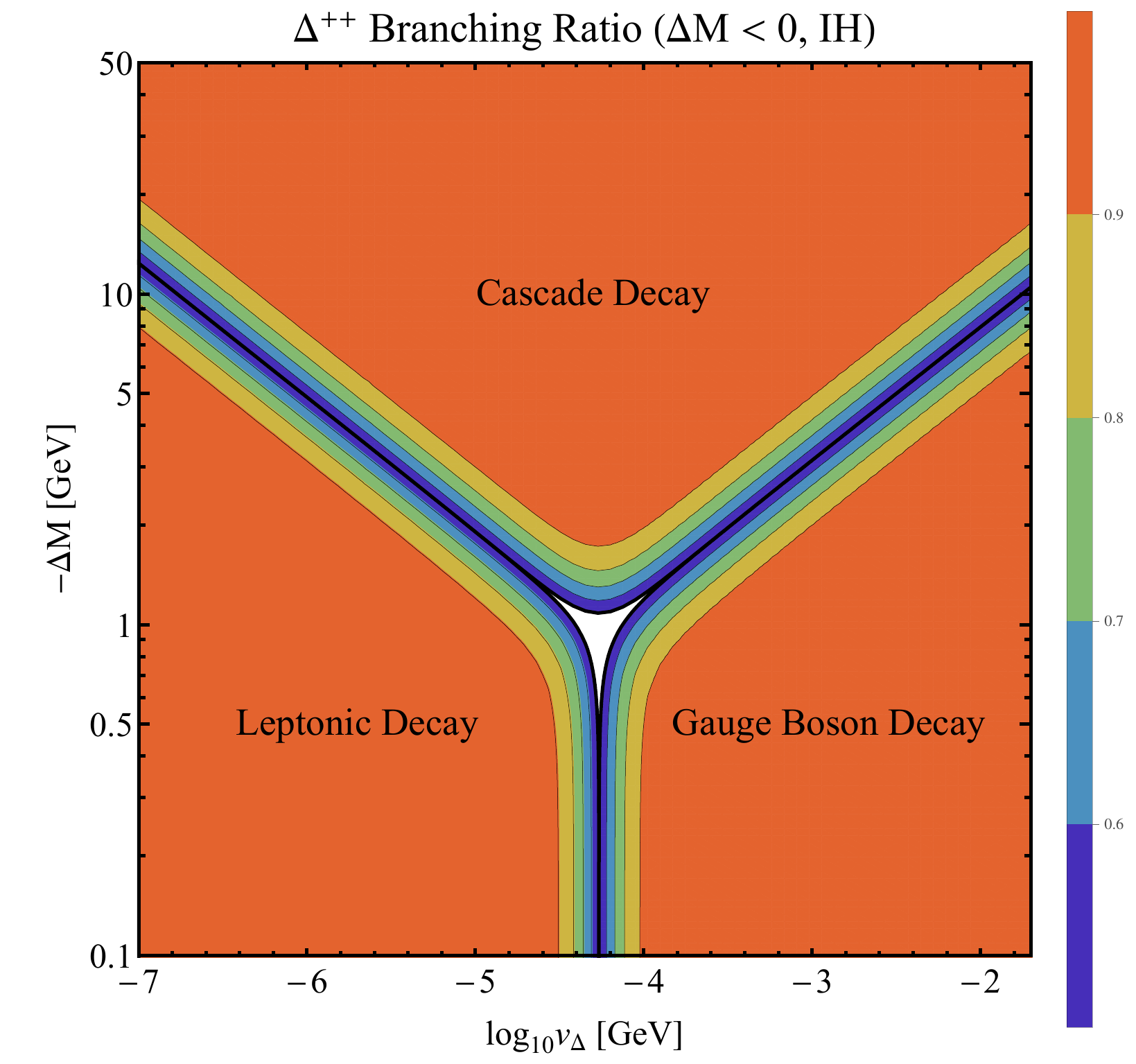}
\includegraphics[scale=0.35]{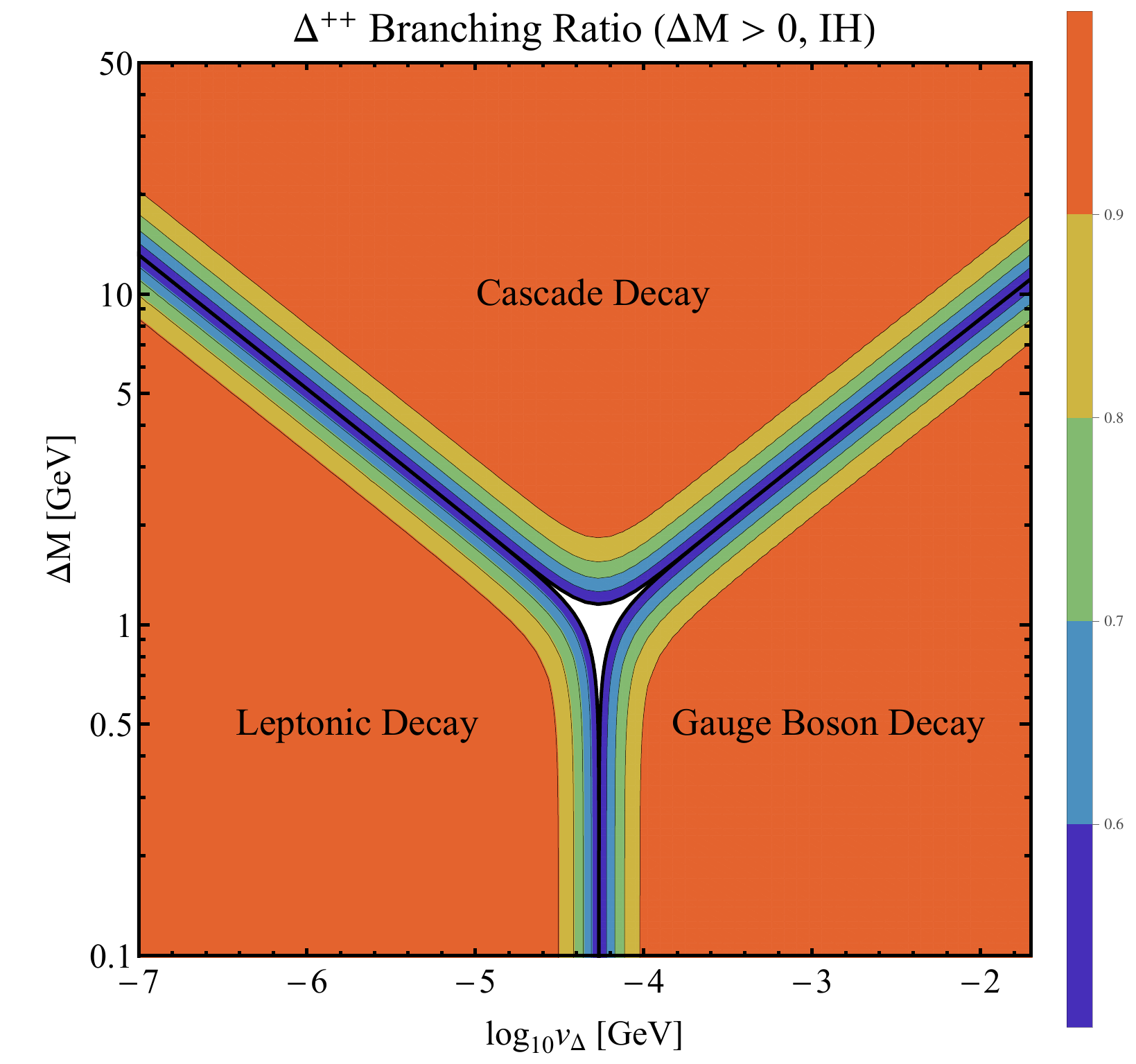}
\caption{Generic decay phase diagram for $\Delta^{\pm \pm}$ decays in the BNT model, with $M_{\Delta^{\pm \pm}} = 400$ GeV. In the top panel we show the scenarios when $\Delta M < 0$ [$Left$] and $\Delta M > 0$ [$Right$] respectively for NH of neutrino masses. In the lower panel the same is shown for IH. Here $\Delta M = M_{\Delta^{\pm \pm}}  - M_{\Delta^{\pm \pm \pm}} $  }
\label{fig:H++_decay_N}
\end{figure}

In Fig.~\ref{fig:H++_decay_N} we present a set of representative decay phase diagrams of $\Delta^{\pm \pm}$ in $\Delta M - v_{\Delta}$ plane for $M_{\Delta^{\pm \pm}} = 400$ GeV. In the top panel we show the scenarios when $\Delta M < 0$ (Left) and $\Delta M > 0$ (Right) respectively for NH of neutrino masses. In the lower panel, the same is shown for IH. The feature of four plots is almost identical. From Eq.~\ref{H++_decay} it is clear that the leptonic decay BR of $\Delta^{\pm \pm}$ falls with $v_{\Delta}$ but the gauge boson decay BR increases with $v_{\Delta}$. The cross-over between leptonic decay dominated region to gauge boson dominated one  happens at $v_{\Delta} = 4.6 \times 10^{-5} \, (5.4 \times 10^{-5})$ GeV for NH (IH) with $\Delta M \sim 0$. Neglecting the loop contribution in the leptonic couplings of Eq.~\ref{FR_Yuk_tot} will shift the cross-over point to a $18 \%$ higher value in $v_{\Delta}$ for both NH and IH. On the other hand, cascade decay channels open up for $\Delta M \neq 0$ and they becomes dominant for $\Delta M \approx 2 - 20$ GeV depending on the exact value of $v_{\Delta}$. Now, a few comments are in order for cascade decay channels. Clearly, for $\Delta M$ below the charged pion mass of 140 MeV the only cascade decay channels open are $\Delta^{\pm \pm} \rightarrow \Delta^{\pm} l^{\pm} \nu_l \,\, (l=e, \mu)$. Once the pion channel is open, it will dwarf the leptonic channels decay width. Then at their respective masses, other charged mesons like kaon channels will be accessible. However, they will always be sub-dominant compared to the pion channel. For $\Delta M > m_{\tau}$ the third lepton channel will be available. Finally, for $\Delta M \sim \mathcal{O}(2$ GeV) the light quarks will cease to be confined, and they can be treated as free particles. So, at this stage we can ignore the mesonic decay channels and replace them by $\Delta^{\pm \pm} \rightarrow \Delta^{\pm} q \overline{q'}$.  

\begin{figure}[!htp]
\includegraphics[scale=0.2]{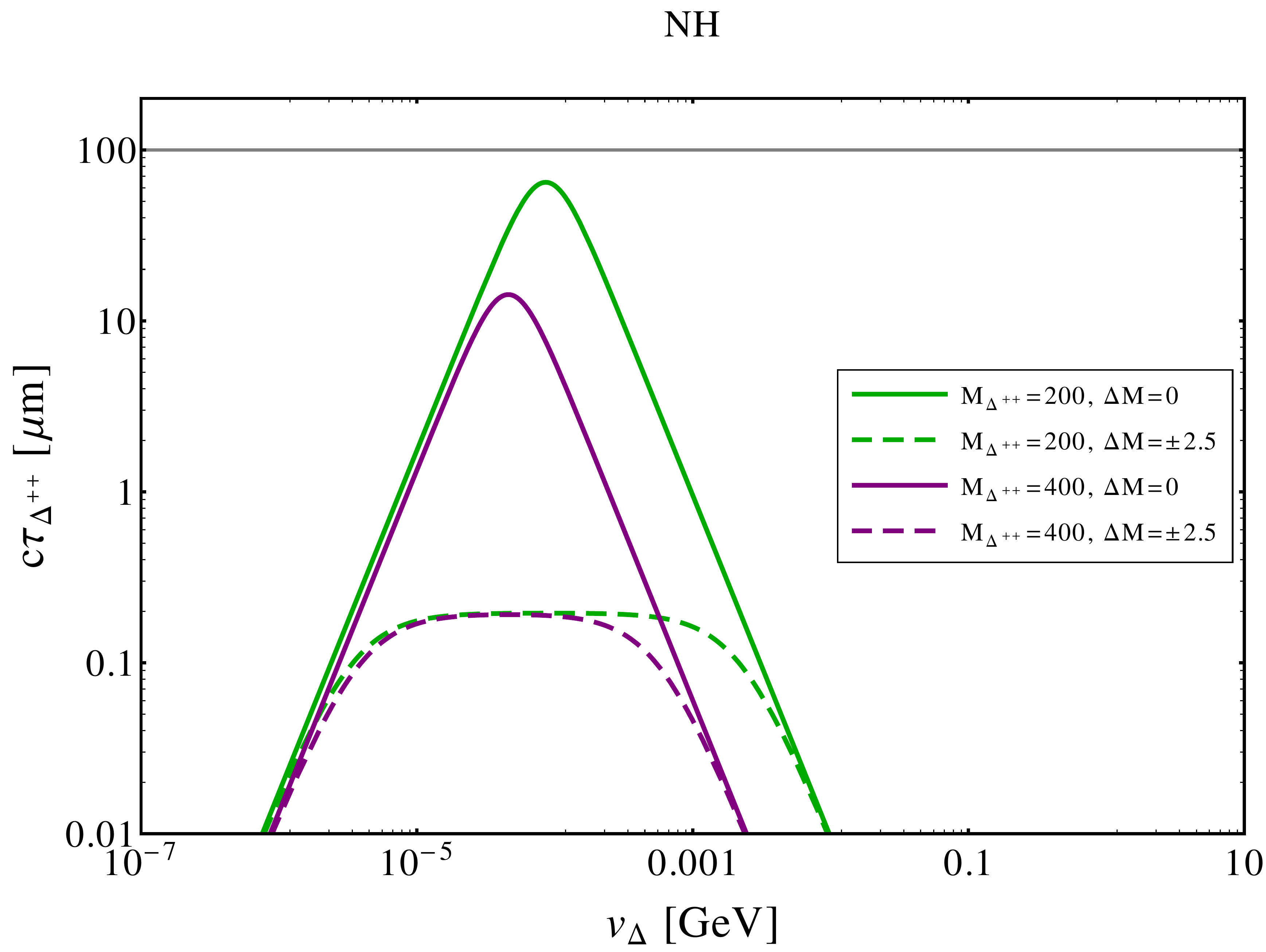}
\includegraphics[scale=0.2]{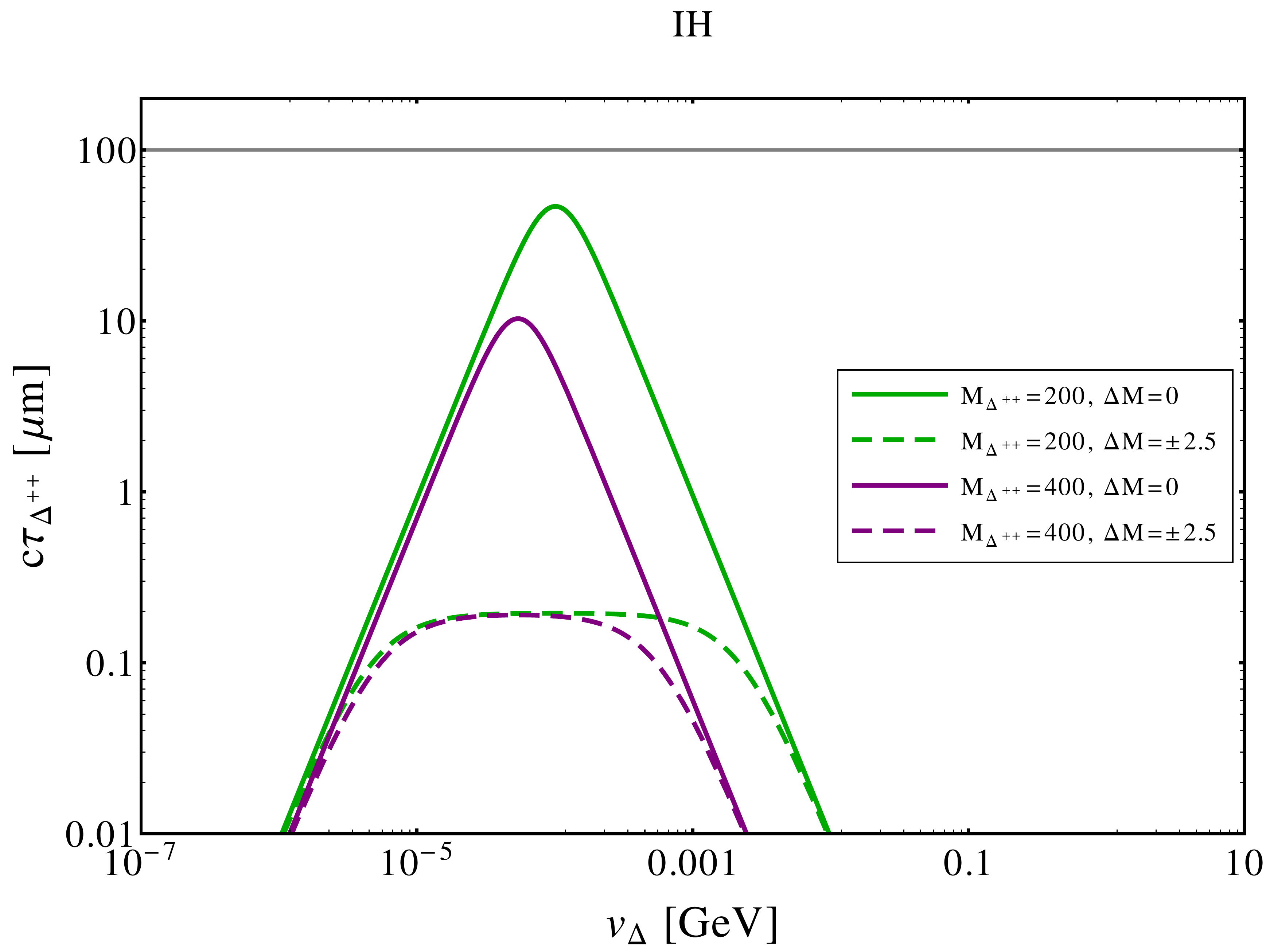}
\caption{Proper decay length of $\Delta^{\pm \pm}$ for different values of $M_{\Delta^{\pm \pm}}$ and $\Delta M$ for both NH [$Left$] and IH [$Right$] of neutrino masses. The gray horizontal lines in both panels refer to the limiting value of $c\tau$, up to which prompt-lepton searches at the LHC remain sensitive.}
\label{fig:H++_pdl}
\end{figure}
\begin{figure}[!htp]
\includegraphics[scale=0.25]{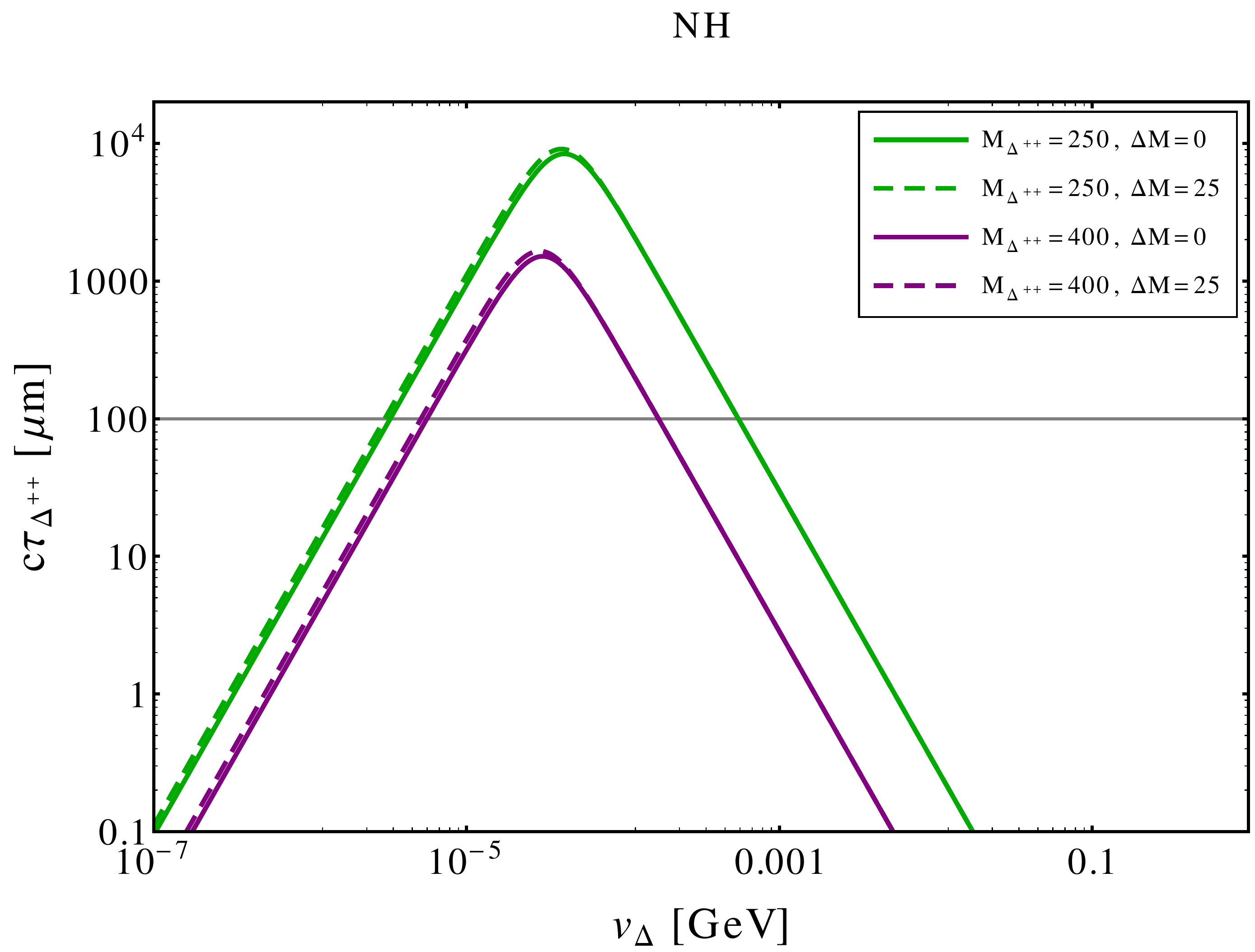}
\includegraphics[scale=0.25]{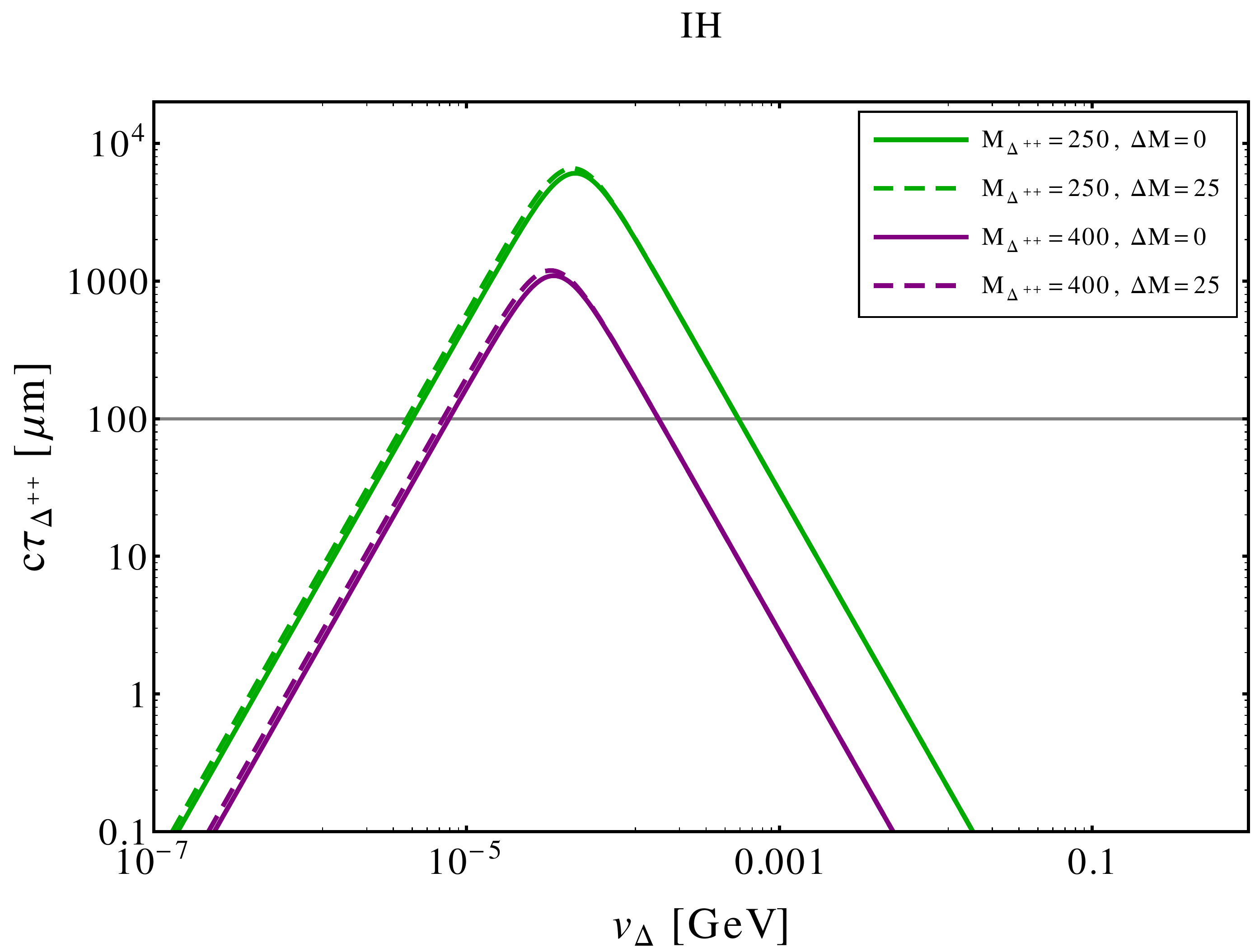}
\caption{Proper decay length of $\Delta^{\pm \pm \pm}$ for different values of $M_{\Delta^{\pm \pm \pm}}$ and $\Delta M$ for both NH [$Left$] and IH [$Right$] of neutrino masses. The gray horizontal lines in both panels refer to the limiting value of $c\tau$, up to which prompt-lepton searches at the LHC remain sensitive.}
\label{fig:H+++_pdl}
\end{figure}

Let us focus now on the total decay width $\Delta^{\pm \pm}$. We have seen above that the total decay width of $\Delta^{\pm \pm}$ depends on neutrino and Higgs quadruplet parameters. In Fig.~\ref{fig:H++_pdl} we present the proper decay length, $c \tau$, of $\Delta^{\pm \pm}$ for four different settings of $M_{\Delta^{\pm \pm}}$ and $\Delta M$ for both NH ($Left$ panel) and IH ($Right$ panel). As seen in the Fig.~\ref{fig:H++_pdl} that $c \tau \gtrsim 10$ $\mu$m is achievable for $M_{\Delta^{\pm \pm}} \lesssim 200$ GeV. A general feature of both plots of the above figure is that the proper decay length is maximum when the cross-over between $ll$ and $WW$ dominant regions happens at $v_{\Delta} \sim 10^{-5} - 10^{-4}$ GeV with $\Delta M =0$. However, the introduction of even a tiny mass-splitting reduces $c \tau$ drastically since the cascade decay channels start dominating. Cascade decay widths are not tiny since they are not proportional to small parameter $v_{\Delta}$ or $m_{\nu}$. In Fig.~\ref{fig:H++_pdl} we show few cases for $\Delta M = \pm 2.5$ GeV to illustrate this behaviour. Given the total decay width of $\Delta^{\pm \pm}$ we obtained it can not be a long-lived charged particle but they can possibly give rise to large displaced vertices. 
To place our calculated $c \tau_{\Delta^{\pm \pm}}$ in some perspective we want to draw the readers attention to the latest CMS search of $\Delta^{\pm \pm}$~\cite{CMS:H++}. This prompt lepton study is sensitive to lepton tracks that start from a distance of $\mathcal{O}(100 \mu m)$ from primary vertex (see Section 4 of the above reference). Also, CMS initiate their displaced vertex searches for a proper decay length of $\mathcal{O}(100 \mu m)$~\cite{CMS_tracker}. We highlight this threshold proper decay length value by gray horizontal lines in Figs.~\ref{fig:H++_pdl} and \ref{fig:H+++_pdl} . Hence, when BRs of $\Delta^{\pm \pm}$ in $ll$ and $WW$ channels are comparable, it may remain beyond the traditional prompt-lepton searches of the LHC for a small range of $\Delta^{\pm \pm}$ mass ($\lesssim 200$ GeV) with $\Delta M \sim 0$. 
 

\begin{figure}[!htp]
\includegraphics[scale=0.25]{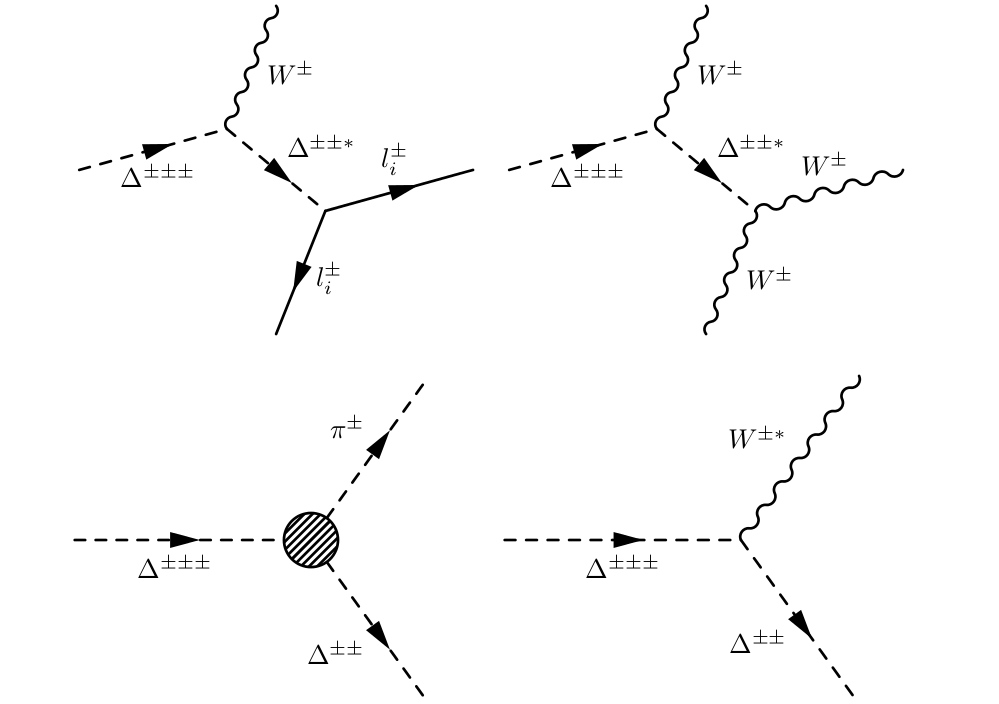}
\caption{Feynman diagrams for decay of $\Delta^{\pm \pm \pm}$. The top two diagrams are for $\Delta M > 0$ and the bottom two diagrams are for $\Delta M <0$}
\label{fig:H+++_decay}
\end{figure}

\begin{figure}[!htp]
\includegraphics[scale=0.35]{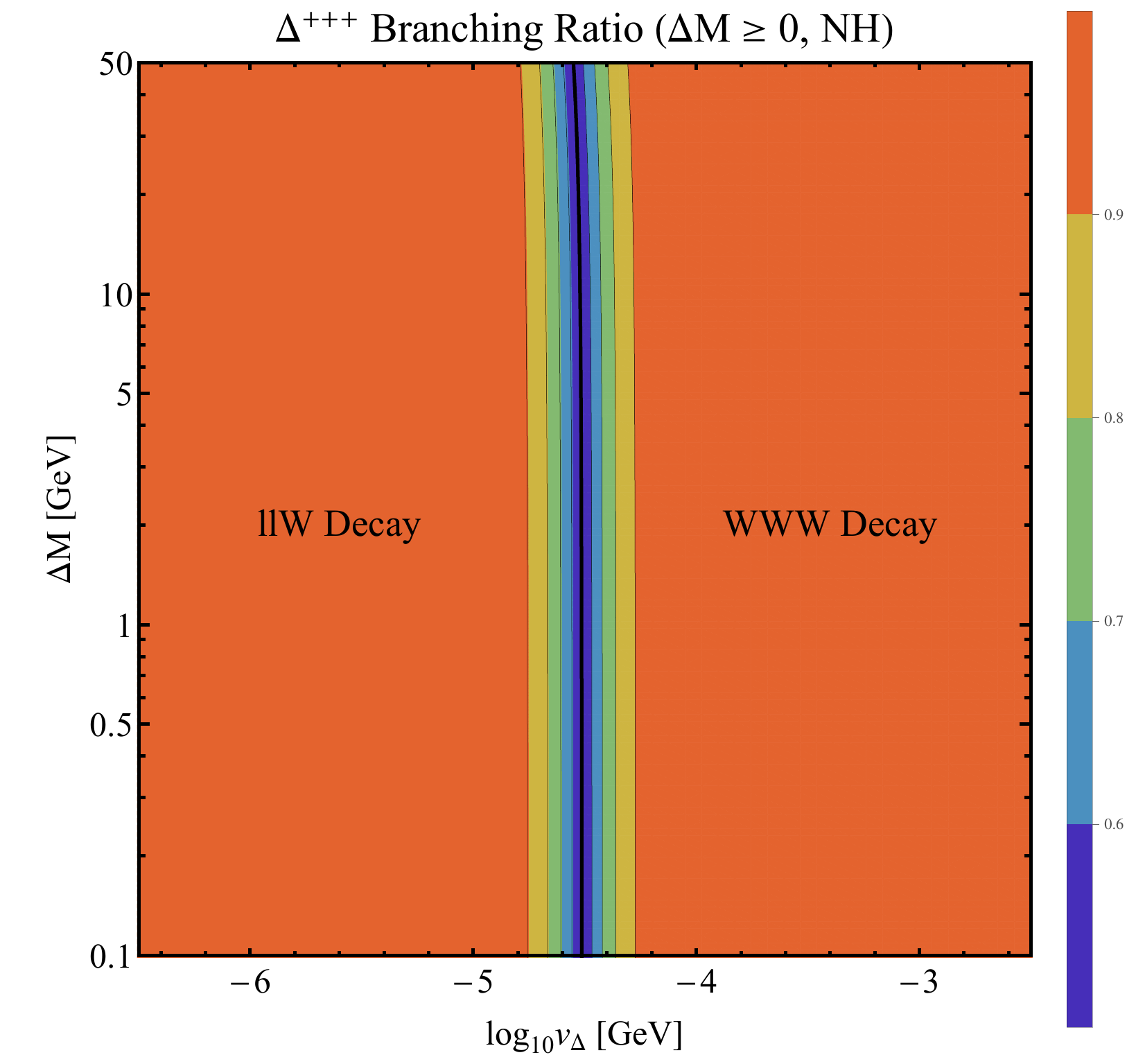}
\includegraphics[scale=0.35]{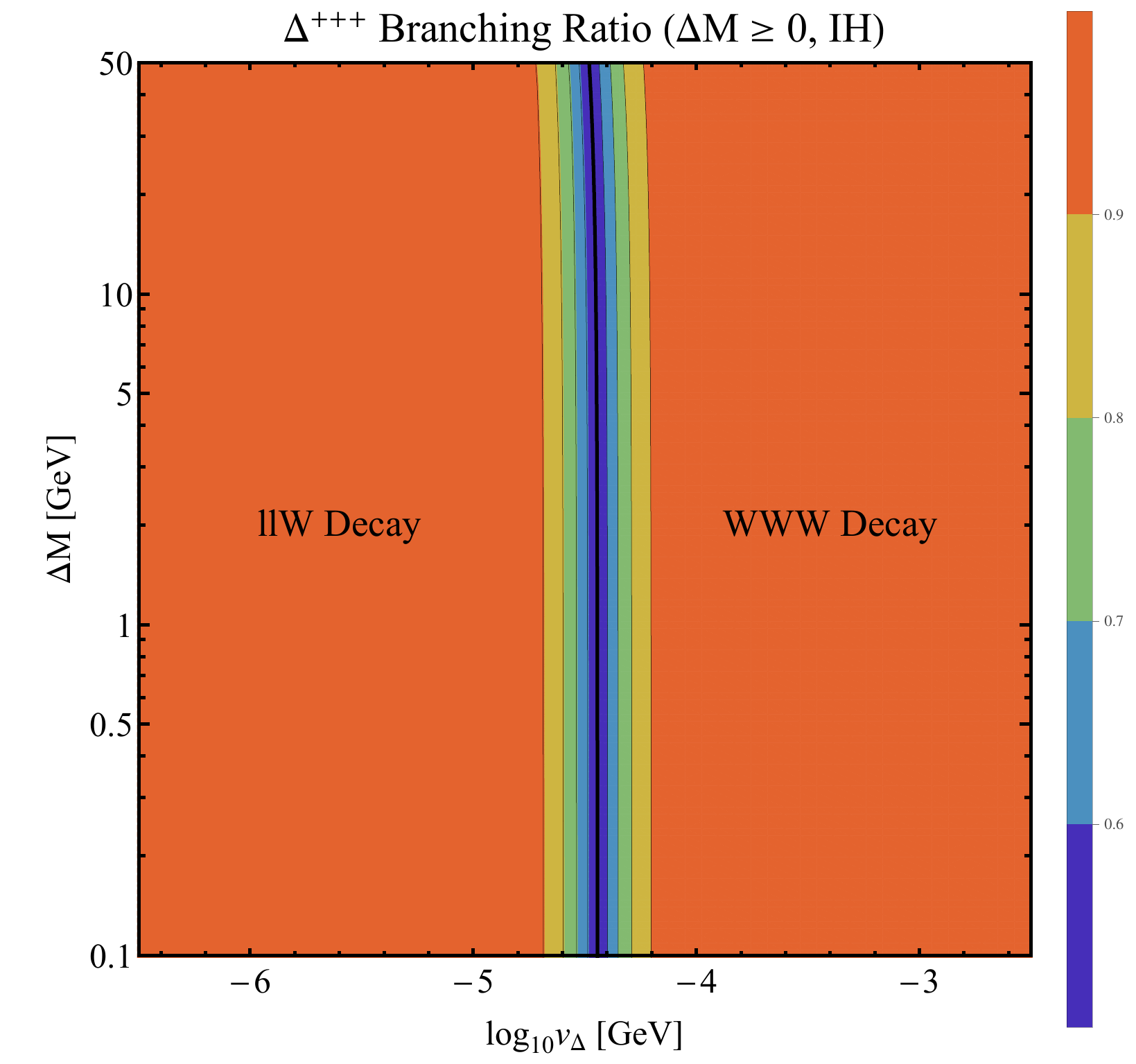}
\caption{Generic decay phase diagram for $\Delta^{\pm \pm \pm}$ decays in the BNT model, with $M_{\Delta^{\pm \pm \pm}} = 400$ GeV and $\Delta M \geq 0$, for both NH [$Left$] and IH [$Right$] of neutrino masses.  }
\label{fig:H+++_decay_N}
\end{figure}

Finally, we investigate various decay channels of the triply-charged Higgs. In the BNT model $\Delta^{\pm \pm \pm}$ can be the lightest (heaviest) particle of the quadruplet for the $\Delta M > 0 \, (\Delta M < 0)$ case. In the first case it can only decay in three-body final states $llW$ or $WWW$ via an off-shell $\Delta^{\pm \pm}$ exchange. In the latter case it will always decay to either $\Delta^{\pm \pm} W^{\pm*}$ or $\Delta^{\pm \pm} \pi^{\pm}$. the relevant Feynman diagrams are presented in Fig.~\ref{fig:H+++_decay}. Decay of $\Delta^{\pm \pm \pm}$, when it is the lightest, is an unique feature of this model. We discuss these decay channels in detail below. On the other hand, for $\Delta M < 0$ the decay of $\Delta^{\pm \pm \pm}$ is very similar to $\Delta^{\pm \pm}$ decay and one can easily convert the results of Eq.~\ref{H++_decay} for this purpose. The decay widths of $\Delta^{\pm \pm \pm}$ for $\Delta M \geq 0$ scenarios are given by 
\bea
\Gamma(\Delta^{\pm \pm \pm} \rightarrow l^{\pm}_i l^{\pm}_j W^{\pm}) & = & \dfrac{g^2}{1536 (1+\delta_{ij}) \pi^3} \dfrac{M_{\Delta^{\pm \pm \pm}} {(m_{\nu})_{ij}^{\text{tot}}}^2}{v_{\Delta}^2} \, J,
 \nn \\
\Gamma(\Delta^{\pm \pm} \rightarrow W^{\pm} W^{\pm} W^{\pm}) &=& \dfrac{3 g^6}{4096 \pi^3} \dfrac{M_{\Delta^{\pm \pm \pm}}^5 v_{\Delta}^2}{M_W^6} \, I,
\label{H+++_decay}
\eea
where $I,J$ are dimensionless integrals, with values $\approx 1$ in the limit $M_{\Delta^{\pm \pm \pm}} \gg M_W$ and $M_{\Delta^{\pm \pm \pm}} \gg \Gamma_{\Delta^{\pm \pm \pm}}$. The decay phase diagram of $\Delta^{\pm \pm \pm}$ is shown in Fig.~\ref{fig:H+++_decay_N} for $M_{\Delta^{\pm \pm \pm}} = 400$ GeV. We see from Fig.~\ref{fig:H+++_decay_N} that $llW$ decays of $\Delta^{\pm \pm \pm}$ dominate for $v_{\Delta} < 3.1 \times 10^{-5} \, (3.6 \times 10^{-5})$ GeV and the $WWW$ decay dominates otherwise for NH (IH). Similar to $\Delta^{\pm \pm}$ decay, neglicting the dimension-5 contribution in the couplings of Eq.~\ref{FR_Yuk_tot} will move the cross-over point by $17 \%$ in $v_{\Delta}$ to the higher side. The mass-splitting has minimal impact on the decay phase diagrams.

Since $\Delta^{\pm \pm \pm}$ decays to three body final states for $\Delta M \geq 0$, its proper decay decay length is expected to be very large as confirmed by Fig.~\ref{fig:H+++_pdl}. For the range of $\Delta^{\pm \pm \pm}$ mass that is not excluded by EWPT, $c \tau$ can be as large as few mm. However, for heavier masses it falls sharply, as expected. Similar to ${\Delta^{\pm \pm}}$, $c \tau$ is maximum for a value of $v_{\Delta}$ where the transition happens from $llW$ dominated decay to $WWW$ dominated decay of $\Delta^{\pm \pm \pm}$. In general, the effect of mass-splitting is marginal since in $\Delta M \geq 0$ case $\Delta^{\pm \pm \pm}$ is the lightest member of the quadruplet and no cascade channel is available. Nonetheless, it can change the decay length marginally in the $llW$ dominated region due to the mass-splitting entering in dimension-5 contribution to Yukawa couplings via $\Delta^0$ and $\Delta^{\pm}$ mass. 
Thus, we can infer beyond any reasonable doubt that for a large range of parameter space where $llW$ and $WWW$ decay widths are commensurable, $\Delta^{\pm \pm \pm}$ will elude any prompt lepton search at the LHC. In contrast, for $\Delta M < 0$ scenario $\Delta^{\pm \pm \pm}$ always decay via cascade and such channels have large decay width, which makes them less interesting. 

\subsection{Searches for $\Delta^{\pm \pm}$ at the LHC}
\label{sec:H++_bound}

The LHC experiments are searching for doubly charged Higgs boson for some time. CMS collaboration has made public their latest Run-II analysis with 12.9 fb$^{-1}$~\cite{CMS:H++} of data. With 36.1 fb$^{-1}$~\cite{ATLAS:H++} of data ATLAS offer similar exclusion limits. 
 Two crucial aspects of the CMS analysis are that they only consider scenarios where $\Delta M = 0$ and also assume that $\Delta^{\pm \pm}$ decays $100\%$ to a particular flavor combination of $l^{\pm} l^{\pm}$. Ref.~\cite{Jana} also use LHC Run-I data to impose bounds on $\Delta^{\pm \pm}$ in the context of the BNT model. However, in a realistic scenario, consistent with available neutrino mass and mixing data, no leptonic channel will have $100 \%$ BR. Hence, the novelty of our analysis is to take into account a benchmark for both NH and IH, as outlined in Section~\ref{sec:Nu_Mass_fit}, and investigate how the limits relax in each case. 

CMS conduct their search for doubly charged Higgs in exactly $3l$ final state for its associated production with a singly charged Higgs. In the BNT model, $\Delta^{\pm \pm}$ can also be produced in association with $\Delta^{\pm \pm \pm}$, which can potentially double the production cross-section. However, for this channel $\Delta^{\pm \pm \pm} \rightarrow l^{\pm} l^{\pm} W^{\pm} \, ( \Delta^{\pm \pm} W^{*\pm})$ decay for $\Delta M > 0 \, (\Delta M < 0)$ case will give rise to extra leptons in the final state and they will not pass the additional lepton-veto criteria of the CMS analysis. In contrast, pair production of $\Delta^{\pm \pm}$ for the $\Delta M > 0$ case mentioned above will be sensitive to this study if one lepton is lost or mistagged, but given the range of $M_{\Delta^{\pm \pm}}$ we are interested in, the occurrence of such events is very unlikely. This is because the decay of $\Delta^{\pm \pm}$ leads to appreciably energetic leptons~\cite{Dutta:2014dba}, which has high tagging efficiency.  

On the other hand, for the $4l$ study CMS does not require any veto on additional leptons. So, for this final state, not only the pair production of $\Delta^{\pm \pm}$ will contribute but also in the $\Delta M < 0$ case the pair production of $\Delta^{\pm \pm \pm}$ will assist. Therefore, the limits drawn from this study will have some asymmetry between $\Delta M <0$ and $\Delta M > 0$ cases. Another important point we need to address for pair productions of the doubly and triply charged Higgs bosons is whether to include PF in deriving the limits or not. As mentioned previously we choose to adopt a conservative approach in this paper and used DY only for our calculation due to large uncertainties associated with the photon PDF. 

\begin{figure}[!htp]
\includegraphics[scale=0.2]{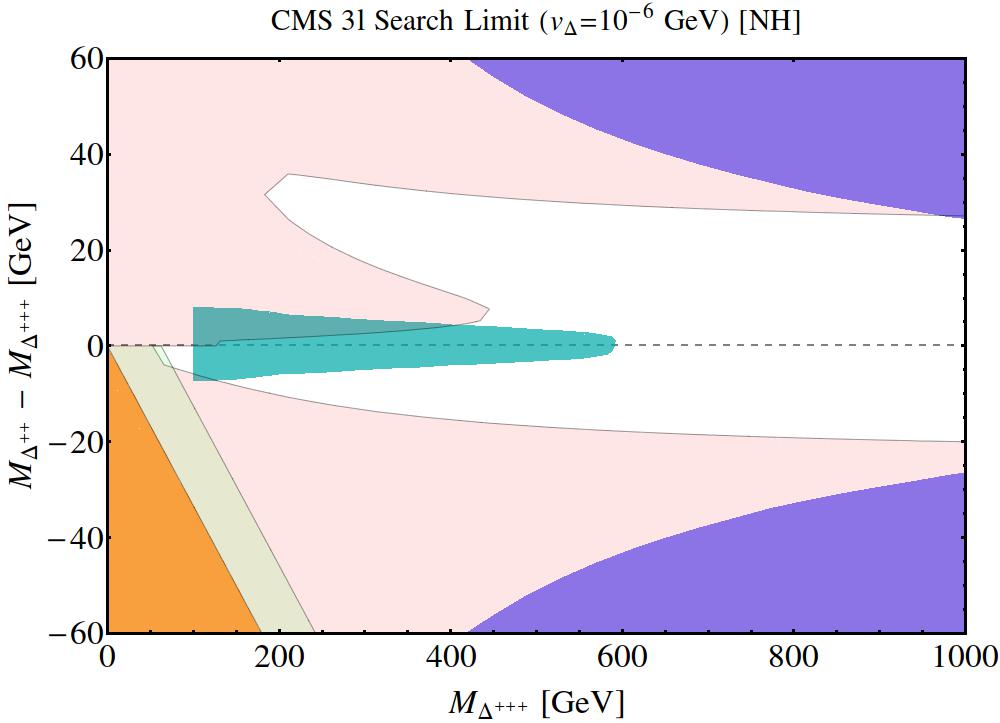}
\includegraphics[scale=0.2]{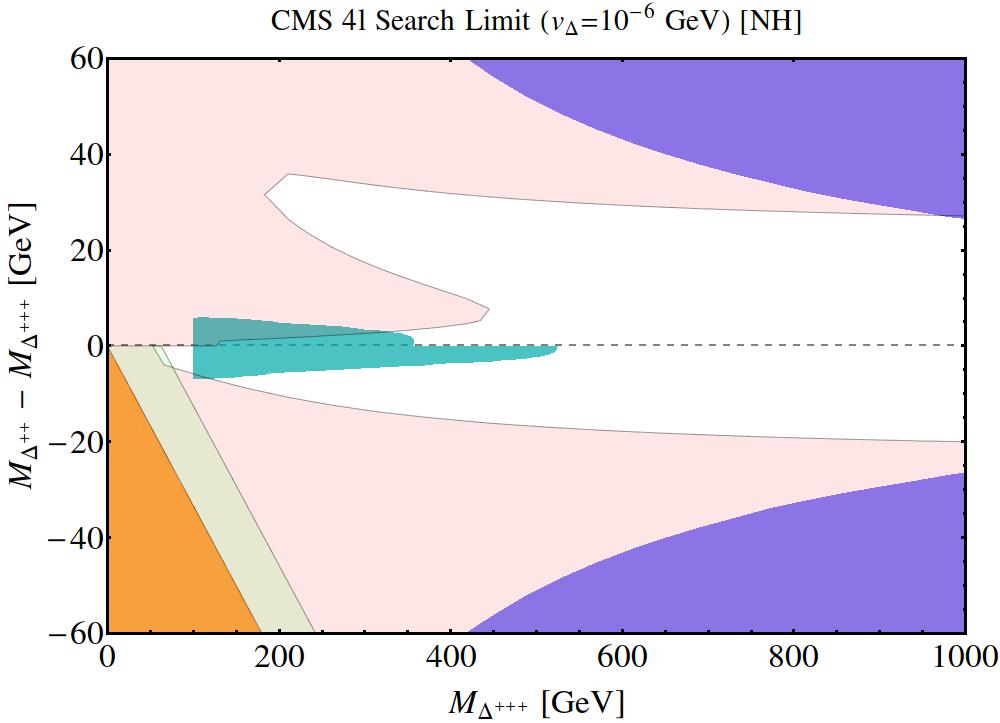}
\\
\includegraphics[scale=0.2]{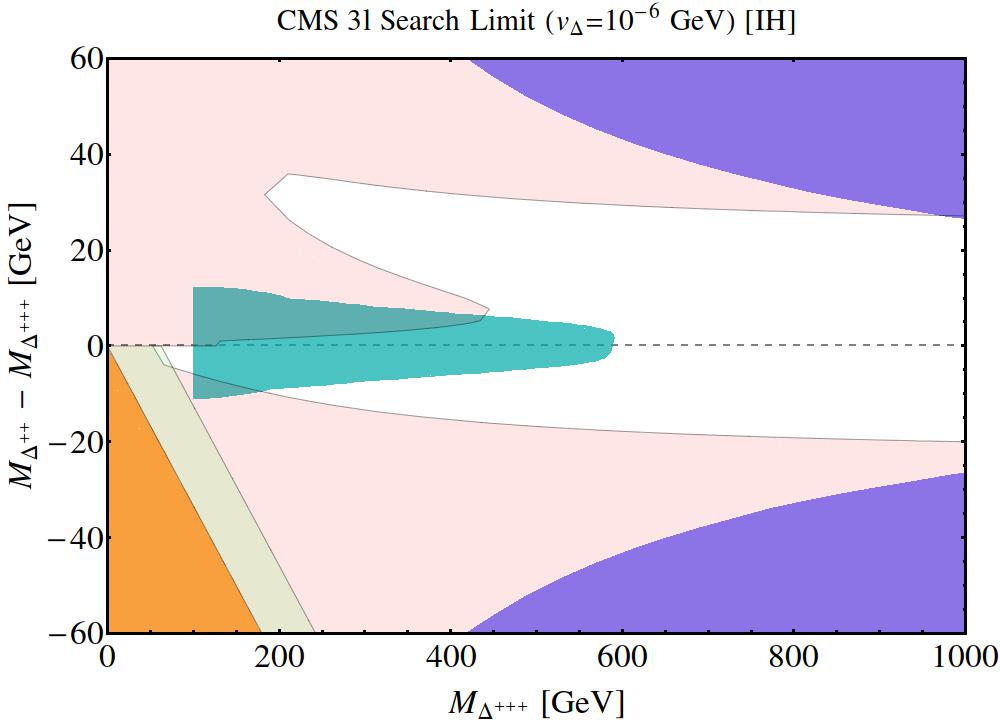}
\includegraphics[scale=0.2]{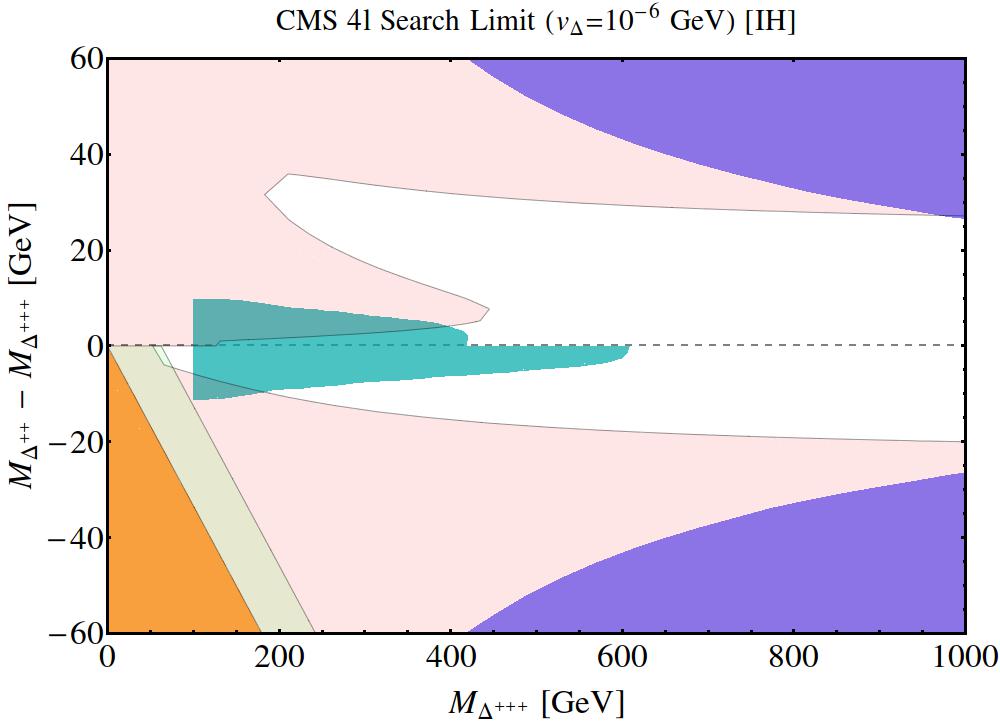}
\caption{Constraints form CMS searches for $\Delta^{\pm \pm}$ using 12.9 fb$^{-1}$ integrated luminosity at $\sqrt{s}=13$  TeV. $v_{\Delta}$ is fixed at $10^{-6}$ GeV so that $\Delta^{\pm \pm}$ decays leptonically  when $\Delta M =0$. We show the limits derived from $3l$ search [Top Left] and $4l$ search [Top Right] for NH by cyan shaded regions. The two figures in the bottom panel are the same for IH. We also impose $c \tau_{\Delta^{\pm \pm}} < 100 \, \mu$m. The bounds derived for NH (IH) are from $\mu \mu \, (ee)$ decay channel. Only DY production is considered in the figure. The other colored regions has the same meaning as Fig.~\ref{fig:EWPT}.}
\label{fig:H++_CMS_3l_ll}
\end{figure}

Also, when $\Delta^{\pm \pm}$ dominantly decays in cascade, it can easily give rise to 3 or 4 leptons in the final state. However, such leptons will come from off-shell $W$ bosons, and the momentum they carry will have an upper bound of $\Delta M$. We have seen from Fig.~\ref{fig:EWPT} that EWPT bound limits $\Delta M \lesssim 30$ GeV for the most part of the range of $M_{\Delta^{\pm \pm}}$ we are studying. We need to juxtapose this limitation with the requirement of the CMS analysis that at least one lepton should have $p_T > 30$ GeV and others should satisfy $p_T > 20$ GeV. Therefore, a tiny amount of cascade events will pass these hard cuts on lepton $p_T$. Furthermore, these soft leptons will not be able to reconstruct the narrow $M_{\Delta^{\pm \pm}}$ mass peak, which is a criterion in the CMS analysis, due to significant momenta will be carried away by missing neutrinos. Hence, we don't consider cascade decay products of $\Delta^{\pm \pm}$ in the subsequent computations. Interestingly, the compressed spectra are very similar to certain supersymmetric scenarios, well studied in the literature~\cite{CompressedSUSY}.

In Fig.~\ref{fig:H++_CMS_3l_ll} we plot the bounds derived from CMS search of Ref.~\cite{CMS:H++}, on top of EWPT excluded regions in $\Delta M - M_{\Delta^{\pm \pm \pm}}$ plane for $v_{\Delta} = 10^{-6}$ GeV. This choice of $v_{\Delta}$ ensures that $\Delta^{\pm \pm}$ decays leptonically when $\Delta M =0$. The exclusion contours from the $3l$ [Left] and $4l$ [Right] final states are shown in the top panel for NH by cyan shaded regions. The bottom panel contains the same for IH. Additionally we require $c \tau_{\Delta^{\pm \pm}} < 100 \, \mu$m so that the leptonic decay products are prompt. As mentioned earlier in the section we consider DY production of $\Delta^{\pm \pm}$ only in the above figure. We should mention here that in Fig.~\ref{fig:H++_CMS_3l_ll} we only show the limits from the flavor combination decay channel that offers the strongest bound. So, for NH and IH we only show bounds derived from $\mu \mu$ and $ee$ channels, respectively. Although $\Delta^{\pm \pm}$ has a large BR to $\tau \tau$ decay for NH, this channel does not impose strong bounds due to poor $\tau$ identification efficiency at the LHC. One may try to combine different channels which will lead to an even stronger bound. However, we don't attempt to do that in this paper. 

\begin{figure}[!htp]
\includegraphics[scale=0.2]{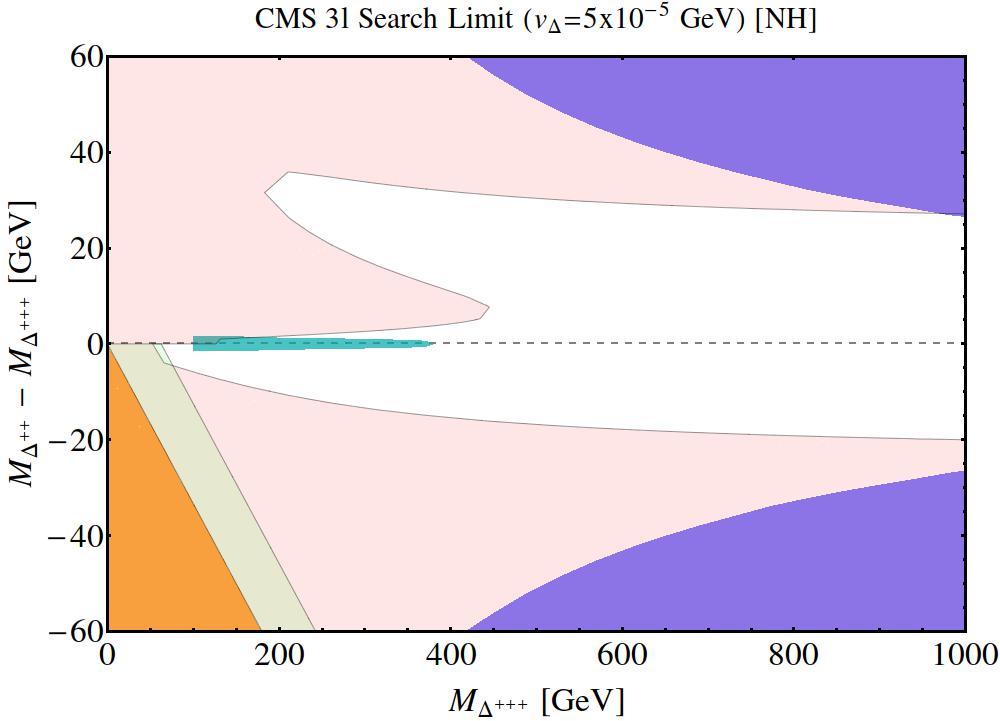}
\includegraphics[scale=0.2]{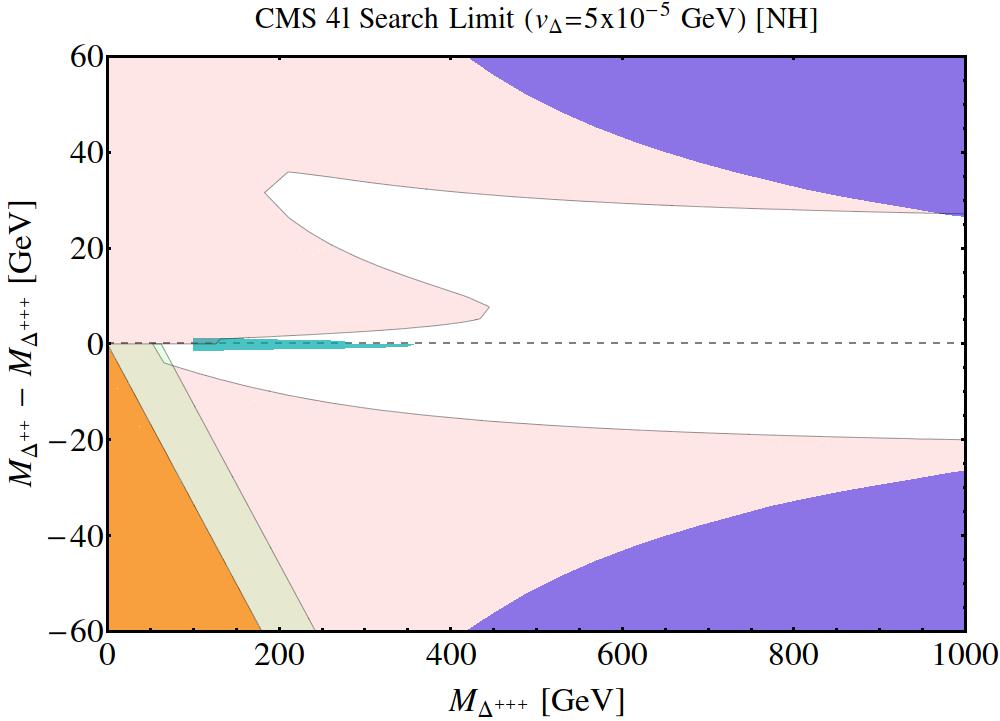}
\\
\includegraphics[scale=0.2]{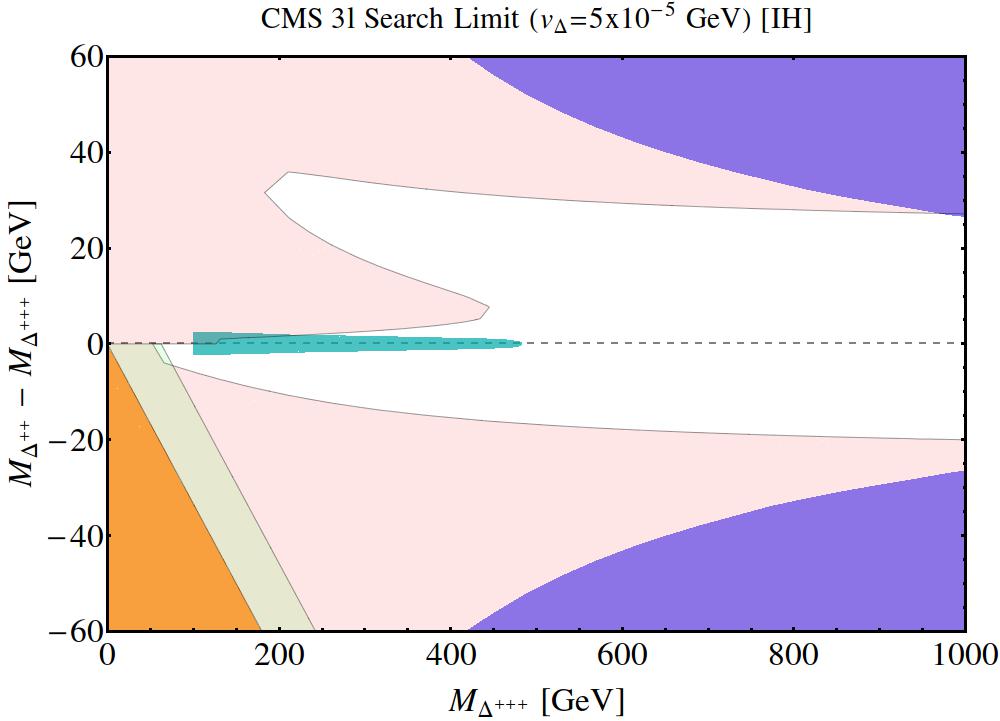}
\includegraphics[scale=0.2]{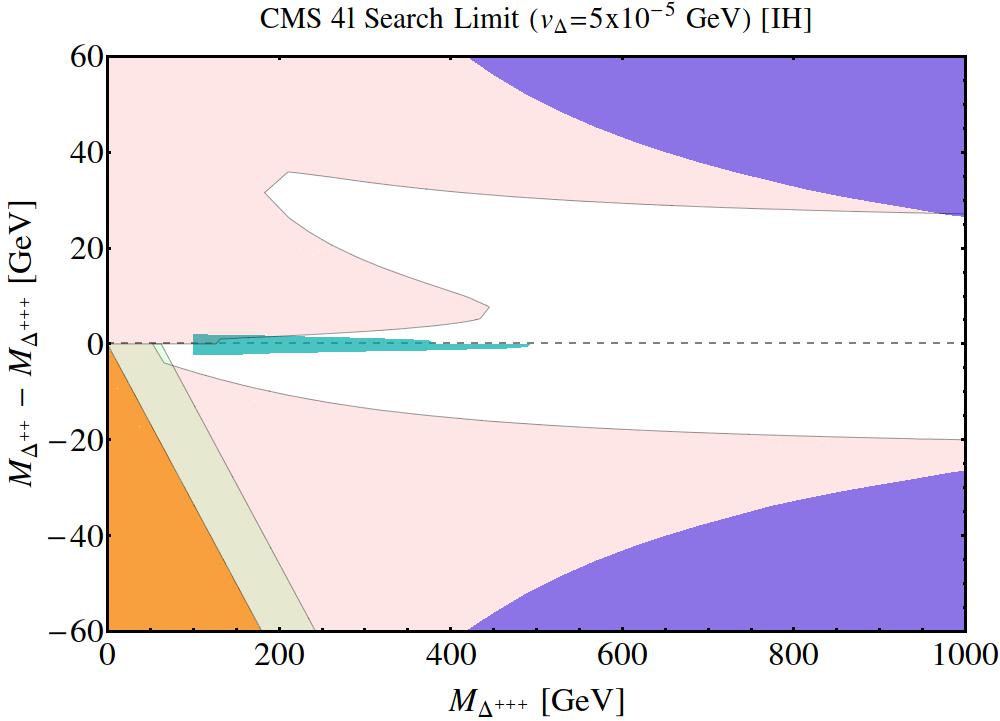}
\caption{Constraints form CMS searches for $\Delta^{\pm \pm}$ using 12.9 fb$^{-1}$ integrated luminosity at $\sqrt{s} = 13$ TeV. $v_{\Delta}$ is fixed at $5 \times 10^{-5}$ GeV so that $\Delta^{\pm \pm}$ decays to a pair of leptons or gauge bosons with equal BR when $\Delta M =0$. We show the limits derived from $3l$ search [Top Left] and $4l$ search [Top Right] for NH by cyan shaded region. The two figures in the bottom panel are the same for IH. We also impose $c \tau_{\Delta^{\pm \pm}} < 100 \, \mu$m. The bounds derived for NH (IH) are from $\mu \mu \, (ee)$ decay channel. Only DY production is considered in the figure. The other colored regions has the same meaning as Fig.~\ref{fig:EWPT}.}
\label{fig:H++_CMS_3l_llWW}
\end{figure}

In general, CMS search for $\Delta^{\pm \pm}$ using 12.9 fb$^{-1}$ integrated luminosity at $\sqrt{s}=13$ TeV bounds $M_{\Delta^{\pm \pm}} \gtrsim 600$ GeV (at 95$\%$ CL) for $\Delta M \lesssim 5$ GeV in the $3l$ final state for both NH and IH. 
For the $4l$ analysis the bounds derived, using DY only, are $M_{\Delta^{\pm \pm}} \gtrsim 600 \, (400)$ GeV for $\Delta M < 0 \, (\Delta M > 0)$ for IH. The bounds for NH are weaker by $\sim 50$ GeV compared to IH. 

Fig.~\ref{fig:H++_CMS_3l_llWW} is the same as Fig.~\ref{fig:H++_CMS_3l_ll} but for $v_{\Delta} = 5 \times 10^{-5}$ GeV. For such a value of $v_{\Delta}$ we have $BR(\Delta^{\pm \pm} \rightarrow l^{\pm} l^{\pm}) \approx BR(\Delta^{\pm \pm} \rightarrow W^{\pm} W^{\pm}) $, when cascade decay channels are not open. As expected, the bounds are relatively weak compared to the previous case. 
Interestingly, the bounds for NH and IH differ appreciably. From the $3l$ analysis we obtain a bound of $M_{\Delta^{\pm \pm}} \gtrsim 400 \, (500)$ GeV for NH (IH), with $\Delta M \sim 0$. Similarly, from the $4l$ final state we get, $M_{\Delta^{\pm \pm}} \gtrsim 350 \, (500)$ GeV for NH (IH), again with $\Delta M \sim 0$. The difference between the NH and IH bounds are due the fact that the cross-over between dominantly $ll$ decay to dominantly $WW$ decay does not happen for the same $v_{\Delta}$ for them. So, for a choice of $v_{\Delta}$ for which $BR(\Delta^{\pm \pm} \rightarrow l^{\pm} l^{\pm}) \approx BR(\Delta^{\pm \pm} \rightarrow W^{\pm} W^{\pm}) $ for IH, the NH BP will be relatively in the $WW$ decay dominated region. 


For a larger value of $v_{\Delta}$ the $WW$ BR will rapidly increase at the expense of $ll$ BR. Hence, the bounds derived from the CMS analysis of Ref.~\cite{CMS:H++} for $v_{\Delta} \gtrsim 10^{-4}$ GeV will be very weak, which will be discussed elsewhere. No dedicated study by CMS or ATLAS exist for $\Delta^{\pm \pm} \rightarrow W^{\pm} W^{\pm}$. However, Ref.~\cite{Kanemura:2014ipa} estimated a bound of $M_{\Delta^{\pm \pm}} > 84$ GeV for such decays using ATLAS Run-I results~\cite{ATLAS:2014kca}.  

\subsection{Signal of $\Delta^{\pm \pm \pm}$ at the LHC}
\label{sec:H+++_LHC}

In the previous section we discussed LHC studies that are searching for $\Delta^{\pm \pm}$. However, $\Delta^{\pm \pm}$ is not exclusive to this model, it may also arise in other models, such as, Georgi-Machacek model \cite{Georgi:1985nv}, Littlest Higgs model \cite{ArkaniHamed:2002qy}, 3-3-1 models \cite{Frampton:1992wt,Pisano:1991ee}, Type II seesaw models~\cite{Type-II}, left-right symmetric models~\cite{lrsm,susylrsm} and radiative neutrino mass models~\cite{Zee}. Discovering/excluding $\Delta^{\pm \pm}$ alone will not identify/falsify the BNT model. In addition, from Figs.~\ref{fig:H++_CMS_3l_ll}, \ref{fig:H++_CMS_3l_llWW}, we have noticed that the LHC can constrain $M_{\Delta^{\pm \pm}}$ for $\Delta M < 5$ GeV only. Hence, to search for  $\Delta^{\pm \pm \pm}$ directly at the LHC is imperative for the validation of the BNT model.   

In this section we present a feasibility study of potential reach of the LHC in search for $\Delta^{\pm \pm \pm}$. We search for $\Delta^{\pm \pm \pm}$ in same-sign (SS) $3l \, (l = e, \mu)$ final state. We have already mentioned that the BNT model is implemented with the {\tt FeynRules$\_$v2.0}~\cite{FR} package. The signal and background events are generated using {\tt MadGraph5$\_$aMC@NLO$\_$v2.5.4} code~\cite{MG5} followed by showering and hadronization  by {\tt PYTHIA\_v8.2}~\cite{pythia} and the detector simulation by {\tt DELPHES\_v3.3}~\cite{delphes}. We produce $\Delta^{\pm \pm \pm}$ by a combination of $pp \rightarrow \Delta^{\pm \pm \pm} \Delta^{\mp \mp \mp} \, + \, \Delta^{\pm \pm } \Delta^{\mp \mp} \, + \, \Delta^{\pm\pm \pm} \Delta^{\mp \mp } $ processes.

The major SM backgrounds for our signal are $t \bar{t} W^{\pm}$+jets. However, $W^{\pm }  Z $+jets and \\$   Z/\gamma^*(\rightarrow l^+l^-) \, Z$+jets may also contribute in case of mis-measurement of the charge of a lepton. The latter backgrounds, in fact, dominate over the former since their production cross-sections are significantly higher. $t \bar{t} Z (\gamma^*)$+jets, $t \bar{t} b \bar{b}$ and $t \bar{t} t \bar{t}$ will also contribute but they are much smaller compared to $t \bar{t} W^{\pm}$~\cite{earlylhc} and we neglect them in our analysis.  All the backgrounds are generated including upto one parton. The MLM scheme~\cite{MLM} for jet-parton matching has been employed to avoid double counting.  For the backgrounds, $W, Z$ bosons and top quarks are decayed in their respective leptonic decay channels with the \texttt{MadSpin}~\cite{MadSpin} module of {\tt MadGraph5}. In contrast, for the  signal samples, the multi-charged Higgs bosons has been decayed within {\tt PYTHIA}. We perform all cross-section calculations at tree-level and do not include any $K$-factor. Therefore, our estimates for signal significance will likely be conservative. We use the default {\tt Delphes 3.3} detector card for various object reconstruction, with jet clustering performed using the anti-kt algorithm. The above detector card employ the following lepton and $b$-quark reconstruction criteria 
\begin{itemize}
\item {\it Lepton identification and efficiency:} electrons and muons are identified for $p_T > 10$ GeV with $|\eta| < 2.4$. While the electron efficiency is 85\% and 95\% for $|\eta| < 1.5$ and $1.5 < |\eta| < 2.4$, respectively, the muon efficiency is kept constant at 95\% over the whole pseudo-rapidity range.
\item {\it Lepton isolation:} lepton isolation is parametrized by $I_{rel} < 0.25 \, (0.12)$ for $\mu \, (e)$, where $I_{rel}$ is the ratio of the sum of transverse momenta of isolation objects (tracks, calorimeter towers, etc) within a $\Delta R = \sqrt{(\Delta \eta)^2 + (\Delta \phi)^2}= 0.5$ cone around a candidate, and the candidate's transverse momentum.
\item {\it $b$-tagging efficiency:} the $b$-tagging efficiency is just above 70\% for transverse momenta between 85 and 250~GeV, with a mistag rate $\lesssim 2 \%$, coming from $u,d,c,s,g$ jets, over the same energy range.
\end{itemize}

Next, using the above reconstructed objects we list the selection cuts used in our SS $3l$ study. They are -
\begin{enumerate}
\item {\it Basic cuts:} The signal and background events are preselected with the requirement of $p_{T_l (j) } > 10 \, (20)$ GeV and $|\eta_{l (j)}| < 2.4 (5)$. The subsequent cuts applied on the pre-selected events are optimized to maximize the signal significance, $S/\sqrt{S+B}$, where $S$ and $B$ denote signal and background rates.
\item {\it $\geq$ 3 SS leptons:} We select events with at least 3 isolated SS light leptons ($e, \mu$).
\item {\it Lepton $p_T$ cuts:} We impose the following stringent $p_T$ cuts on the selected SS leptons, $p_{T_{l_1}}> 30$ GeV, $p_{T_{l_2}}> 30$ GeV and $p_{T_{l_3}}> 20$ GeV.
\item {\it Missing energy cut:} The missing energy cut is not very effective for the signal process after applying the hard lepton $p_T$ cuts. The $p_T$ cuts force the QCD radiation into a regime where jets produce a fair amount of missing energy as well. Hence, we enforce a nominal $\met > 30$ GeV.
\item {\it $Z$-veto:} If leptons having a charge opposite of that of the three tagged leptons are present in an event, we veto such an event if any opposite-sign same flavor lepton pair combination satisfy 80 GeV $< M_{l^{\pm}l^{\mp}} < $100 GeV.   
\item {\it $b$-veto:} We veto any events with one or more identified $b$-tagged jets, with $p_T > 20$ GeV and $|\eta| < 2.5$.
\end{enumerate}


\begin{table}[!htp] 
\begin{tabular}{|c |c |c |c |c |c|} 
\hline  
     $(M_{\Delta^{\pm \pm \pm}}, \Delta M, v_{\Delta}) $  & Selection  & Signal  & $W Z$+ jets & $Z l^+ l^-$+ jets  & $t \bar{t} W $+ jets
     \\
GeV  & Cuts  & [fb] & [fb] & [fb]   & [fb]
\\          
          \hline    \hline  
                     
\multirow{5}{*}{ $(400,0, 10^{-6})$}  &  Basic cuts     & $23.35 \pm 0.1044$   & $1167 \pm 1.948$  & $155.5 \pm 0.2596$ & $24.41 \pm 0.0446$ \\
    \hline 
    & $\geq 3$ SS leptons & $1.670 \pm 0.0279$ & $0.0975 \pm 0.0178$ & $0.0347 \pm 0.0039  $ & $0.0044 \pm 0.0006$ \\
    \hline
     & Lepton $p_T$ cuts & $1.443 \pm 0.0260$ & $0.0227 \pm 0.0086$ & $0.0087 \pm 0.0019  $ & $0.0017 \pm 0.0004$ \\
     \hline
      & $Z$-veto & $1.2847 \pm 0.0245$ & $0.0130 \pm 0.0065$ & $0.0039 \pm 0.0013  $ & $0.0015 \pm 0.0003$ \\
      \hline
      & $b$-veto & $1.1946 \pm 0.0236$ & $0.0130 \pm 0.0065$ & $0.0039 \pm 0.0013  $ & $0.0003 \pm 0.0002$ \\
      \hline
\end{tabular} 
\caption{Summary of the signal and the background cross-sections and corresponding statistical errors at our chosen benchmark point, after each kinematical cut, for NH of neutrino masses. The LHC center of mass energy is 14 TeV. In the first row, all background cross-sections are presented after decaying top quarks and $W, Z$ bosons in their respective leptonic channels within \texttt{MadSpin}.} 
\label{tab:SS3l_X-sec}
\end{table}

Table~\ref{tab:SS3l_X-sec} gives the signal and background cross-sections at $\sqrt{s}= 14$ TeV after applying each cut listed above, accompanied by corresponding statistical errors. For the signal we choose a BP with $(M_{\Delta^{\pm \pm \pm}}, \Delta M, v_{\Delta}) = $ (400, 0, $10^{-6}$) GeV for NH of neutrino masses. $v_{\Delta}$ is chosen to be $10^{-6}$ GeV to ensure BR$(\Delta^{\pm \pm \pm} \rightarrow l^{\pm} l^{\pm} W^{\pm}) =1$, when $\Delta^{\pm \pm \pm}$ is the lightest member of the quadruplet. Here we use 14 TeV of center of mass energy as opposed to 13 TeV used in previous sub-sections. This is due to the fact that we intend to estimate the discovery potential of $\Delta^{\pm \pm \pm}$ not only at an immediately achievable integrated luminosity of 100 fb$^{-1}$, but also at high luminosity of 3000 fb$^{-1}$. The LHC is expected to run at 14 TeV for that high luminosity benchmark. Elevating the center of mass energy to 14 TeV for our simulation leads to an increase in overall cross-section of $pp \rightarrow \Delta^{\pm \pm \pm} \Delta^{\mp \mp \mp} \, + \, \Delta^{\pm \pm } \Delta^{\mp \mp } \, + \, \Delta^{\pm \pm} \Delta^{\mp \mp } $ processes by $\sim 20 \%$. Clearly, Table~\ref{tab:SS3l_X-sec} indicates that the final state we are studying is almost devoid of SM background for our chosen BP. We don't show the the effect of $\met$ cut in the above cut-flow table since both signal and background has $\sim 100 \%$ efficiency for that cut.
\begin{figure}[!htp]
\includegraphics[scale=0.35]{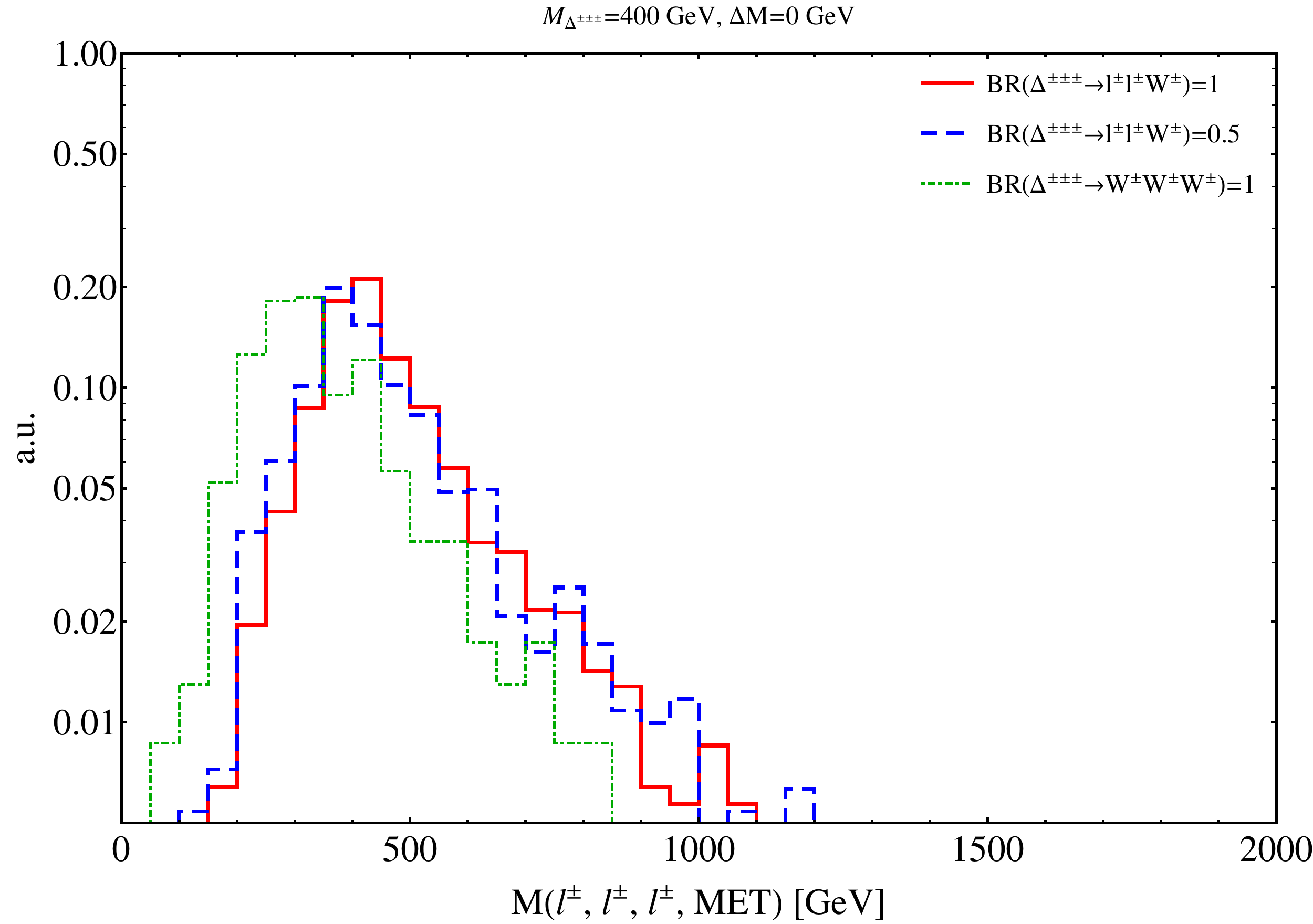}
\caption{The invariant mass of the three leading SS light leptons and $\met$ for the signal, after all the kinematic cuts. We keep $(M_{\Delta^{\pm \pm \pm}},\Delta M)=$(400, 0) GeV fixed for three distinct BR scenarios. The BP is chosen for NH of neutrino masses.}
\label{fig:MlllMET}
\end{figure}

Fig.~\ref{fig:MlllMET} shows the invariant mass of the three leading SS leptons and $\met$ for the signal with $(M_{\Delta^{\pm \pm \pm}, \Delta M})=$(400, 0) GeV for NH. We set $v_{\Delta} = 10^{-6}$ GeV, $6 \times 10^{-5}$ GeV, $5 \times 10^{-3}$ GeV to achieve BR$(\Delta^{\pm \pm \pm} \rightarrow l^{\pm} l^{\pm} W^{\pm}) =1$, BR$(\Delta^{\pm \pm \pm} \rightarrow l^{\pm} l^{\pm} W^{\pm}) =0.5$, and BR$(\Delta^{\pm \pm \pm} \rightarrow W^{\pm} W^{\pm} W^{\pm}) =1$, respectively. While we see a peak close to but not exactly at  $M_{\Delta^{\pm \pm \pm}}$ for the first two cases, the peak is shifted significantly to a lower mass for the third case due to a large fraction of momentum carried by neutrinos coming from  three $W$ decays.

\begin{figure}[!htp]
\includegraphics[scale=0.2]{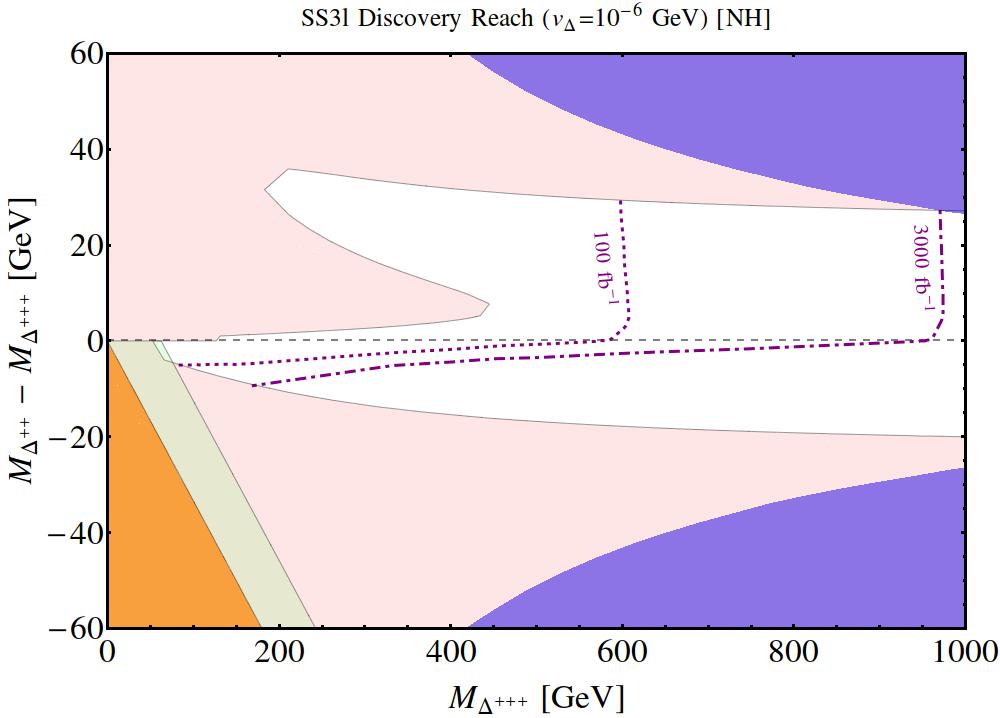}
\includegraphics[scale=0.2]{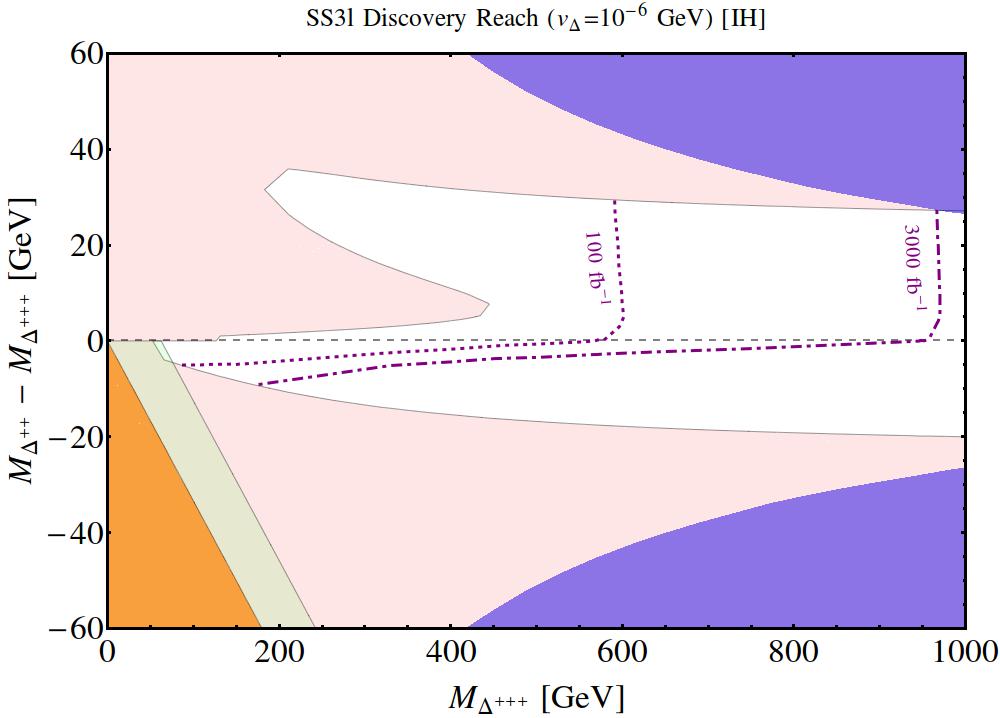}
\caption{ Discovery reach ($5 \sigma$) of $\Delta^{\pm \pm \pm}$ at the LHC at $\sqrt{s} = 14$ TeV for integrated luminosities 100 fb$^{-1}$ and 3000 fb$^{-1}$. We show mass reach for both NH [Left] and IH [Right] of neutrino masses.$v_{\Delta}$ is set at $10^{-6}$ GeV to ensure BR$(\Delta^{\pm \pm \pm} \rightarrow l^{\pm} l^{\pm} W^{\pm}) =1$ for $\Delta M > 0$. The other colored regions has the same meaning as Fig.~\ref{fig:EWPT}.}
\label{fig:H+++_llW}
\end{figure}

In Fig.~\ref{fig:H+++_llW} we present $5 \sigma$ discovery reaches of $\Delta^{\pm \pm \pm}$ at $\sqrt{s} = 14$ TeV for integrated luminosities 100 fb$^{-1}$ and 3000 fb$^{-1}$. We show mass reach for both NH [Left panel] and IH [Right panel] of neutrino masses for $v_{\Delta} = 10^{-6}$ GeV. Also, for this value of $v_{\Delta}$, $c \tau_{\Delta^{\pm \pm \pm}} \lesssim 100 \, \mu$m is definitely satisfied (cf. Fig.~\ref{fig:H+++_pdl}). The difference in mass reaches for NH and IH are minimal. We find that at $5 \sigma$ level $M_{\Delta^{\pm \pm \pm}}$ can be probed upto $\sim 600$ GeV for 100 fb$^{-1}$, and $\sim 950$ GeV with 3000 fb$^{-1}$.
  
\begin{figure}[!htp]
\includegraphics[scale=0.2]{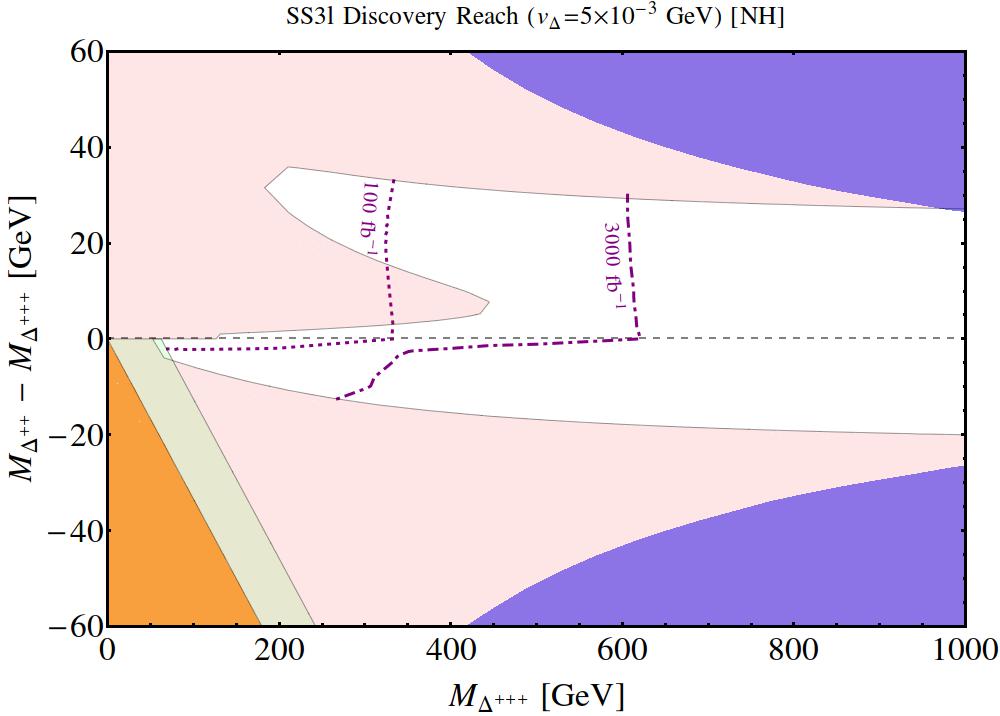}
\includegraphics[scale=0.2]{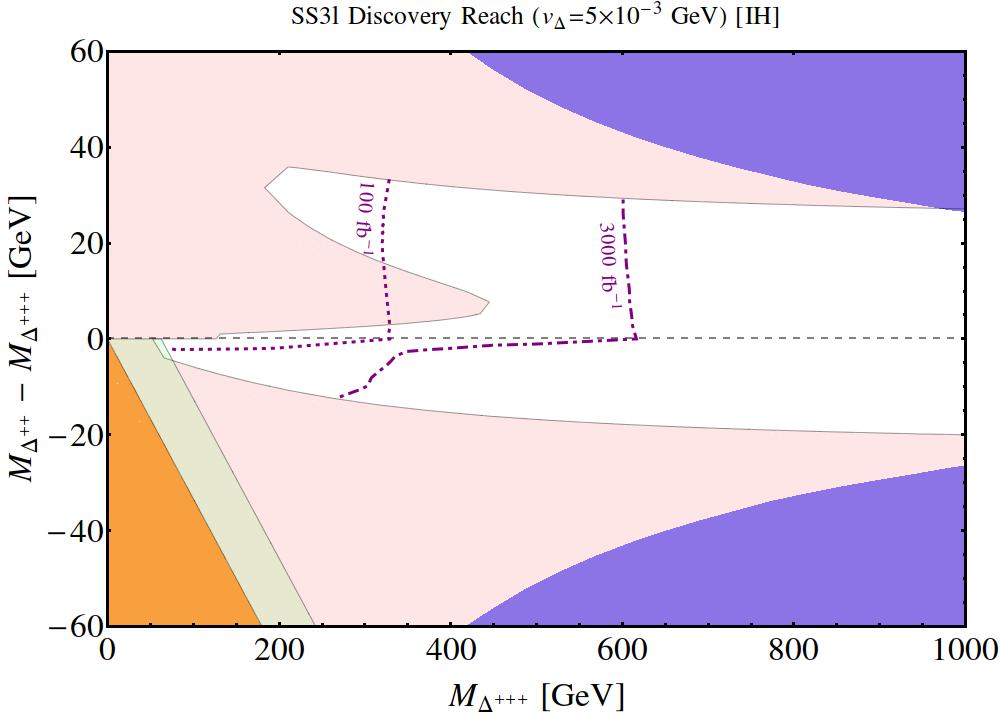}
\caption{ Discovery reach ($5 \sigma$) of $\Delta^{\pm \pm \pm}$ at the LHC at $\sqrt{s} = 14$ TeV for integrated luminosities 100 fb$^{-1}$ and 3000 fb$^{-1}$. We show mass reach for both NH [Left] and IH [Right] of neutrino masses.$v_{\Delta}$ is set at $5 \times 10^{-3}$ GeV to ensure BR$(\Delta^{\pm \pm \pm} \rightarrow W^{\pm} W^{\pm} W^{\pm}) =1$ for $\Delta M > 0$. The other colored regions has the same meaning as Fig.~\ref{fig:EWPT}.}
\label{fig:H+++_WWW}
\end{figure}

Fig.~\ref{fig:H+++_WWW} is the same as Fig.~\ref{fig:H+++_llW} but for $v_{\Delta} = 5 \times 10^{-3}$ GeV that simultaneously ensures  BR$(\Delta^{\pm \pm \pm} \rightarrow W^{\pm} W^{\pm} W^{\pm}) =1$ for $\Delta M > 0$, and $c \tau_{\Delta^{\pm \pm \pm}} \lesssim 100 \, \mu$m (cf. Fig.~\ref{fig:H+++_pdl}). The discovery potentials of $M_{\Delta^{\pm \pm \pm}}$ at the LHC are $\sim 325$ GeV and $\sim 600$ GeV with 100 fb$^{-1}$ and 3000 fb$^{-1}$ of integrated luminosities, respectively. We don't show a separate plot for BR$(\Delta^{\pm \pm \pm} \rightarrow l^{\pm} l^{\pm} W^{\pm})=$BR$(\Delta^{\pm \pm \pm} \rightarrow W^{\pm} W^{\pm} W^{\pm}) = 0.5$ cases as most of the parameter space that can be probed at 100 fb$^{-1}$ will possess  $c \tau_{\Delta^{\pm \pm \pm}} \gtrsim 100 \, \mu$m and will not respond to our prompt lepton search strategy. Nonetheless, 3000 fb$^{-1}$ of integrated luminosity will offer a discovery reach for $M_{\Delta^{\pm \pm \pm}} \sim 500-900$ GeV for $\Delta M \geq 0$. One important point to notice is that we cover the entire $\Delta M \geq 0$ range allowed by EWPT in all cases.   

One common feature of both Fig.~\ref{fig:H+++_llW} and Fig.~\ref{fig:H+++_WWW} is that our SS $3l$ search strategy is sensitive to a mass-splitting of $\lesssim 10$ GeV when $\Delta^{\pm \pm \pm}$ is the heaviest member of the quadruplet. In those scenarios cascade decay of $\Delta^{\pm \pm \pm}$ will give rise to soft leptons that won't pass through our strong lepton $p_T$ cuts. A dedicated study is needed with boosted topologies for this kind of mass spectra, similar in flavor to compressed supersymmetric spectra studies~\cite{CompressedSUSY}. One might use the use of Bayesian optimization techniques, as recently outlined in Ref.~\cite{Alves:2017ued}, for a systematic study of compressed spectra. 

Finally, a comment is in order to distinguish NH and IH scenarios. The best way to distinguish them is to probe different flavor combinational leptonic decay channels of $\Delta^{\pm \pm}$. We refer the reader to Ref.~\cite{Perez} for a detailed study on this, also including the impact of Dirac and Majorana phases. However, as it is made clear earlier in our analysis of Section~\ref{sec:H++_bound}, any search of $\Delta^{\pm \pm}$ is futile for $\Delta M \gtrsim 5$ GeV in the context of this model. Our SS $3l$ search of this section is, on the other hand, can probe all $\Delta M \geq 0$ mass spectra but the total signal yield for both NH and IH are very similar. For example, for our chosen BP of $(M_{\Delta^{\pm \pm \pm}},\Delta M, v_{\Delta})= (400,0,10^{-6})$ GeV, we expect to produce 119 and 111 events at 100 fb$^{-1}$ for NH and IH. 
However, one needs to classify these SS $3l$ events in different lepton flavor combinations to compare NH and IH more meticulously. A comparison, in that spirit, is presented in Table~\ref{tab:NHvIH} for the above BP. The experimentally measured neutrino mixing angles implies that the heaviest neutrino mass state contains a tiny fraction of $\nu_e$ for NH. Thus, one would expect very few events involving $e$ compared to $\mu$, as reflected in Table~\ref{tab:NHvIH}. In contrast, for IH the more massive neutrino mass states have large $\nu_e$ and $\nu_{\mu}$ components. Therefore, a comparable number of $e$ and $\mu$ events are expected in this case, which can again be noticed from Table~\ref{tab:NHvIH}. Although the lepton flavor combinations of SS $3l$ final state events are more or less reflective of neutrino mixing hierarchies, one should also keep in mind that $e, \mu$ identification efficiencies, and energy resolutions differ, but they are expected to have minimal impact on our analysis due to strong $p_T$ cuts used.

\begin{table}[!htp]
\begin{tabular}{|c|c |c |c |c| c|}
\hline
SS $3l$ & $eee$ & $ee \mu$ & $e \mu \mu$ & $\mu \mu \mu$  & Total events
\\
\hline
\hline
NH & 1 & 9 & 62 & 47 & 119
\\
\hline
IH & 31 & 54 & 14 & 12 & 111
\\
\hline
\end{tabular}
\caption{Neutrino mass hierarchy dependency in SS $3l$ signal in llW dominant region.}
\label{tab:NHvIH}
\end{table}

\section{Conclusions}
\label{sec:conclusions}

We study various phenomenological implications of a dimension-7 neutrino mass generation mechanism, as proposed in the BNT model~\cite{Babu:2009aq}, in this paper. The model contains an isospin 3/2 scalar quadruplet ($\Delta$) and two vector-like iso-triplet leptons ($\Sigma_{1, 2}$), in addition to the SM field content. We reiterate the claim of Ref.~\cite{Babu:2009aq} that one can get light neutrino masses, consistent with observed oscillation parameters, with $\mathcal{O}$(TeV) scale new physics. Although the dimension-7 operator develops neutrino masses at tree level, the model can not prevent dimension-5 operator contributions to the same at loop level. In fact, one needs to set $M_{\Sigma} \lesssim 1$ TeV to probe dimension-7 operator contribution explicitly but such choice of parameters lead us to a very computationally expensive regime without any new insight into the Higgs sector of the model. Hence, we integrate out $\Sigma_{1,2}$ by setting $M_{\Sigma} = 5$ TeV and work with the resulting effective Lagrangian. For this choice of $M_{\Sigma}$, $(m_{\nu})_{ij}^{\text{loop}}$ and $(m_{\nu})_{ij}^{\text{tree}}$ are comparable for the range of $M_{\Delta}$ accessible to the ongoing run of the LHC. Loop contributions will dominate for higher values of $M_{\Sigma}$.

One novel feature of our paper is a high precision electroweak study of the model. It is well known that the EW $\rho$ parameter constrains the induced VEV obtained by the quadruplet, $v_{\Delta} \lesssim 1$ GeV. However, we probe the model more closely for its contribution to the oblique parameters and estimate the impact of them on quadruplet mass spectrum. Over the range of $M_{\Delta}$ that is accessible to the LHC, the most robust constraint comes from the $T$ parameter, which is controlled by the mass-splitting, $\Delta M$, between the quadruplet members. We find that EWPT limits $\Delta M \lesssim 30$ GeV, which in turn give rise to compressed spectra over a vast area of the parameter space. Due to the softness of decay products in a compressed scenario, a significant part of the parameter space will remain beyond the reach of the LHC, when $\Delta M < 0$. 

Next, we investigate the unique signatures of the model at the LHC. The presence of multi-charged scalars of the model can potentially enhance or suppress the $h \rightarrow \gamma \gamma$ decay rate depending on the sign of the coupling. Using 36 fb$^{-1}$ data from both CMS and ATLAS we deduce that $h \rightarrow \gamma \gamma$ can exclude regions parameter space not ruled out by EWPT, albeit for $\mathcal{O}(1)$ values of $\lambda_3$. For smaller $\lambda_3$ it does not add anything to EWPT. We also consider the bounds form $\mu \rightarrow e \gamma$ LFV process on our parameter space and derive a lower bound on $v_{\Delta} \sim \mathcal{O}$(1 eV) form $M_{\Delta^{\pm \pm}} \lesssim 1$ TeV. Mass-splitting between $\Delta$ components, or ordering of neutrino masses, has a negligible impact on the above limit.   

We also examine the BRs and proper decay lengths of $\Delta^{\pm \pm}$ and $\Delta^{\pm \pm}$ in detail, along with their consequences at the LHC, for the whole range of $\Delta M$ allowed by EWPT. We find that for $\Delta^{\pm \pm}$ cascade decays start to dominate for $\Delta M \sim 2-20 $ GeV for both signs of $\Delta M$. In contrast, for $\Delta^{\pm \pm \pm}$ no cascade decay is available when $\Delta M \geq 0$, but it always decays in cascade for $\Delta M < 0$. A large $c \tau$ is achievable for both $\Delta^{\pm \pm}$ and $\Delta^{\pm \pm \pm}$ when $v_{\Delta} \sim 10^{-5} - 10^{-4}$ GeV and $\Delta M \sim 0$. In this region the leptonic and gauge bosonic decay rate of $\Delta^{\pm \pm}$ are comparable, and $c \tau$ can be as large as $10 \, \mu$m, which is still within the realm of prompt lepton searches at the LHC. Similarly, for $\Delta^{\pm \pm \pm}$ a transition from $llW$ dominated decay to $WWW$ dominated decay happens around that region and $c \tau \gtrsim 100 \, \mu$m is feasible for $M_{\Delta^{\pm \pm \pm}} \lesssim 500$ GeV, as a result force this region to be insensitive to prompt lepton searches at the LHC. However, when cascade decay opens up proper decay length increases rapidly, and brings $\Delta^{\pm \pm \pm}$ within the reach of the LHC.

A strong bound on $M_{\Delta^{\pm \pm}}$ can be derived from $3l$ and $4l$ searches performed by the CMS collaboration with 12.9 fb$^{-1}$ data. The strongest bounds are obtained when BR$(\Delta^{\pm \pm} \rightarrow l^{\pm} l^{\pm}) =1$, which we ensure by setting $v_{\Delta}= 10^{-6}$ GeV. We extract the limits from the leptonic decay channel that provides the best sensitivity for a particular ordering of neutrino masses, which is $\mu \mu$ for NH and $ee$ for IH for our chosen neutrino mass and mixing benchmark values. Using the CMS $3l$ analysis we constrain $M_{\Delta^{\pm \pm}} \gtrsim 600$ GeV. 
The limits on $\Delta^{\pm \pm}$ mass falls sharply as BR$(\Delta^{\pm \pm} \rightarrow l^{\pm} l^{\pm})$ deviates from 1. Moreover, the above bounds are sensitive for $|\Delta M| < 5$ GeV only.  

Finally, we perform a feasibility study to examine the discovery reach of $\Delta^{\pm \pm \pm}$  at the LHC. A search for $\Delta^{\pm \pm \pm}$ is necessary, independent of $\Delta^{\pm \pm }$ searches conducted by the LHC experiments, to validate the BNT model, as $\Delta^{\pm \pm}$ is not unique to this model. Furthermore, the LHC multi-lepton searches for $\Delta^{\pm \pm}$ is not sensitive for large mass-gap. In contrast, a direct search for $\Delta^{\pm \pm \pm}$ can cover the whole range of $\Delta M $, allowed by EWPT for $\Delta M > 0$. A simple set of cuts, led by hard cuts on $p_T$ of leptons, is sufficient to isolate SS $3l$ signature that can arise from $\Delta^{\pm \pm \pm}$ decay. With 3 ab$^{-1}$ of integrated luminosity, the LHC can discover $\Delta^{\pm \pm \pm}$ for a mass up to 950 (600) GeV in the $llW \, (WWW)$ decay dominant regions for both NH and IH of neutrino masses.

Nevertheless, the search strategy used in our analysis will not be effective for $\Delta M < 0$ scenarios. In these cases $\Delta^{\pm \pm \pm}$ will predominantly decay via cascade and the decay products will not pass hard lepton $p_T$ cuts we used here. A dedicated analysis is needed to probe such a mass spectra in the flavor of compressed SUSY spectra studies.


\begin{acknowledgments}
We thank Kaladi Babu, Teruki Kamon, Luca Pernie, Bhupal Dev and Xerxes Tata for useful comments. TG has been supported in part by the United
States Department of Energy Grant Number de-sc 0016013, and is now supported by NSF CAREER grant PHY-1250573. SN and SJ have been  supported in part by the US Department of Energy Grant No. de-sc 0016013. The work of SJ is also supported in part by the Fermilab Distinguished Scholars Program. SJ thanks the Fermilab Theoretical Physics Department for warm hospitality during the completion of this work.

\end{acknowledgments}

\section*{Appendix}
\appendix

\section{Expansion of Lagrangians in tensor notation}
\label{sec:AA}

The field $\Delta$ has component fields: $\Delta = (\Delta^{+++}, \Delta^{++}, \Delta^{+}, \Delta^0)^T$. In tensor notation $\Delta$ is a total symmetric tensor $\Delta_{ijk}$, with three indices $i,j,k$ taking values 1 and 2. Therefore, we can write various components of $\Delta$ as 
\beq
\Delta_{111}=\Delta^{+++}, \, \, \,\Delta_{112}=\dfrac{\Delta^{++}}{\sqrt{3}}, \, \, \,\Delta_{122}=\dfrac{\Delta^{+}}{\sqrt{3}},\, \, \, \Delta_{222}=\Delta^{0}.   
\label{Delta_tensor}
\eeq 
$\Sigma_{1,2}$ are symmetric tensors alike, with two indices, and they can be written in tensor notation as 
\beq
\Sigma_{i_{11}}=\Sigma^{++}_i, \, \, \,\Sigma_{i_{12}}=\dfrac{\Sigma^{+}_i}{\sqrt{2}}, \, \, \,\Sigma_{i_{22}}= \Sigma^0_i \, \, \, \, \, \, \, (i=1,2).   
\label{Sigma_tensor}
\eeq 
Hence, in terms of component fields  the last term of the scalar potential of Eq.~\ref{V_H_Delta} is given by
\beq
H^3\Delta^* = H_a H_b H_c \Delta^{*^{abc}} = \phi^{+^3} \Delta^{---} + 3 \, \phi^{+^2} \phi^0 \,  \dfrac{ \Delta^{--}}{\sqrt{3}} + 3 \, \phi^{+} \phi^{0^2} \, \dfrac{ \Delta^{-}}{\sqrt{3}} + \phi^{0^3} \Delta^{0}.
\eeq
Similarly, the Yukawa terms in Eq.~\ref{Lsig} can be expanded as
\bea
\overline{{L_{iL}}^c} H^* \Sigma_1 \, = \, \big(\overline{{L_{iL}}^c}\big)_a \, H^{*^b} \Sigma_{1_{bc}} \epsilon^{ac} &  \, = \, & \overline{{\nu_{iL}}^c} \, \bigg( \phi^- \dfrac{\Sigma^+_1}{\sqrt{2}} + \phi^{0^*} \Sigma^0_1  \bigg) - \overline{{l^-_{iL}}^{c}} \, \bigg( \phi^-  \Sigma^{++}_1 + \phi^{0^*} \dfrac{\Sigma^+_1}{\sqrt{2}}  \bigg),
\\ \nn
\overline{\Sigma_2} \Delta L_{iL} \, = \, \overline{\Sigma_2}^{ab} \Delta_{abc} \big(L_{iL}\big)_d \, \epsilon^{cd} & \, = \, & \bigg(\Sigma^{--}_2 \Delta^{+++}+ 2 \dfrac{\Sigma^-_2}{\sqrt{2}} \dfrac{\Delta^{++}}{\sqrt{3}}+\Sigma^{0^*}_2 \dfrac{\Delta^+}{\sqrt{3}}\bigg) \, l^-_{iL} 
\\ 
& & \, \, \, \, \, \, \,- \, \bigg(\Sigma^{--}_2 \dfrac{\Delta^{++}}{\sqrt{3}} + 2 \dfrac{\Sigma^-_2}{\sqrt{2}} \dfrac{\Delta^{+}}{\sqrt{3}} + \Sigma^{0^*}_2 \Delta^0 \bigg) \, \nu_{iL},
\eea
where $\epsilon^{ab} = \begin{pmatrix}
0 & 1 \\
-1 & 0
\end{pmatrix}$, is a totally anti-symmetric tensor. Finally, we present the effective Lagrangian of Eq.~\ref{Leff} in terms of component fields 
\bea
\overline{{L_{iL}}^c} L_{jL} H^* \Delta & = & \big( \overline{{L_{iL}}^c}\big)_a {L_{jL}}_{a'} {H^*}^b \Delta_{bcd} \epsilon^{ac} \epsilon^{a'd} 
\nn \\ 
 & = &  \overline{{\nu_{iL}}^c} \nu_{jL} \, \bigg( \phi^- \dfrac{\Delta^+}{\sqrt{3}} + {\phi^0}^* \Delta^0 \bigg) - \overline{{l^-_{iL}}^{c}} \nu_{jL} \, \bigg( \phi^- \dfrac{\Delta^{++}}{\sqrt{3}} + {\phi^0}^* \dfrac{\Delta^+}{\sqrt{3}} \bigg) 
\nn \\
 & - & \overline{{\nu_{iL}}^c} l^-_{jL} \, \bigg( \phi^- \dfrac{\Delta^{++}}{\sqrt{3}} + {\phi^0}^* \dfrac{\Delta^+}{\sqrt{3}} \bigg) + \overline{{l^-_{iL}}^{c}} l^-_{jL} \, \bigg( \phi^- \Delta^{+++} + {\phi^0}^* \dfrac{\Delta^{++}}{\sqrt{3}} \bigg). 
\eea

\section{Feynman Rules relevant for $\Delta^{\pm \pm}$ and $\Delta^{\pm \pm \pm}$ interactions}
\label{sec:FR}
The couplings relevant for production and decay of doubly- and triply- charged scalars are shown in Table~\ref{tab:FR}. A factor 2 included whenever 2 identical particles are in the vertex. Two such examples are $\Delta^{\pm\pm} W^{\mp} W^{\mp}$ and $\Delta^{\pm\pm} l^{\mp}_{i} l^{\mp}_{j}$ (for $i=j$). Also, we consider the $CP$-violating phases of the PMNS matrix to be 0 for our BPs. Hence, for our study $(m_{\nu})_{ij}^{\text{tot}}=(m_{\nu})_{ji}^{\text{tot}}$, and a factor of 2  is included for $\Delta^{\pm\pm} l^{\mp}_{i} l^{\mp}_{j}$ (for $i \neq j$) as well.

\begin{table}[!htp]
 \renewcommand{\arraystretch}{0.8}
\begin{center}
\begin{tabular}{|c|c|}
\hline 
Vertex & Couplings \\ 
\hline 
$A^{\mu} \Delta^{\pm\pm\pm} \Delta^{\mp\mp\mp}$ & $-3ie(p_{2} -p_{3})_{\mu} $ \\ 
$A^{\mu} \Delta^{\pm\pm} \Delta^{\mp\mp}$ & $-2ie(p_{2} - p_{3})_{\mu} $ \\ 
$Z^{\mu} \Delta^{\pm\pm\pm} \Delta^{\mp\mp\mp}$ & $-\dfrac{3ie \cos{2 \theta_{w}}}{\sin{2 \theta_{w}}}(p_{2} - p_{3})_{\mu} $\\ 
$Z^{\mu} \Delta^{\pm\pm} \Delta^{\mp\mp}$ & $-\dfrac{2 i e (\cos{2 \theta_{w}} - 1/2)}{\sin{2 \theta_{w}}}(p_{2} - p_{3})_{\mu} $\\ 
$W^{\mu\mp} \Delta^{\mp\mp} \Delta^{\pm\pm\pm} $ & $-i\sqrt{\dfrac{3}{2}} g (p_{2} - p_{3})_{\mu} $ \\ 
$W^{\mu\mp} \Delta^{\mp} \Delta^{\pm\pm} $ & $-i\sqrt{{2}} g (p_{2} - p_{3})_{\mu} $ \\
$\Delta^{\pm\pm} W^{\mp} W^{\mp}$ & $\sqrt{6} g^{2}v_{\Delta} g_{\mu \nu}$ \\ 
$\Delta^{\pm\pm} l^{\mp}_{i} l^{\mp}_{j}$ & $\dfrac{2}{ \sqrt{6}} \dfrac{ (m_{\nu})_{ij}^{\text{tot}}}{D}$ \\ 
\hline 
\end{tabular} 
\end{center}
\caption{Feynman Rules relevant for production and decay of doubly and triply-charged scalars of the BNT Model. Here $p_i$ stands for the 4-momentum of the i-th particle at the vertex, with the convention that all the particle momenta are coming into the vertex. For brevity $\cos{2\theta_W} \, (\sin 2 \theta_W)$ has been abbreviated as $c_{2W} \, (s_{2W})$. In the last interaction $D$ is given in Eq.~\ref{deno}}.
\label{tab:FR}
\end{table}



\end{document}